\documentclass[reqno,draft]{amsart}
\usepackage{amssymb}
\usepackage{upref}

\newcommand{\bbN}{{\mathbb{N}}}
\newcommand{\bbR}{{\mathbb{R}}}

\newcommand{\bbZ}{{\mathbb{Z}}}
\newcommand{\bbC}{{\mathbb{C}}}

\newcommand{\calB}{{\mathcal B}}
\newcommand{\calC}{{\mathcal C}}
\newcommand{\calD}{{\mathcal D}}
\newcommand{\calE}{{\mathcal E}}
\newcommand{\calF}{{\mathcal F}}

\newcommand{\calK}{{\mathcal K}}
\newcommand{\calL}{{\mathcal L}}
\newcommand{\calM}{{\mathcal M}}

\newcommand{\calU}{{\mathcal U}}
\newcommand{\calV}{{\mathcal V}}

\newcommand{\Green}{{\mathcal G}}

\newcommand{\g}{n}
\newcommand{\dott}{\,\cdot\,}
\newcommand{\hatt}{\widehat}  
\newcommand{\Pinf}{P_\infty}
\newcommand{\Div}{\operatorname{Div}}
\newcommand{\mini}{\wedge}
\newcommand{\maxi}{\vee}
\newcommand{\no}{\nonumber}
\newcommand{\lb}{\label}
\newcommand{\f}{\frac}
\newcommand{\ul}{\underline}

\newcommand{\ti}{\widetilde} 
\newcommand{\Oh}{O}

\newcommand{\humu}{{ \hat{\underline{\mu} }}}
\newcommand{\hunu}{{\underline{\hat{\nu}}}}
\newcommand{\hmu}{{\hat{\mu} }}

\newcommand{\hnu}{{\hat{\nu}}}
\newcommand{\hulam}{{ \hat{\underline{\lambda} }}}
\newcommand{\hlam}{{\hat{\lambda} }}
\newcommand{\huunu}{\underline{\underline{\hat{\nu}}}}
\newcommand{\uz}{{\underline{z}}}
\newcommand{\umu}{{\underline{\mu}}}
\newcommand{\uxi}{{\underline{\Xi}}}
\newcommand{\al}{\alpha}
\newcommand{\ual}{{\underline{\alpha}}}
\newcommand{\ua}{{\underline{A}}}
\newcommand{\hua}{{ \underline{\hatt{A} }}}
\newcommand{\Pinfp}{{P_{\infty_+}}}
\newcommand{\Pinfm}{{P_{\infty_-}}}

\newcommand{\Pinfpm}{{P_{\infty_\pm}}}
\newcommand{\Pzeropm}{{P_{0_\pm}}}

\newcommand{\res}{\operatornamewithlimits{res}}

\renewcommand{\Im}{\text{\rm Im}}

\DeclareMathOperator{\CH}{CH}
\DeclareMathOperator{\sCH}{s-CH}
\DeclareMathOperator{\shCH}{s-\hatt {CH}}
\allowdisplaybreaks
\numberwithin{equation}{section}

\newtheorem{theorem}{Theorem}[section]
\newtheorem{lemma}[theorem]{Lemma}

\newtheorem{hypothesis}[theorem]{Hypothesis}
\theoremstyle{definition}

\newtheorem{example}[theorem]{Example}

\theoremstyle{remark}
\newtheorem{remark}[theorem]{Remark}

\newcommand{\abs}[1]{\lvert#1\rvert}

\begin{document}
\title[The Camassa--Holm hierarchy]{Algebro-Geometric Solutions of the \\
 Camassa--Holm hierarchy}
\author[F.\ Gesztesy]{Fritz Gesztesy}
\address{Department of Mathematics,
University of Missouri,
Columbia, MO 65211, USA}
\email{fritz@math.missouri.edu}
\urladdr{http://www.math.missouri.edu/people/fgesztesy.html}
\author[H.\ Holden]{Helge Holden}
\address{Department of Mathematical Sciences,
Norwegian University of
Science and Technology, NO--7491 Trondheim, Norway}
\email{holden@math.ntnu.no}
\urladdr{http://www.math.ntnu.no/\~{}holden/}
\thanks{Research supported in part by the Research Council of Norway.}
\date{April 29, 2001}

\begin{abstract}
We provide a detailed treatment of the Camassa--Holm (CH) hierarchy with
special emphasis on its algebro-geometric solutions. In analogy to other
completely integrable hierarchies of soliton equations such as the KdV or
AKNS hierarchies, the CH hierarchy is recursively constructed by means of
basic polynomial formalism invoking a spectral parameter. Moreover, we 
study Dubrovin-type equations for auxiliary divisors and associated trace
formulas, consider the corresponding algebro-geometric initial value
problem, and derive the theta function representations of
algebro-geometric solutions of the CH hierarchy.
\end{abstract}

\maketitle

%
%
\section{Introduction}\lb{chs1}

Very recently, the Camassa--Holm (CH) equation, also known as the
dispersive shallow water equation, as isolated, for instance, in
\cite{CamassaHolm:1993} and \cite{CamassaHolmHyman:1994}, 
\begin{equation}
4u_t-u_{xxt}-2u u_{xxx}-4u_x u_{xx}+24u u_x=0, \quad (x,t)\in\bbR^2
\lb{ch1.1}
\end{equation}
(chosing a scaling of $x,t$ that's convenient for our purpose),
with $u$ representing the fluid velocity in $x$-direction, received
considerable attention. Actually, \eqref{ch1.1} represents the limiting
case $\kappa\to 0$ of the general Camassa--Holm equation,
\begin{equation}
4v_t-v_{xxt}-2v v_{xxx}-4v_x v_{xx}+24v v_x+4\kappa v_x=0, \quad
\kappa\in\bbR, \; (x,t)\in\bbR^2. \lb{ch1.2}
\end{equation}
However, in our formalism the general Cammassa--Holm equation \eqref{ch1.2}
just represents a linear combination of the first two equations in the CH
hierarchy and hence we consider without loss of generality \eqref{ch1.1} as
the first nontrivial element of the Camassa--Holm hierarchy. Alternatively,
one can transform 
\begin{equation}
v(x,t) \mapsto u(x,t)=v(x-(\kappa/2)t,t)+(\kappa/4) \lb{ch1.3}
\end{equation}
and thereby reduce \eqref{ch1.2} to \eqref{ch1.1}.

Various aspects of local existence, global existence, and uniqueness of 
solutions of \eqref{ch1.1} are treated in \cite{ConstantinEscher:1998},
\cite{ConstantinEscher2:1998}, \cite{ConstantinEscher3:1998},
\cite{ConstantinMolinet:2000}, \cite{MarsdenRatiuShkoller:2000},
\cite{MarsdenShkoller:2000}, \cite{Shkoller:1998}, \cite{Shkoller:2000},
\cite{XinZhang:2000},  wave breaking phenomena are discussed in
\cite{Constantin:2000}, \cite{ConstantinEscher1:1998}, \cite{ConstantinEscher:2000}.
Soliton-type solutions (called ``peakons'') were extensively studied due to 
their unusual non-meromorphic (peak-type) behavior, which features a
discontinuity in the $x$-derivative of $u$ with existing left and right
derivatives of opposite sign at the peak. In this context we refer,
for instance, to \cite{AlberCamassaFedorovHolmMarsden:2001}, 
\cite{AlberCamassaHolmMarsden:1994},
\cite{AlberFedorov:2000}, \cite{AlberFedorov:2001},
\cite{AlberMiller:2001}, \cite{BealsSattingerSzmigielski:1999},
\cite{BealsSattingerSzmigielski:2000},
\cite{BealsSattingerSzmigielski:2001}, \cite{CamassaHolm:1993},
\cite{CamassaHolmHyman:1994}. Integrability aspects such as infinitely
many conservation laws, (bi-)Hamiltonian formalism, B\"acklund
transformations, infinite dimensional symmetry groups, etc., are
discussed, for instance,  in \cite{CamassaHolm:1993},
\cite{CamassaHolmHyman:1994}, \cite{FisherSchiff:1999}, 
\cite{Fuchssteiner:1996} (see also \cite{FuchssteinerFokas:1981}),
\cite{Schiff:1996}. The general CH equation \eqref{ch1.2} is shown to give 
rise to a geodesic flow of a certain right invariant metric on the Bott-Virasoro 
group in \cite{Misiolek:1998}. In the case $\kappa=0$, the CH equation \eqref{ch1.1} 
corresponds to the geodesic flow on the group of orientation preserving
diffeomorphisms of the  circle. This follows from the Lie-Poisson structure 
established in \cite{CamassaHolmHyman:1994} and is also remarked upon in
\cite{Misiolek:1998}. That the equations define a smooth vector field was first 
observed by Shkoller in the case of periodic \cite{Shkoller:1998} and Dirichlet
\cite{Shkoller:2000} boundary conditions, which directly leads to the
corresponding local existence theory. Scattering data and their evolution under  the
CH flow are determined in \cite{BealsSattingerSzmigielski:1998} and  intimate
relations with the classical moment problem and the finite Toda  lattice are worked
out in \cite{BealsSattingerSzmigielski:1999}, \cite{BealsSattingerSzmigielski:2000},
and \cite{BealsSattingerSzmigielski:2001}. The case of spatially periodic solutions,
the corresponding inverse spectral problem, isospectral classes of solutions, and
quasi-periodicity of solutions with respect to time are discussed in
\cite{Constantin:1997}, \cite{Constantin:1998a}, \cite{Constantin:1998}, and
\cite{ConstantinMcKean:1999}. Moreover, algebro-geometric solutions of \eqref{ch1.1}
and their properties are studied in \cite{Alber:2000},
\cite{AlberCamassaFedorovHolmMarsden:1999}, 
\cite{AlberCamassaFedorovHolmMarsden:2001}, \cite{AlberCamassaGekhtman:2000},
\cite{AlberCamassaHolmMarsden:1994}, \cite{AlberCamassaHolmMarsden:1995},
\cite{AlberFedorov:2000}, \cite{AlberFedorov:2001} (connections as well as
differences between the  latter references and our own approach to algebro-geometric
solutions will be outlined in the following paragraph). Moreover, even though the
following very recent developments  are not directly related to the principal topic
of this paper, they put the CH  equation in a broader context: In
\cite{DullinGottwaldHolm:2001}, a  basic integrable shallow water equation,
originally introduced in \cite{CamassaHolm:1993}, is analyzed in detail. It combines
the linear dispersion  of the KdV equation with the nonlinear/nonlocal dispersion of
the CH equation  and contains the KdV and CH equations (as well as an equation
studied by  Fornberg and Whitham \cite{FornbergWitham:1978}) as special limiting
cases.  Finally, the three-dimensional viscous Camassa-Holm equations, their
connection with the  Navier-Stokes equations, estimates for the Hausdorff and
fractal  dimension of the associated global attractor, and turbulence theory
according to Kolmogorov, Landau, and Lifshitz, are discussed in
\cite{FoiasHolmTiti:2001}.

Our own approach to algebro-geometric solutions of the CH hierarchy 
differs from the ones pursued in \cite{Alber:2000}, 
\cite{AlberCamassaFedorovHolmMarsden:1999},
\cite{AlberCamassaFedorovHolmMarsden:2001}, \cite{AlberCamassaGekhtman:2000},
\cite{AlberCamassaHolmMarsden:1994}, \cite{AlberCamassaHolmMarsden:1995},
\cite{AlberFedorov:2000}, \cite{AlberFedorov:2001} in several aspects and we will
outline some of the differences next. Following previous treatments of the 
KdV, AKNS, Toda, and Boussinesq hierarchies and the sine-Gordon and massive 
Thirring models (cf., e.g., \cite{BullaGesztesyHoldenTeschl:1997},
\cite{DicksonGesztesyUnterkofler1:2000}, \cite{DicksonGesztesyUnterkofler:2000}, 
\cite{EnolskiiGesztesyHolden:1999}, \cite{GesztesyHolden:1999}, 
\cite{GesztesyHolden:2000}, \cite{GesztesyHolden2:1997},
\cite{GesztesyHolden:1999b}, \cite{GesztesyRatnaseelan:1996},
\cite{GesztesyRatnaseelanTeschl:1996}), we develop a systematic polynomial
recursion formalism for the CH hierarchy and its algebro-geometric
solutions. In contrast to the treatments in 
\cite{AlberCamassaFedorovHolmMarsden:2001}, \cite{AlberFedorov:2000},
and \cite{AlberFedorov:2001}, we rely on a zero-curvature approach $U_t-V_x=[V,U]$
(as the compatibility requirement for the system $\Psi_x=U\Psi$, $\Psi_t=V\Psi$) as
opposed to their Lax formalism. However, we incorporate important features of the
recursion formalism developed in \cite{AlberCamassaHolmMarsden:1994} into our
zero-curvature approach. Our treatment is comprehensive and self-contained in the
sense that it includes Dubrovin-type equations for auxiliary divisors on the
associated compact  hyperelliptic curve, trace formulas, and theta function
represenations of solutions, the usual ingredients of such a formalism. Moreover,
while
\cite{AlberCamassaFedorovHolmMarsden:2001}, \cite{AlberFedorov:2000},
\cite{AlberFedorov:2001} focus on solutions of the  CH equation itself, we
simultaneously derive theta function formulas for  solutions of any equation of the
CH hierarchy. The key element in our formalism is the solution $\phi$ of a
Riccati-type equation associated with the zero-curvature representation of the CH
equation \eqref{ch1.1}. Roughly speaking, $\phi=-z\psi_2/\psi_1$, where
$\Psi=(\psi_1,\psi_2)^t$ and $z$ denotes a spectral parameter in $U$ and $V$ (cf.\
\eqref{ch2.40} for more details). $\phi$ is then used to introduce appropriate
auxiliary divisors on the underlying hyperelliptic curve, the Baker-Akhizer vector
in the stationary case, etc. Combining $\phi$ with the polynomial recursion
formalism for the CH hierarchy then leads to Dubrovin-type differential equations
and trace formulas for $u$ in terms of auxiliary divisors. Explicit theta function
representations for symmetric functions of (projections of) these auxiliary divisors
then yield the theta function representations for any algebro-geometric solution $u$
of the CH hierarchy. Here our strategy differs somewhat from that employed in
\cite{AlberCamassaFedorovHolmMarsden:2001}, \cite{AlberFedorov:2000},
\cite{AlberFedorov:2001} for the CH equation. While the latter references also
employ the trace formula for $u$ in terms of (projections of) auxiliary divisors,
they subsequently rely on generalized theta functions and generalized Jacobians
(going back to investigations of Clebsch and Gordan \cite{ClebschGordan:1866}),
whereas we stay within the traditional framework familar from the KdV, AKNS, Toda
hierarchies, etc. Finally, we point out a novel feature of our treatment of the CH
hierarchy that appears to be with out precedent. In Theorems \ref{theorem-ch3.10} and
\ref{theorem-ch4.10} we formulate and solve the algebro-geometric initial value
problem for the stationary and time-dependent CH hierarchy, in the following sense.
Starting from the initial value problem for auxiliary divisors induced by the
Dubrovin-type equations, we define $u$ in terms of the trace formula involving the
(projections of) auxiliary divisors and then prove directly that $u$ so defined
satisfies the corresponding (stationary, resp., time-dependent) equation of the CH
hierarchy.

Without going into further details, we note that our constructions extend in a
straightforward manner to a closely related hierarchy of completely integrable
nonlinear evolution equations, the Dym hierarchy. For different approaches to
algebro-geometric solutions of the latter we refer to
\cite{AlberCamassaFedorovHolmMarsden:2001}, \cite{AlberCamassaHolmMarsden:1995},
\cite{AlberLutherMiller:2000}, \cite{Dmitrieva:1993}, and \cite{Novikov:1999}.

In Section \ref{chs2} we develop the basic polynomial recursion formalism
that  defines the CH hierarchy using a zero-curvature approach. Section
\ref{chs3} then treats the stationary CH hierarchy and its
algebro-geometric solutions. The corresponding
time-dependent results are the subject of Section \ref{chs4}. Appendix
\ref{A} summarizes the necessary results needed from the theory of compact
Riemann surfaces and also serves to establish the notation used throughout
this paper. Appendix \ref{B} contains a few technical results concerning
the polynomial recursion formalism and associated high-energy expansions.
Finally, Appendix \ref{C} provides a detailed discussion of elementary
symmetric functions associated with Dirichlet divisors and their
representations in terms of theta functions associated with the underlying 
hyperelliptic curve. It contains several core results needed in our derivation of
algebro-geometric solutions of the CH hierarchy. The results of this appendix apply
to a  variety of soliton equations and hence are of independent interest.

%
%
\section[Fundamentals of the CH hierarchy]{The CH hierarchy,
recursion relations, and hyperelliptic curves}  \label{chs2}

In this section we provide the basic construction of a completely
integrable hierarchy of nonlinear evolution equations in which the
Camassa--Holm equation, or dispersive shallow water equation, is the first
element in the hierarchy (the higher-order CH equations will turn out to
be nonlocal with respect to $x$). We will use a zero-curvature approach and
combine it with a polynomial recursion formalism containing a spectral
parameter.

Throughout this section we will suppose the following hypothesis.
\begin{hypothesis}\lb{hypo-ch1.1}
In the stationary case we assume that
\begin{equation}
u\in C^\infty(\bbR), \; \f{d^m u}{dx^m}\in L^\infty(\bbR), \; 
m\in\bbN_{0}. \lb{ch2.1}
\end{equation}
In the time-dependent case we suppose 
\begin{align}
&u(\dott,t)\in C^\infty(\bbR), \; \f{\partial^m u}{\partial
x^m}(\dott,t)\in L^\infty(\bbR), \;  m\in\bbN_{0}, \, t\in\bbR, \no \\
&u(x,\dott), u_{xx}(x,\dott)\in C^1(\bbR), \; x\in\bbR. \lb{ch2.2}
\end{align}
\end{hypothesis}

We start by formulating the basic polynomial setup taken essentially
{}from \cite{AlberCamassaHolmMarsden:1994}.
One defines $\{f_\ell\}_{\ell\in\bbN_0}$ recursively by
\begin{align}
f_{0}&=1, \no \\
f_{\ell,x}&=-2\Green\big(2(4u-u_{xx})f_{\ell -1,x}
+(4u_{x}-u_{xxx})f_{\ell -1}  \big), \quad \ell\in\bbN, \lb{ch2.3} 
\end{align}
where $\Green$ is given by
\begin{equation}
\Green\colon L^\infty(\bbR)\to L^\infty(\bbR), \quad
(\Green v)(x)=\frac14 \int_{\bbR} dy\, e^{-2\abs{x-y}} v(y),
\quad x\in\bbR, \; v\in L^\infty(\bbR). \lb{ch2.4} 
\end{equation}
One observes that $\Green$ is the 
resolvent of minus the one-dimensional Laplacian at energy parameter
equal to $-4$, that is,
\begin{equation}
    \Green=\bigg(-\frac{d^2}{dx^2}+4\bigg)^{-1}. \lb{ch2.5} 
\end{equation}
The first coefficient reads
\begin{equation}
f_1=-2u+c_1,\lb{ch2.6} 
\end{equation}
where $c_1$ is an integration constant.  Subsequent coefficients
are non-local with respect to $u$. At each level a new integration
constant,  denoted by $c_{\ell}$, is introduced.
Moreover, we introduce coefficients  
$\{g_\ell\}_{\ell\in\bbN_0}$ and $\{h_\ell\}_{\ell\in\bbN_0}$ by
\begin{align}
g_\ell&=f_{\ell}+ \f12 f_{\ell,x}, \quad \ell\in\bbN_0, 
\lb{ch2.7} \\
h_{\ell} & = (4u-u_{xx})f_{\ell} - g_{\ell+1,x}, 
\quad \ell\in\bbN_{0}. \lb{ch2.8} 
\end{align}
Explicitly, one computes
\begin{align}
f_{0}&=1, \no \\
f_{1}&=-2u+c_{1}, \no \\
f_{2}&=2u^2+2\Green\big(u_{x}^2+8u^2\big)+c_1(-2 u)+c_2, 
\no \\
g_{0}&=1,\no \\
g_{1}&=-2u-u_{x}+c_{1}, \lb{ch2.9} \\
g_{2}&=2u^2+2uu_x+2\Green\big(u_x^2
+ u_x u_{xx}+8u u_x+8u^2\big) \no \\
& \quad +c_1(-2u-u_x)+c_2,\no
\\ h_{0}&=4u+2u_{x}, \no\\
h_{1}&=-2u_x^2-4u u_{x}-8u^2 \no \\
&\quad-2\Green\big(u_x u_{xxx}+u_{xx}^2+2u_x u_{xx}+8uu_{xx}
+8u_x^2+16uu_x \big) \no \\
& \quad +c_1(4u+2u_{x}), \, \text{  etc.}\no
\end{align}

For later use it is convenient also to introduce the corresponding
homogeneous coefficients $\hat f_\ell$, $\hat g_\ell$, and $\hat h_\ell$ defined
by the vanishing of the integration constants $c_k$, $k=1,\dots,\ell$, 
\begin{align}
 \hat f_0&=f_0=1, \quad  \hat{f}_\ell=f_\ell\big|_{c_k=0, \,
k=1,\dots,\ell}, 
\quad \ell\in\bbN, \lb{ch2.10} \\
\hat g_{0}&=g_{0}=1,\quad\hat{g}_\ell=g_\ell\big|_{c_k=0, \, k=1,\dots,\ell}, \quad \ell\in\bbN, 
 \lb{ch2.11} \\
\hat h_0&=h_0=(4u+2u_{x}), \quad 
\hat{h}_\ell=h_\ell\big|_{c_k=0, \, k=1,\dots,\ell}, 
\quad \ell\in\bbN. \lb{ch2.12}
\end{align}
Hence,
\begin{equation}
f_\ell=\sum_{k=0}^{\ell}c_{\ell-k}\hat{f}_{k},\quad
g_\ell=\sum_{k=0}^{\ell}c_{\ell-k}\hat{g}_{k},\quad
h_\ell=\sum_{k=0}^{\ell}c_{\ell-k}\hat{h}_{k},  \quad
\ell\in\bbN_0, \lb{ch2.13}
\end{equation}
defining
\begin{equation}
c_0=1. \lb{ch2.13a}
\end{equation}
Next, given Hypothesis \ref{hypo-ch1.1}, one introduces the $2\times 2$ 
matrix
\begin{align}
U(z,x)&=\begin{pmatrix}-1 &1\\
z^{-1}(4u(x)-u_{xx}(x)) &1 \end{pmatrix}, \quad 
x\in\bbR, \lb{ch2.14} 
\end{align}
and for each $n\in\bbN_{0}$ the following $2\times 2$  matrix $V_n$
\begin{equation}
V_{n}(z,x)=
\begin{pmatrix}-G_{n}(z,x)& F_{n}(z,x)\\
    z^{-1} H_{n}(z,x) &G_{n}(z,x)
\end{pmatrix}, \quad n\in\bbN_0, \, z\in\bbC\setminus\{0\}, 
\, x\in\bbR. \lb{ch2.15} 
\end{equation}

Postulating the zero-curvature condition
\begin{equation}
-V_{n,x}(z,x)+[U(z,x),V_n(z,x)]=0 \lb{ch2.16}
\end{equation}
one finds
\begin{align}
F_{n,x}(z,x)&=2 G_{n}(z,x)-2F_n(z,x), \lb{ch2.17} \\
 zG_{n,x}(z,x)&=(4u(x)-u_{xx}(x))F_{n}(z,x)-H_n(z,x), \lb{ch2.18}
\\ H_{n,x}(z,x)&=2H_{n}(z,x)-2(4u(x)-u_{xx}(x)) G_n(z,x).
\lb{ch2.19}
\end{align}
{}From \eqref{ch2.17}--\eqref{ch2.19} one infers that
\begin{equation}
\f{d}{dx}\det(V_n(z,x))=-\f{1}{z}\f{d}{dx}\bigg(zG_{n}(z,x)^2+F_n(z,x)
H_n(z,x)\bigg)=0, \lb{ch2.20}
\end{equation}
and hence
\begin{equation}
zG_{n}(z,x)^2+F_n(z,x) H_n(z,x)=Q_{2n+1}(z), \lb{ch2.21}
\end{equation}
where the polynomial $Q_{2n+1}$ of degree $2n+1$ is $x$-independent.
Actually it turns out that it is more convenient to define
\begin{equation} 
R_{2n+2}(z)=zQ_{2n+1}(z)=\prod_{m=0}^{2n+1}(z-E_m), \quad E_0=0, \, 
E_1,\dots,E_{2n+1} \in\bbC \lb{ch2.22}
\end{equation}
so that \eqref{ch2.21} becomes
\begin{equation}
z^2G_{n}(z,x)^2+zF_n(z,x) H_n(z,x)=R_{2n+2}(z). \lb{ch2.23}
\end{equation}

Next one makes the ansatz that $F_{n}$, $H_{n}$, and $G_{n}$ are
polynomials of degree $n$, related to the coefficients 
$f_\ell$, $h_\ell$, and $g_\ell$ by
\begin{align}
F_{n}(z,x)&=\sum_{\ell=0}^n f_{n-\ell}(x)z^\ell, \lb{ch2.24} \\
G_{n}(z,x)&=\sum_{\ell=0}^{n}g_{n-\ell}(x)z^\ell, \lb{ch2.25} \\
H_{n}(z,x)&=\sum_{\ell=0}^n h_{n-\ell}(x)z^\ell. \lb{ch2.26}
\end{align}
Insertion of
\eqref{ch2.24}--\eqref{ch2.26} into \eqref{ch2.17}--\eqref{ch2.19} then yields the 
recursion relations \eqref{ch2.3}--\eqref{ch2.4} and \eqref{ch2.7} for
$f_{\ell}$ and $g_{\ell}$ for $\ell=0,\dots,n$. In the case $n\in\bbN$ we 
obtain  the recursion \eqref{ch2.8} for $h_{\ell}$ for $\ell=0,\dots,n-1$ 
and 
\begin{equation}
    h_{n}=(4u-u_{xx})f_{n}. \lb{ch2.27}
\end{equation}
(When $n=0$ one gets directly $h_{0}=(4u-u_{xx})$.) Moreover,
taking $z=0$ in \eqref{ch2.23} yields
\begin{equation}
f_n(x)h_n(x)=\prod_{m=1}^{2n+1} E_m. \lb{ch2.27a} 
\end{equation}
In addition, one finds
\begin{equation}
    h_{n,x}(x)-2h_{n}(x)+2(4u(x)-u_{xx}(x))g_{n}(x)=0, 
    \quad n\in\bbN_{0}. \lb{ch2.28}
\end{equation}
Using the relations \eqref{ch2.27} and \eqref{ch2.7} permits one to write
\eqref{ch2.28} as
\begin{align}
\sCH_{n}(u)=(u_{xxx}-4u_{x})f_{n}-2(4u-u_{xx})f_{n,x}=0, 
    \quad n\in\bbN_{0}. \lb{ch2.29}
\end{align}
Varying $n\in\bbN_0$ in \eqref{ch2.29} then defines the stationary CH hierarchy. 
We record the first few equations explicitly,
\begin{align}
\sCH_0(u)&=u_{xxx}-4u_{x}=0, \no \\
\sCH_1(u)&=-2u u_{xxx}-4u_{x}u_{xx}+24u u_{x}+ c_{1}(u_{xxx}-4u_{x})=0,
\lb{ch2.30} \\
\sCH_2(u)&=2u^2 u_{xxx}-8u u_{x} u_{xx}-40u^2 u_{x} 
+2(u_{xxx}-4u_x)\Green\big(u_{x}^2+8u^2\big) \no \\ 
&\quad
-8(4u-u_{xx})\Green\big(u_{x} u_{xx}+8u u_{x}\big)
\no \\ &\quad +c_1(-2u u_{xxx}-4u_{x}u_{xx}+24u u_{x})
+c_{2}(u_{xxx}-4u_{x})=0,  
\, \text{  etc.} \no
\end{align}

By definition, the set of solutions of \eqref{ch2.29}, with $n$ ranging in
$\bbN_0$, represents the class of algebro-geometric CH
solutions. If $u$ satisfies one of the stationary CH equations in
\eqref{ch2.29} for a particular value of $n$, then it satisfies 
infinitely many
such equations of order higher than $n$ for certain choices of integration
constants $c_\ell$. At times it will be convenient to abbreviate
algebro-geometric stationary CH solutions $u$ simply as CH {\it
potentials}.

For later use we also introduce the corresponding homogeneous polynomials
$\hatt F_\ell$, $\hatt G_{\ell}$, and $\hatt H_\ell$ defined by 
\begin{align}
\hatt F_\ell(z)&=F_\ell(z)\big|_{c_k=0, \, k=1,\dots,\ell}= 
\sum_{k=0}^{\ell} \hat f_{\ell-k}z^k, \quad \ell=0,\dots,n, \lb{ch2.31} \\
\hatt G_{\ell}(z)&=G_{\ell}(z)\big|_{c_k=0, \, k=1,\dots,\ell}= 
\sum_{k=0}^{\ell} \hat g_{\ell-k}z^k, \quad \ell=0,\dots,n, \lb{ch2.32} \\
\hatt H_\ell(z)&=H_\ell(z)\big|_{c_k=0, \, k=1,\dots,\ell}= 
\sum_{k=0}^{\ell} \hat h_{\ell-k}z^k, \quad \ell=0,\dots,n-1, 
\lb{ch2.33} \\
\hatt H_n(z)&=(4u-u_{xx})\hat f_{n}
+\sum_{k=1}^{n} \hat h_{n-k}z^k. \lb{ch2.34}
\end{align}

In accordance with our notation introduced in
\eqref{ch2.10}--\eqref{ch2.12} and \eqref{ch2.31}--\eqref{ch2.34}, the
corresponding homogeneous stationary CH equations are then defined by
\begin{align}
\shCH_n (u)  =\sCH_n (u) \big|_{c_\ell=0,\ \ell=1,\dots,n}=0, 
\quad n\in\bbN_{0}. \label{ch2.35}
\end{align}

Using equations \eqref{ch2.17}--\eqref{ch2.19} one can also derive
individual differential equations for $F_n$ and $H_n$. Focusing on
$F_n$ only, one obtains
\begin{align}
&F_{n,xxx}(z,x)-4\big(z^{-1}(4u(x)-u_{xx}(x))+1\big)F_{n,x}(z,x)
\no \\
& -2z^{-1}(4u_{x}(x)-u_{xxx}(x))F_{n}(z,x)=0. \lb{ch2.37}    
\end{align}
This is of course consistent with \eqref{ch2.24} and \eqref{ch2.3} 
(applying $\Green^{-1}$ to \eqref{ch2.3}). Multiplying \eqref{ch2.37} 
with $F_n$ and integrating the result yields
\begin{equation}
F_{n,xx}F_n-2^{-1}F_{n,x}^2-2F_n^2-2z^{-1}(4u-u_{xx})F_n^2=C(z), 
\lb{ch2.38}
\end{equation}
for some $C(z)$, constant with respect to $x$. Differentiating
\eqref{ch2.17}, inserting \eqref{ch2.18} into the resulting equation, and
comparing with \eqref{ch2.17} and \eqref{ch2.23} then yields
\begin{equation}
C(z)=-2z^{-2}R_{2n+2}(z). \lb{ch2.39}
\end{equation}
Thus,\begin{align}
&-(z^2/2)F_{n,xx}(z,x)F_n(z,x)+(z^2/4)F_{n,x}(z,x)^2+z^2 F_n(z,x)^2 
\no \\
& +z(4u(x)-u_{xx}(x))F_n(z,x)^2=R_{2n+2}(z). \lb{ch2.39a}
\end{align}

Next, we turn to the time-dependent CH hierarchy. Introducing a    
deformation parameter $t_n\in\bbR$ into $u$ (i.e., replacing 
$u(x)$ by $u(x,t_n)$), the definitions \eqref{ch2.14}, 
\eqref{ch2.15}, and \eqref{ch2.24}--\eqref{ch2.26} of $U$, $V_n$, and 
$F_n$, $G_n$,
and $H_n$, respectively, still apply. The corresponding zero-curvature 
relation reads
\begin{equation}
U_{t_n}(z,x,t_n)-V_{n,x}(z,x,t_n)+[U(z,x,t_n),V_n(z,x,t_n)]=0, 
\quad n\in\bbN_0,\lb{ch2.40}
\end{equation}
which results in the following set of equations
\begin{align}
4u_{t_n}(x,t_n)-&u_{xxt_{n}}(x,t_n)-
H_{n,x}(z,x,t_{n})+2H_{n}(z,x,t_{n}) \no \\
-2(4u(x,t_n)&-u_{xx}(x,t_n))G_{n}(z,x,t_{n})=0, \lb{ch2.41} \\
F_{n,x}(z,x,t_n)&=2G_{n}(z,x,t_n)-2F_n(z,x,t_n), \lb{ch2.42} \\
zG_{n,x}(z,x,t_n)&=(4u(x,t_n)-u_{xx}(x,t_n))F_n(z,x,t_n)
-H_{n}(z,x,t_n).  \lb{ch2.43}
\end{align}
Inserting the polynomial expressions for $F_n$, $H_n$,
and $G_{n}$ into \eqref{ch2.42} and  \eqref{ch2.43}, respectively, first
yields  recursion  relations \eqref{ch2.3} and \eqref{ch2.7} for
$f_{\ell}$ and $g_{\ell}$ for $\ell=0,\dots,n$. In the case $n\in\bbN$ we 
obtain {}from \eqref{ch2.41} the recursion for 
$h_{\ell}$ for $\ell=0,\dots,n-1$ and 
\begin{equation}
    h_{n}=(u-u_{xx})f_{n}. \lb{ch2.44}
\end{equation}
(When $n=0$ one gets directly $h_{0}=(u-u_{xx})$.)
In addition, one finds
\begin{align}
(&4u_{t_n}(x,t_n)-u_{xxt_n}(x,t_n)-h_{n,x}(x,t_n)+2h_{n}(x,t_n)\no \\
& -2(4u(x,t_n)-u_{xx}(x,t_n))g_{n}(x,t_n)=0, 
    \quad n\in\bbN_{0}. \lb{ch2.45}
\end{align}
Using relations \eqref{ch2.27} and \eqref{ch2.44} permits one to write 
\eqref{ch2.45} as
\begin{align}
\CH_{n}(u)=4u_{t_n}-u_{xxt_n}+(u_{xxx}-4u_{x})f_{n}-2(4u-u_{xx})f_{n,x}=0,
\quad n\in\bbN_{0}. \lb{ch2.46}
\end{align}
Varying $n\in\bbN_0$ in \eqref{ch2.46} then defines the time-dependent 
CH hierarchy. We record the first few equations explicitly,
\begin{align}
\CH_0(u)&=4u_{t_{0}}-u_{xxt_{0}}+u_{xxx}-4u_{x}=0, \no \\
\CH_1(u)&=4u_{t_{1}}-u_{xxt_{1}} -2u u_{xxx}-4u_{x}u_{xx}
+24u u_{x}+c_{1}(u_{xxx}-4u_{x})=0, \no \\
\CH_2(u)&=4u_{t_{2}}-u_{xxt_{2}}+2u^2 u_{xxx}-8u u_{x} u_{xx}
-40u^2 u_{x} \lb{ch2.47} \\
& \quad +2(u_{xxx}-4u_x)\Green\big(u_{x}^2+8u^2\big) 
-8(4u-u_{xx})\Green\big(u_{x} u_{xx}+8u u_{x} \big) \no \\ 
&\quad +c_1(-2u u_{xxx}-4u_{x}u_{xx}+24u u_{x})
+c_{2}(u_{xxx}-4u_{x})=0, \, \text{  etc.} \no
\end{align}
Similarly, one introduces the corresponding homogeneous CH hierarchy by 
\begin{align}
&\widehat{\CH}_n(u)=\CH_n(u)\big|_{c_\ell=0,\, \ell=1,\dots,n}=0, 
\quad n\in\bbN_{0}. \lb{ch2.48}
\end{align}
Up to an inessential scaling of the $(x,t_1)$ variables,
$\widehat{\CH}_1(u)=0$ represents {\it the Camassa--Holm} equation as
discussed in \cite{CamassaHolm:1993}, \cite{CamassaHolmHyman:1994}. 

We note that our zero-curvature approach is similar (but not identical) to
that sketched in \cite{Schiff:1996}. This is in contrast to almost all
other treatments of the CH equation where a Lax equation approach appears
to be preferred.

Our recursion formalism was introduced under the assumption of a
sufficiently smooth function $u$ in Hypothesis \ref{hypo-ch1.1}. The
actual  existence of smooth global solutions of the initial value problem
associated with the CH hierarchy \eqref{ch2.47} is a nontrivial issue and
various aspects of it are discussed, for instance, in \cite{Constantin:2000},
\cite{ConstantinEscher2:1998}, \cite{ConstantinEscher3:1998},
\cite{ConstantinEscher1:1998}, \cite{ConstantinMolinet:2000},
\cite{MarsdenRatiuShkoller:2000}, \cite{MarsdenShkoller:2000}, 
\cite{Shkoller:1998}, \cite{Shkoller:2000}, \cite{XinZhang:2000}.

%
%
\section{The stationary CH formalism} \lb{chs3}

This section is devoted to a detailed study of the stationary CH equation
and its algebro-geometric solutions. Our principal tool will be a
combination of the polynomial recursion formalism introduced in Section
\ref{chs2} and a meromorphic function $\phi$ (the solution of a
Riccati-type equation associated with the zero-curvature representation of
\eqref{ch1.1}) on a hyperelliptic curve $\calK_n$ defined in terms of the
polynomial $R_{2n+2}$. 

For major parts of this section we suppose 
\begin{equation}
u\in C^\infty(\bbR), \; \f{d^m u}{dx^m}\in L^\infty(\bbR), \; 
m\in\bbN_{0}, \label{ch3.0}
\end{equation}
and assume  \eqref{ch2.3}, \eqref{ch2.4}, \eqref{ch2.7}, 
\eqref{ch2.8}, \eqref{ch2.14}--\eqref{ch2.16}, \eqref{ch2.22}, 
\eqref{ch2.23}, \eqref{ch2.24}--\eqref{ch2.26},  
\eqref{ch2.27}--\eqref{ch2.29}, keeping $n\in\bbN_0$ fixed.

Returning to \eqref{ch2.23} we infer {}from \eqref{ch2.25} and 
\eqref{ch2.9} that $R_{2n+2}(z)=zQ_{2n+1}(z)$ is a
monomial of degree $2n+2$ of the form
\begin{equation}
R_{2n+2}(z)=\prod_{m=0}^{2n+1}(z-E_m), \quad E_0=0,\, E_1,\dots,
E_{2n+1}\in\bbC. \lb{ch3.1}
\end{equation}
Computing
\begin{align}
\det(wI_2-iV_n(z,x)) & =  w^2 - \det(V_n(z,x)) \no \\
& =   w^2 + G_{n}(z,x)^2
+\f{1}{z}F_n(z,x)H_n(z,x) \lb{ch3.2} \\
& = w^2 +\f{1}{z^2}R_{2n+2}(z), \no
\end{align}
that is,
\begin{equation}
R_{2n+2}(z)=z^2G_n(z,x)^2+zF_n(z,x)H_n(z,x) \lb{ch3.2a}
\end{equation}
(with $I_2$ the identity matrix in $\bbC^2$), we are led to
introduce the (possibly singular)
hyperelliptic  curve $\calK_n$ of arithmetic genus $n$ defined by
\begin{equation}
\calK_n \colon \calF_n(z,y)=y^2-R_{2n+2}(z)=0. \lb{ch3.3}
\end{equation}
In the following we will occasionally impose further constraints on
the zeros $E_m$ of $R_{2n+2}$ introduced in \eqref{ch3.1} and assume that 
\begin{equation}
E_0=0, \; E_1,\dots,E_{2n+1}\in\bbC\setminus\{0\}. \lb{ch3.3a}
\end{equation}
We compactify $\calK_n$ by adding two points at infinity,  $\Pinfp$, 
$\Pinfm$, with  $\Pinfp\neq \Pinfm$, still denoting
its projective closure by  $\calK_n$.  Hence $\calK_n$
becomes a two-sheeted Riemann surface
of arithmetic genus $n$.  Points $P$ on $\calK_n\setminus\{\Pinfpm\}$ are
denoted by $P=(z,y)$, where $y(\dott)$ denotes the meromorphic
function on $\calK_n$ satisfying $\calF_n(z,y)=0$. For additional facts
on $\calK_{n}$ and further notation freely employed throughout this
paper, the reader may want to consult Appendix \ref{A}. 

For notational simplicity we will usually tacitly assume that $n\in\bbN$.
(The case $n=0$ is explicitly treated in Example~\ref{example-ch3.9}.)

In the following the roots of the polynomials $F_n$ and $H_n$ will play a
special role and hence we introduce on $\bbC\times\bbR$
\begin{equation}
F_n(z,x)=\prod_{j=1}^n(z-\mu_j(x)), \quad
H_n(z,x)=h_{0}(x)\prod_{j=1}^n(z-\nu_j(x)). \lb{ch3.5}
\end{equation}
Moreover, we introduce
\begin{align}
\hat\mu_j(x)&=(\mu_j(x),-\mu_j(x)G_{n}(\mu_j(x),x))\in\calK_n,
\quad j=1,\dots,n, \; x\in\bbR, \lb{ch3.6}\\
\hat\nu_j(x)&=(\nu_j(x),\nu_j(x)G_{n}(\nu_j(x),x))\in\calK_n,
\quad j=1,\dots,n, \; x\in\bbR, \lb{ch3.7}
\end{align}
and
\begin{equation}
P_0 = (0,0). \lb{ch3.8}
\end{equation}
The branch of $y(\dott)$ near $\Pinfpm$ is fixed according to
\begin{equation}
\lim_{\substack{\abs{z(P)}\to\infty\\P\to\Pinfpm}}\f{y(P)}{z(P)
G_n(z(P),x)}=\mp 1. \lb{ch3.8a}
\end{equation}
Due to assumption \eqref{ch3.0}, $u$ is smooth and bounded, and hence  
$F_n(z,\dott)$ and $H_{n}(z,\dott)$ share the same property. 
Thus, one also concludes 
\begin{equation}
\mu_j, \nu_k \in C^\infty(\bbR)\cap L^\infty(\bbR), \; j,k=1,\dots,n,  
\lb{ch3.9}
\end{equation}
by \eqref{ch3.5}. 

Next, define the fundamental meromorphic function $\phi(\dott,x)$ on
$\calK_n$ by
\begin{align}
\phi(P,x)&=\f{y-zG_{n}(z,x)}{F_n(z,x)} \lb{ch3.11} \\
&=\f{zH_{n}(z,x)}{y+z G_{n}(z,x)}, \quad
 P=(z,y)\in\calK_n, \, x\in\bbR. \lb{ch3.12}
\end{align}
Assuming \eqref{ch3.3a}, the divisor $(\phi(\dott,x))$ of $\phi(\dott,x)$
is given by
\begin{equation}
(\phi(\dott,x))=\calD_{P_0\hat{\ul\nu}(x)}
-\calD_{\Pinfp\hat{\ul\mu}(x)}, \lb{ch3.13}
\end{equation}
taking into account \eqref{ch3.8a}. 
Here we abbreviated
\begin{equation}
\hat{\ul\mu}=\{\hat\mu_1,\dots,\hat\mu_n\}, \, 
\hat{\ul\nu}=\{\hat\nu_1,\dots,\hat\nu_n\} \in \sigma^n\calK_n. \lb{ch3.14}
\end{equation}
Given $\phi(\dott,x)$, one defines the associated vector 
$\Psi(\dott,x,x_0)$ on $\calK_n\setminus\{\Pinfp,\Pinfm,P_0\}$ by
\begin{equation}
\Psi(P,x,x_0)=\begin{pmatrix} \psi_1(P,x,x_0) \\ \psi_2(P,x,x_0)
\end{pmatrix}, \quad
P\in\calK_n\setminus\{\Pinfp,\Pinfm,P_0\}, \; (x,x_0)\in\bbR^2,
\lb{ch3.15}
\end{equation}
where
\begin{align}
\psi_1(P,x,x_0)&=\exp\left(-(1/z) \int_{x_0}^x dx'\,
\phi(P,x')-(x-x_0)\right), \lb{ch3.16}\\
\psi_2(P,x,x_0)&=-\psi_1(P,x,x_0) \phi(P,x)/z. \lb{ch3.17}
\end{align}
Although $\Psi$ is formally the analog of the stationary Baker-Akhiezer
vector of the stationary CH hierarchy when compared to analogous
definitions in the context of the KdV or AKNS hierarchies, its actual
properties in a neighborhood of its essential singularity will feature
characteristic differences to standard Baker-Akhiezer vectors (cf.\
Remark \ref{remark-ch3.5}). We summarize the fundamental properties of
$\phi$ and
$\Psi$ in the following result.

\begin{lemma} \lb{lemma-ch3.1}
Suppose \eqref{ch3.0}, assume the $n$th stationary CH
equation \eqref{ch2.29} holds, and let $P=(z,y)\in\calK_n\setminus
\{\Pinfp,\Pinfm,P_0\}$, $(x,x_0)\in\bbR^2$. Then
$\phi$ satisfies the Riccati-type equation
\begin{equation}
\phi_x(P,x)-z^{-1}\phi(P,x)^2-2\phi(P,x)+4u(x)-u_{xx}(x)=0, 
\lb{ch3.18}
\end{equation}
as well as
\begin{align}
\phi(P,x)\phi(P^*,x)&=-\f{zH_{n}(z,x)}{F_n(z,x)}, \lb{ch3.19}\\
\phi(P,x)+\phi(P^*,x)&=-2\f{zG_{n}(z,x)}{F_n(z,x)}, \lb{ch3.20}\\
\phi(P,x)-\phi(P^*,x)&=\f{2y}{F_n(z,x)}, \lb{ch3.21}
\end{align}
while $\Psi$ fulfills
\begin{align}
&\Psi_x(P,x,x_0)=U(z,x)\Psi(P,x,x_0), \lb{ch3.22} \\
&-y \Psi(P,x,x_0)=zV_n(z,x)\Psi(P,x,x_0), \lb{ch3.23} \\
&\psi_1(P,x,x_0)=\bigg(\f{F_n(z,x)}{F_n(z,x_0)}\bigg)^{1/2}
\exp\bigg(-(y/z)\int_{x_0}^x dx'F_n(z,x')^{-1} \bigg), \lb{ch3.24} \\
&\psi_1(P,x,x_0)\psi_1(P^*,x,x_0)=\f{F_n(z,x)}{F_n(z,x_0)},\lb{ch3.25} \\
&\psi_2(P,x,x_0)\psi_2(P^*,x,x_0)=-\f{H_n(z,x)}{zF_n(z,x_0)},\lb{ch3.26}
\\
&\psi_1(P,x,x_0)\psi_2(P^*,x,x_0)+\psi_1(P^*,x,x_0)\psi_2(P,x,x_0)
=2\f{G_{n}(z,x)}{F_n(z,x_0)}, \lb{ch3.27} \\
&\psi_1(P,x,x_0)\psi_2(P^*,x,x_0)-\psi_1(P^*,x,x_0)\psi_2(P,x,x_0)
=\f{2y}{zF_n(z,x_0)}. \lb{ch3.28}
\end{align}
In addition, as long as the zeros of $F_n(\dott,x)$ are all simple for 
$x\in\Omega$, $\Omega\subseteq\bbR$ an open interval, $\Psi(\dott,x,x_0)$, 
$x,x_0\in\Omega$, is meromorphic on $\calK_n\setminus\{P_0\}$. 
\end{lemma}
\begin{proof}
Equation \eqref{ch3.18} follows  using the definition \eqref{ch3.11} 
of $\phi$ as well as
relations \eqref{ch2.17}--\eqref{ch2.19}. The other relations, 
\eqref{ch3.19}--\eqref{ch3.21},
are easy consequences of $y(P^*)=-y(P)$ in addition to 
\eqref{ch2.17}--\eqref{ch2.19}. 
By \eqref{ch3.15}--\eqref{ch3.17}, $\Psi$ is meromorphic on 
$\calK_n\setminus\{\Pinfpm\}$ away
{}from the poles $\hat \mu_j(x')$ of $\phi(\dott,x')$. By
\eqref{ch2.17}, \eqref{ch3.6}, and \eqref{ch3.11},
\begin{equation}
-\f{1}{z}\phi(P,x') \underset{P\to\hat \mu_j(x')}{=} \frac{\partial}{\partial x'}
\ln (F_n(z,x')) + \Oh(1)\text{  as $z\to\mu_j(x')$}, \lb{ch3.29}
\end{equation}
and hence $\psi_1$ is meromorphic on $\calK_n\setminus\{\Pinfpm\}$ by
\eqref{ch3.16} as long as the zeros of $F_n (\dott,x)$ are all simple. 
This follows {}from \eqref{ch3.16} by restricting
$P$ to a sufficiently small neighborhood $\calU_j$ of
$\{\hmu_j(x^{\prime})\in\calK_{n}\mid x^{\prime}\in\Omega,
\,x^{\prime}\in [x_{0},x] \}$ such that $\hmu_k(x^{\prime})\notin
\calU_j$ for all $x^{\prime}\in [x_{0},x]$ and all
$k\in\{1,\dots,n\}\setminus\{j\}$.
Since $\phi$ is meromorphic on $\calK_n$ by
\eqref{ch3.11}, $\psi_2$ is meromorphic on
$\calK_n\setminus\{\Pinfpm\}$ by \eqref{ch3.17}.
The remaining properties of $\Psi$ can be verified by using the  definition
\eqref{ch3.15}--\eqref{ch3.17} as well as relations 
\eqref{ch3.18}--\eqref{ch3.21}.  
In particular, equation \eqref{ch3.24} follows by
inserting the definition of $\phi$, \eqref{ch3.11}, into
\eqref{ch3.16}, using \eqref{ch2.17}.
\end{proof}

Next, we derive 
Dubrovin-type equations for $\mu_j$ and $\nu_j$. Since in the remainder of
this section we will frequently assume $\calK_n$ to be nonsingular, we
list all restrictions on $\calK_n$ in this case, 
\begin{equation}
E_0=0, \; E_m\in\bbC\setminus\{0\}, \; E_m\neq E_{m'} \text{ for } 
m\neq m', \, m,m'=1,\dots,2n. \lb{ch3.30}
\end{equation}

\begin{lemma} \lb{lemma-ch3.2}  
Suppose $\eqref{ch3.0}$ and the $n$th stationary CH equation
$\eqref{ch2.29}$ holds subject to the constraint $\eqref{ch3.30}$ on an open
interval $\ti\Omega_\mu\subseteq\bbR$. Moreover,  suppose that the zeros
$\mu_j$, $j=1,\dots,n$, of $F_n(\dott)$ remain distinct and nonzero on
$\ti\Omega_\mu$. Then $\{\hat\mu_j\}_{j=1,\dots,n}$, defined by
\eqref{ch3.6},  satisfies the following first-order system of differential
equations
\begin{equation}
\mu_{j,x}(x)=2\f{y(\hat\mu_j(x))}{\mu_{j}(x)}
\prod_{\substack{\ell=1\\ \ell\neq j}}^n(\mu_j(x)-\mu_\ell(x))^{-1},
\quad j=1, \dots, n, \, x\in \ti\Omega_\mu. \lb{ch3.31}
\end{equation}
Next, assume $\calK_n$ to be nonsingular and introduce the initial
condition
\begin{equation}
\{\hat\mu_j(x_0)\}_{j=1,\dots,n}\subset\calK_n \lb{ch3.32}
\end{equation}
for some $x_0\in\bbR$, where $\mu_j(x_0)\neq 0$, $j=1,\dots,n$, are
assumed to be distinct. Then there exists an open interval
$\Omega_\mu\subseteq\bbR$, with $x_0\in\Omega_\mu$, such that the initial
value problem \eqref{ch3.31}, \eqref{ch3.32} has a unique solution
$\{\hat\mu_j\}_{j=1,\dots,n}\subset\calK_n$ satisfying
\begin{equation}
\hat\mu_j\in C^\infty(\Omega_\mu,\calK_n),\quad j=1, \dots, n,
\lb{ch3.33}
\end{equation}
and $\mu_j$, $j=1,\dots,n$, remain distinct and nonzero on $\Omega_\mu$.\\
For the zeros $\{\nu_j\}_{j=1,\dots,n}$ of $H_n(\dott)$ similar
statements hold with $\mu_j$ and $\Omega_\mu$ replaced by $\nu_j$ and
$\Omega_\nu$, etc. In particular, $\{\hat\nu_j\}_{j=1,\dots,n}$, defined
by \eqref{ch3.7}, satisfies the system 
\begin{align}
\nu_{j,x}(x)&=2
\f{(4u(x)-u_{xx}(x))y(\hat\nu_j(x))}{(4u(x)+2u_{x}(x))\nu_{j}(x)}
\prod_{\substack{\ell=1\\ \ell\neq j}}^n(\nu_j(x)-\nu_\ell(x))^{-1}, 
\lb{ch3.34} \\
&\hspace*{4.65cm} j=1, \dots, n, \, x\in \Omega_\nu. \no
\end{align}
\end{lemma}
\begin{proof}
We only prove  equation \eqref{ch3.31} since the proof of \eqref{ch3.34}
follows in an identical manner. Inserting
$z=\mu_j$ into  equation \eqref{ch2.17}, one concludes {}from
\eqref{ch3.6},
\begin{equation}
F_{n,x}(\mu_j)=-\mu_{j,x}
\prod_{\substack{\ell=1\\ \ell\neq j}}^n(\mu_j-\mu_\ell)
=2 G_{n}(\mu_j)=-2y(\hat \mu_j)/\mu_{j}, \lb{ch3.35}
\end{equation}
proving \eqref{ch3.31}. The smoothness assertion \eqref{ch3.33}
is clear as long as $\hat \mu_j$ stays away {}from the
branch points $(E_m,0)$. In case $\hat \mu_j$ hits such a
branch point, one can use the local chart around $(E_m,0)$
(with local coordinate $\zeta=\sigma (z-E_m)^{1/2}$, $\sigma\in\{1,-1\}$) 
to verify \eqref{ch3.33}.
\end{proof}

Combining the polynomial approach in Section \ref{chs2} with
\eqref{ch3.5} readily yields trace formulas for the CH invariants.
For simplicity we just record the simplest case.

\begin{lemma} \lb{lemma-ch3.3}
Suppose \eqref{ch3.0}, assume the $n$th stationary CH
equation \eqref{ch2.29} holds, and let $x\in\bbR$. Then 
\begin{align}
u(x)&=\f12\sum_{j=1}^n\mu_{j}(x)-\f14\sum_{m=1}^{2n+1} E_{m}. \lb{ch3.36}
\end{align}
\end{lemma}
\begin{proof}
Equation \eqref{ch3.36} follows by considering the
coefficient of $z^{n-1}$ in 
$F_n$ in \eqref{ch2.24} which yields
\begin{equation}
    u=\f12\sum_{j=1}^n \mu_{j}+\f{c_{1}}{2}. \lb{ch3.37}
\end{equation}
The constant $c_{1}$ can be determined by considering the coefficient 
of the term $z^{2n+1}$ in \eqref{ch2.23}, which results in
\begin{equation}
c_1=-\f12 \sum_{m=0}^{2n+1} E_m. \lb{ch3.38}
\end{equation}
\end{proof}

Next we turn to asymptotic properties of $\phi$ and $\psi_j$, $j=1,2$.

\begin{lemma} \lb{lemma-ch3.4}
Suppose \eqref{ch3.0}, assume the $n$th stationary CH
equation \eqref{ch2.29} holds, and let
$P=(z,y)\in\calK_n\setminus\{\Pinfp, \Pinfm,P_0\}$, $x\in\bbR$. Then
\begin{align}
&\phi(P,x)\underset{\zeta\to 0}{=}\begin{cases} 
-2\zeta^{-1}-2u(x)+u_x(x)+\Oh(\zeta), & P\to\Pinfp, \\
2u(x)+u_x(x)+\Oh(\zeta), & P\to\Pinfm, \end{cases} 
\quad \zeta=z^{-1},  \lb{ch3.38a} \\
&\phi(P,x)\underset{\zeta\to 0}{=}\f{\Big(\prod_{m=1}^{2n+1}
E_m\Big)^{1/2}}{f_n(x)}\zeta +\Oh(\zeta^2),
\quad  P\to P_0, \quad \zeta=z^{1/2}, \lb{ch3.38b} 
\end{align}
and
\begin{align}
&\psi_1(P,x,x_0)\underset{\zeta\to 0}{=}\exp(\pm(x-x_0))(1+\Oh(\zeta)),
\quad P\to\Pinfpm, \quad \zeta=1/z, \lb{ch3.38c} \\
&\psi_2(P,x,x_0)\underset{\zeta\to 0}{=}\exp(\pm(x-x_0))
 \begin{cases}  -2+\Oh(\zeta), & P\to\Pinfp, \\
(2u(x)+u_x(x))\zeta +\Oh(\zeta^2), & P\to\Pinfm, \end{cases}
\lb{ch3.38d} \\ 
& \hspace*{9.96cm} \zeta=1/z, \no \\
& \psi_1(P,x,x_0)\underset{\zeta\to
0}{=}\exp\Bigg(-\f{1}{\zeta}\int_{x_0}^x dx' \f{\Big(\prod_{m=1}^{2n+1}
E_m \Big)^{1/2}}{f_n(x')}+\Oh(1)\Bigg), \quad P\to P_0, \lb{ch3.38e} \\
& \hspace*{9.65cm} \zeta=z^{1/2}, \no \\
& \psi_2(P,x,x_0)\underset{\zeta\to
0}{=}\Oh\big(\zeta^{-1}\big)\exp\Bigg(-\f{1}{\zeta}\int_{x_0}^x dx'
\f{\Big(\prod_{m=1}^{2n+1} E_m
\Big)^{1/2}}{f_n(x')}+\Oh(1)\Bigg), \quad P\to P_0, \no \\
& \hspace*{9.8cm} \zeta=z^{1/2}. \lb{ch3.38f} 
\end{align} 
\end{lemma}
\begin{proof}
The existence of the asymptotic expansions of $\phi$ in terms of the
appropriate local coordinates $\zeta=1/z$ near $\Pinfpm$ and
$\zeta=z^{1/2}$ near $P_0$ is clear {}from the explicit form of $\phi$ in
\eqref{ch3.11}. Insertion of the polynomials $F_n$, $G_n$, and $H_n$ into
\eqref{ch3.11} then, in principle, yields the explicit expansion
coefficients in \eqref{ch3.38a} and \eqref{ch3.38b}. However, a more
efficient way to compute these coefficients consists in utilizing the
Riccati-type equation \eqref{ch3.18}. Indeed, inserting the ansatz
\begin{equation}
\phi\underset{z \to\infty}{=}\phi_1 z+\phi_0+\Oh\big(z^{-1}\big)
\lb{ch3.38g}
\end{equation}
into \eqref{ch3.18} and comparing the leading powers of $1/z$ immediately
yields the first line in \eqref{ch3.38a}. Similarly, the ansatz
\begin{equation}
\phi\underset{z \to\infty}{=}\phi_0+\phi_1 z^{-1}+\Oh\big(z^{-2}\big)
\lb{ch3.38h}
\end{equation}
inserted into \eqref{ch3.18} then yields the second line in
\eqref{ch3.38a}. Finally, the ansatz
\begin{equation}
\phi\underset{z\to 0}{=}\phi_1 z^{1/2} + \phi_2 z +\Oh\big(z^{3/2}\big) 
\lb{ch3.38i}
\end{equation}
inserted into \eqref{ch3.18} yields \eqref{ch3.38b}. 
\eqref{ch3.38c}--\eqref{ch3.38f} then follow {}from \eqref{ch3.16},
\eqref{ch3.17}, \eqref{ch3.38a}, and \eqref{ch3.38b}. 
\end{proof}

\begin{remark} \lb{remark-ch3.5}
We note the unusual fact that $P_0$, as opposed to $\Pinfpm$, is the
essential singularity of $\psi_j$, $j=1,2$. What makes matters worse is the
intricate $x$-dependence of the leading-order exponential term in
$\psi_j$, $j=1,2$, near $P_0$, as displayed in \eqref{ch3.38e},
\eqref{ch3.38f}. This is in sharp contrast to standard Baker-Akhiezer
functions that feature a linear behavior with respect to $x$ in connection
with their essential singularities of the type
$\exp\big(c(x-x_0)\zeta^{-1}\big)$ near $\zeta=0$.  
\end{remark}

Introducing
\begin{align}
\ul{\hatt B}_{Q_0} & \colon
\hatt\calK_n\setminus \{\Pinfp,\Pinfm\}\to\bbC^n, \lb{ch3.38m}  \\
&P\mapsto \ul{\hatt B}_{Q_0}(P)=\big({\hatt B}_{Q_0,1},\dots,
{\hatt B}_{Q_0,n}\big)=\bigg(\int_{Q_0}^P \eta_2,\dots,\int_{Q_0}^P
\eta_n, \int_{Q_0}^P \ti\omega^{(3)}_{\Pinfp,\Pinfm}\bigg) \no 
\end{align}
and
\begin{align}
\ul{\hatt\beta}_{Q_0} & \colon
\sigma^n\big(\hatt\calK_n\setminus\{\Pinfp,\Pinfm\}\big)\to\bbC^n,
\lb{ch3.38n} \\ 
&\calD_{\ul Q}\mapsto \ul{\hatt\beta}_{Q_0}
(\calD_{Q_0})=\sum_{j=1}^n \ul{\hatt B}_{Q_0}(Q_j), \quad 
\ul Q=\{Q_1,\dots,Q_n\}\in\sigma^n\hatt\calK_n\setminus \{\Pinfp,\Pinfm\},
\no
\end{align}
choosing identical paths of integration {}from $Q_0$ to $P$ in all
integrals in \eqref{ch3.38m} and \eqref{ch3.38n}, one obtains the following
result, which indicates a characteristic difference between the CH
hierarchy and other completely integrable systems such as the KdV and 
AKNS hierarchies.

\begin{lemma} \lb{lemma-ch3.6}
Assume \eqref{ch3.30} and suppose that
$\{\hat\mu_j\}_{j=1,\dots,n}$ satisfies the stationary Dubrovin equations
\eqref{ch3.31} on an open interval $\Omega_\mu\subseteq\bbR$ such that
$\mu_j$, $j=1,\dots,n$, remain distinct and nonzero on $\Omega_\mu$.
Introducing the associated divisor $\calD_{\humu}\in\sigma^n\hatt\calK_n$,
$\humu=\{\hat\mu_1,\dots,\hat\mu_n\}\in\sigma^n\hatt\calK_n$, one computes
\begin{equation}
\f{d}{dx}\ul \al_{Q_0} (\calD_{\humu(x)})=-\f{2}{\Psi_n(\ul\mu(x))}\ul
c(1), \quad x\in\Omega_\mu. \lb{ch3.38j}
\end{equation}
In particular, the Abel map does not linearize the divisor
$\calD_{\humu(\dott)}$ on $\Omega_\mu$. In addition,
\begin{align}
&\f{d}{dx}\sum_{j=1}^n \int_{Q_0}^{\hat\mu_j(x)}
\eta_1=-\f{2}{\Psi_n(\ul\mu (x))}, \quad x\in\Omega_\mu, \lb{ch3.38o} \\
&\f{d}{dx}\ul{\hatt\beta}(\calD_{\ul{\hat\mu}(x)})=2(0,\dots,0,1), \quad
x\in\Omega_\mu. \lb{ch3.38p}
\end{align}
\end{lemma}
\begin{proof}
Let $x\in\Omega_\mu$. Then, using
\begin{equation}
\f{1}{\mu_j}=\f{\prod_{\substack{p=1\\p\neq j}}^n \mu_p}{\prod_{m=1}^n
\mu_m}=-\f{\Phi^{(j)}_{n-1}(\ul\mu)}{\Psi_n(\ul\mu)}, \quad j=1,\dots,n, 
\lb{ch3.38k}
\end{equation}
(cf.\ \eqref{functpsi}, \eqref{functphi}) one obtains 
\begin{align}
\f{d}{dx}\bigg(\sum_{j=1}^n \int_{Q_0}^{\hat\mu_j} \ul\omega\bigg)
&=\sum_{j=1}^n \mu_{j,x}\sum_{k=1}^n \ul c(k)\f{\mu_j^{k-1}}{y(\hat\mu_j)}
=2\sum_{j=1}^n\sum_{k=1}^n \ul
c(k)\f{\mu_j^{k-2}}{\prod_{\substack{\ell=1\\ \ell\neq
j}}^n(\mu_j-\mu_\ell)} \no \\
&=-\f{2}{\Psi_n(\ul\mu)}\sum_{j=1}^n\sum_{k=1}^n \ul c(k)
\f{\mu_j^{k-1}}{\prod_{\substack{\ell=1\\ \ell\neq
j}}^n(\mu_j-\mu_\ell)}\Phi^{(j)}_{n-1}(\ul\mu) \no \\
&=-\f{2}{\Psi_n(\ul\mu)}\sum_{j=1}^n\sum_{k=1}^n \ul c(k)
(U_n(\ul\mu))_{k,j} (U_n(\ul\mu))^{-1}_{j,1} \no \\
&=-\f{2}{\Psi_n(\ul\mu)}\sum_{k=1}^n \ul c(k) \delta_{k,1}
=-\f{2}{\Psi_n(\ul\mu)} \ul c(1), \lb{ch3.38l}
\end{align}
using \eqref{B18} and \eqref{B19}. \eqref{ch3.38o} is just a special case
of \eqref{ch3.38j} and \eqref{ch3.38p} follows as in \eqref{ch3.38l} using
\eqref{A5}.
\end{proof}
The analogous results hold for the corresponding divisor
$\calD_{\hunu (x)}$ associated with $\phi(\dott,x)$.

The fact that the Abel map does not provide the proper change of
variables to linearize the divisor $\calD_{\humu(x)}$ in the CH context is
in sharp contrast to standard integrable soliton equations such as the KdV
and AKNS hierarchies (cf.\ also Remark \ref{remark-ch3.5}). The change of
variables 
\begin{equation}
x \mapsto \tilde x= \int^x dx'\,\Psi_n(\ul\mu(x'))^{-1} \lb{ch3.38L}
\end{equation}
linearizes the Abel map $\ul A_{Q_0} (\calD_{\hat{\ul {\tilde\mu}}(\tilde
x)})$, $\tilde\mu_j(\tilde x)=\mu_j(x)$, $j=1,\dots,n$. These facts are
well-known and discussed (by different methods) by Constantin and McKean
\cite{ConstantinMcKean:1999}, Alber \cite{Alber:2000}, Alber, Camassa, Fedorov,
Holm, and Marsden \cite{AlberCamassaFedorovHolmMarsden:2001}, and Alber and Fedorov
\cite{AlberFedorov:2000}, \cite{AlberFedorov:2001}. The intricate relation between
the variables $x$ and $\tilde x$ is detailed in \eqref{ch3.38ab}. Our approach
follows a route similar to Novikov's treatment of the Dym equation
\cite{Novikov:1999}. 

Next we turn to representations of $\phi$ and $u$ in
terms of the Riemann theta function associated with $\calK_n$, assuming
$\calK_n$ to be nonsingular. In the following, the notation established in
Appendices \ref{A}--\ref{C} will be freely employed. In fact, given the
preparatory work collected in Appendices \ref{A}--\ref{C}, the proof of
Theorem \ref{theorem-ch3.7} below will be reduced to a few lines.

We choose a fixed base point $Q_0$ on $\calK_n\setminus\{\Pinfp,P_0\}$.
Let $\omega^{(3)}_{\Pinfp,P_0}$ be a normal differential of the third
kind holomorphic on $\calK_n\setminus\{\Pinfp,P_0\}$ with simple poles at
$\Pinf$ and $P_0$ and residues $1$ and $-1$, respectively  (cf.\
\eqref{a36}--\eqref{a37d}),
\begin{align}
\omega^{(3)}_{\Pinfp,P_0} &=\f{\prod_{j=1}^n (z-\lambda_j)dz}{y}, 
\lb{ch3.38q} \\
\omega^{(3)}_{\Pinfp,P_0} & \underset{\zeta\to 0}{=}
(\zeta^{-1}+\Oh(1))d\zeta \text{ as $P\to\Pinfp$}, \lb{ch3.38r} \\
\omega^{(3)}_{\Pinfp,P_0} &\underset{\zeta\to
0}{=}(-\zeta^{-1}+\Oh(1))d\zeta \text{ as $P\to P_0$}, \lb{ch3.38s}
\end{align}
where
\begin{equation}
\zeta=1/z \text{ for } P \text{ near } \Pinfp, \quad
\zeta=\sigma z^{1/2} \text{ for } P \text{ near } P_0, 
\; \sigma \in\{1,-1\}.  \lb{ch3.38t} 
\end{equation}
Moreover,
\begin{align}
&\int_{a_j} \omega^{(3)}_{\Pinfp,P_0} =0, \quad j=1,\dots,n, 
\lb{ch3.38u} \\
&\int_{Q_0}^P \omega^{(3)}_{\Pinfp,P_0}
\underset{\zeta\to 0}{=}\ln(\zeta)+e_0+\Oh(\zeta)
\text{ as $P\to\Pinfp$}, \lb{ch3.38v} \\
&\int_{Q_0}^P \omega^{(3)}_{\Pinfp,P_0}
 \underset{\zeta\to 0}{=}-\ln(\zeta)+d_0 + \Oh(\zeta)
\text{ as $P\to P_0$} \lb{ch3.38w}
\end{align}
for some constants $e_{0}, d_{0}\in\bbC$. We also record
\begin{equation}
\ul A_{Q_0}(P)-\ul A_{Q_0}(\Pinfpm)\underset{\zeta\to 0}{=}\pm \ul U\zeta 
+\Oh(\zeta^2) \text{ as } P\to \Pinfpm, \quad \ul U=\ul c (n). 
\lb{ch3.38x}
\end{equation}
In the following it will be convenient to introduce the abbreviations  
\begin{equation}
\uz(P,\ul Q) =\ul\Xi_{Q_0}-\ul A_{Q_0}(P)
+\ul\alpha_{Q_0}(\calD_{\ul Q}), \quad P\in\calK_n, \; \ul Q=\{Q_1,\dots,
Q_n\} \in \sigma^n\calK_n,\lb{ch3.38y} 
\end{equation}
and analogously,
\begin{equation}
\hat{\uz}(P,\ul Q) =\ul{\hatt\Xi}_{Q_0}-\ul {\hatt A}_{Q_0}(P)
+\ul{\hatt \alpha}_{Q_0}(\calD_{\ul Q}), \quad 
 P\in\hatt\calK_n, \; \ul Q=\{Q_1,\dots,
Q_n\} \in \sigma^n\hatt\calK_n. \lb{ch3.38ya} 
\end{equation}

\begin{theorem}\lb{theorem-ch3.7} 
Suppose $u\in C^\infty(\Omega)$, $u^{(m)}\in L^\infty(\Omega)$,
$m\in\bbN_0$, and assume the $n$th stationary CH equation \eqref{ch2.29}
holds on $\Omega$ subject to the constraint \eqref{ch3.30}. Moreover, let
$P\in\calK_n\setminus\{\Pinfp,P_0\}$ and $x\in\Omega$, where
$\Omega\subseteq\bbR$ is an open interval. In addition, suppose that
$\calD_{\hat{\ul\mu}(x)}$, or equivalently,
$\calD_{\hat{\ul\nu}(x)}$,  is nonspecial for $x\in\Omega$. Then $\phi$
and $u$ admit the representations 
\begin{align}
\phi (P,x) &= -2\frac{\theta (\ul z(\Pinfp,\humu(x)))\theta (\ul
z(P,\hunu(x)))}{\theta(\ul z(\Pinfp,\hunu(x)))\theta
(\ul z(P,\humu(x)))} \exp \bigg(-\int_{Q_0}^P
\omega_{\Pinfp,P_0}^{(3)}+e_0\bigg), \lb{ch3.38z} \\ 
u(x)&=\f{1}{2}\sum_{j=1}^n \lambda_j -\f{1}{4}\sum_{m=1}^{2n+1} E_m  
 +\f{1}{2}\sum_{j=1}^n U_j \f{\partial}{\partial w_j}\ln\bigg(
\f{\theta\big(\uz(\Pinfp,\hat{\ul\mu}(x))+\ul w\big)} 
{\theta\big(\uz(\Pinfm,\hat{\ul\mu}(x))+\ul w\big)}
\bigg)\bigg|_{\ul w=0}. \lb{ch3.38aa}
\end{align}
Moreover, let $\ti\Omega\subseteq\Omega$ be such that $\mu_j$,
$j=1,\dots,n$, are nonvanishing on $\ti\Omega$, Then, the constraint 
\begin{align}
2(x-x_0)&=-2 \int_{x_0}^x \f{dx'}{\prod_{k=1}^n \mu_k(x')}
\sum_{j=1}^n \bigg(\int_{a_j}\ti\omega^{(3)}_{\Pinfp,\Pinfm}\bigg)c_j(1)
\no \\ 
& \quad +\ln\bigg(\f{\theta\big(\ul z(\Pinfp,\humu(x))\big)
\theta\big(\ul z(\Pinfm,\humu(x_0))\big)} 
{\theta\big(\ul z(\Pinfm,\humu(x))\big)
\theta\big(\ul z(\Pinfp,\humu(x_0))\big)}\bigg), 
\quad x,x_0\in\ti\Omega \lb{ch3.38ab}
\end{align}
holds, with
\begin{align}
\ul{\hat z} (\Pinfpm,\humu (x)))&=\ul{\hatt\Xi}_{Q_0}-\ul{\hatt
A}_{Q_0}(\Pinfpm) +\ul{\hatt\alpha}_{Q_0}(\calD_{\humu (x)}) \no \\
&=\ul{\hatt\Xi}_{Q_0}-\ul{\hatt
A}_{Q_0}(\Pinfpm) +\ul{\hatt\alpha}_{Q_0}(\calD_{\humu (x_0)}) 
-2\int_{x_0}^x \f{dx'}{\Psi_n(\ul\mu(x')}\,\ul c(1), \lb{ch3.38ac} \\
& \hspace*{7.95cm} x\in\ti\Omega. \no
\end{align}
\end{theorem}
\begin{proof}
First we temporarily assume that 
\begin{equation}
\mu_j(x)\neq\mu_{j'}(x), \; \nu_k(x)\neq\nu_{k'}(x) 
\text{ for $j\neq j'$, $k\neq k'$ and 
$x\in\ti\Omega\subseteq\Omega$}, \lb{ch3.38ad}
\end{equation}
where $\ti\Omega$ is open. Since by \eqref{ch3.13},
$\calD_{P_0\hat{\ul\nu}(x)}\sim\calD_{\Pinfp\hat{\ul\mu}(x)}$, and 
$\Pinfm=(\Pinfp)^*\notin\{\hmu_1(x),\dots,\hmu_n(x)\}$ by hypothesis, one
can apply Theorem \ref{taa20} to conclude that
$\calD_{\hunu(x)}\in\sigma^n\calK_n$ is nonspecial. This argument is of
course symmetric with respect to $\humu(x)$ and $\hunu(x)$. Thus,
$\calD_{\humu(x)}$ is nonspecial if and only if $\calD_{\hunu(x)}$ is. 
The representation \eqref{ch3.38z} for $\phi$, subject to
\eqref{ch3.38ad}, then follows by combining \eqref{ch3.13},
\eqref{ch3.38a}, \eqref{ch3.38b}, and Theorem \ref{taa17a} since
$\calD_{\hat{\ul\mu}}$ and $\calD_{\hat{\ul\nu}}$ are nonspecial. The
representation \eqref{ch3.38aa} for $u$ on
$\ti\Omega$ follows {}from the trace formula \eqref{ch3.36} and
\eqref{g.62} (taking $k=1$). By continuity, \eqref{ch3.38z} and
\eqref{ch3.38aa} extend {}from $\ti\Omega$ to $\Omega$. Assuming 
$\mu_j\neq 0$, $j=1,\dots,n$, in addition to \eqref{ch3.38ad}, the
constraint \eqref{ch3.38ab} follows by combining \eqref{ch3.38o},
\eqref{ch3.38p}, and \eqref{g.61c}. Equation \eqref{ch3.38ac} is clear
{}from \eqref{ch3.38j}. Again the extra  assumption \eqref{ch3.38ad} can 
be removed by continuity and hence \eqref{ch3.38ab} and \eqref{ch3.38ac}
extend to $\ti\Omega$.
\end{proof}

\begin{remark} \lb{remark-ch3.7a}
While the stationary CH solution $u$ in \eqref{ch3.38aa} is of course a 
meromorphic quasi-periodic function with respect to the new variable
$\tilde x$ in \eqref{ch3.38L}, $u$ may exhibit a rather intricate
behavior with respect to the original variable $x$. Generically, $u$ has
an infinite number of branchpoints of the type 
\begin{equation}
u(x)\underset{x\to x_0}{=}\Oh((x-x_0)^{2/3}) \lb{ch3.38M}
\end{equation}
and 
\begin{equation}
\tilde x-\tilde x_0\underset{x\to x_0}{=}\Oh((x-x_0)^{1/3}). \lb{ch3.38N}
\end{equation}
Moreover, real-valued bounded stationary CH solutions fall into two 
categories and
are either smooth quasi-periodic functions in $x$, or else \eqref{ch3.38M}
and \eqref{ch3.38N} hold at infinitely many points (depending on whether or
not $\Psi_n(\ul\mu)$ is zero-free, cf.\ \eqref{ch3.38L}), as
discussed in \cite{AlberCamassaFedorovHolmMarsden:2001}, \cite{AlberFedorov:2000},
\cite{AlberFedorov:2001}). We note that \eqref{ch3.38ab} relates the variables $x$
and $\tilde x$.
\end{remark}

\begin{remark} \lb{remark-ch3.8}
We emphasized in Remark \ref{remark-ch3.5} that $\Psi$ in
\eqref{ch3.15}--\eqref{ch3.17} markedly differs from  standard
Baker-Akhiezer vectors. Hence one cannot expect the usual theta
function representation of $\psi_j$, $j=1,2$, in terms of ratios of theta
functions times an exponential term containing a meromorphic differential
with a pole at the essential singularity of $\psi_j$ multiplied by
$(x-x_0)$. However, combining \eqref{A1a} and \eqref{g.62}, one computes
\begin{align}
F_n(z,x)&=z^n+\sum_{\ell=0}^{n-1} \Psi_{n-\ell}(\ul\lambda)z^\ell \no \\
&=z^n+\sum_{k=1}^n \bigg(\Psi_{n+1-k}(\ul\mu(x)) \no \\
& \hspace*{2cm} -\sum_{j=1}^n c_j(k)
\f{\partial}{\partial w_j}\ln\bigg(\f{\theta(\ul
z(\Pinfp,\humu(x))+\ul w)}{\theta(\ul
z(\Pinfm,\humu(x))+\ul w)}\bigg)\bigg|_{\ul w=0} z^{k-1}\bigg) \no \\
&=\prod_{j=1}^n (z-\lambda_j)-\sum_{j=1}^n\sum_{k=1}^n c_j(k)
\f{\partial}{\partial w_j}\ln\bigg(\f{\theta(\ul
z(\Pinfp,\humu(x))+\ul w)}{\theta(\ul
z(\Pinfm,\humu(x))+\ul w)}\bigg)\bigg|_{\ul w=0} z^{k-1}, \lb{ch3.38ae} 
\end{align}
and hence obtains a theta function representation of $\psi_1$ upon
inserting \eqref{ch3.38ae} into \eqref{ch3.24}. The corresponding theta
function representation of $\psi_2$ is then clear from \eqref{ch3.17} and
\eqref{ch3.38z}. 
\end{remark}

Next we briefly consider the trivial case $n=0$ excluded in Theorem
\ref{theorem-ch3.7}.

\begin{example}\lb{example-ch3.9}
Assume $n=0$, $P=(z,y)\in\calK_0\setminus\{\Pinfp, \Pinfm,P_0\}$, and let
$(x,x_0)\in\bbR^2$. Then 
\begin{align}
&\calK_0\colon \calF_0(z,y)=y^2-R_2(z)=y^2-z(z-E_1)=0, 
\quad E_0=0, \, E_1\in\bbC, \no\\ 
&u(x)=-E_1/4, \lb{ch3.39b} \\
&\phi(P,x)=y-z=-\f{E_1z}{y+z}, \no \\
&\psi_1(P,x,x_0)=\exp(-(y/z)(x-x_0)), \no \\
&\psi_2(P,x,x_0)=(1-(y/z))\exp(-(y/z)(x-x_0)). \no 
\end{align}
Actually, the general solution of $\sCH_0(u)=u_{xxx}-4u_x=0$ is given by
\begin{equation}
u(x)=a_1e^{2x}+a_2e^{-2x}-(E_1/4), \quad a_j\in\bbC, \, j=1,2. \lb{ch3.39f}
\end{equation}
However, the requirement $u^{(m)}\in L^\infty (\bbR)$, $m\in\bbN_0$,
according to \eqref{ch3.0}, necessitates the choice $a_1=a_2=0$ and hence
leads to \eqref{ch3.39b}. The latter corresponds to the trace
formula \eqref{ch3.36} in the special case $n=0$. 
\end{example}

Finally, we will show that solvability of the Dubrovin equations 
\eqref{ch3.31} on $\Omega_\mu\subseteq\bbR$ in fact implies equation
\eqref{ch2.29}  on $\Omega_\mu$. 

\begin{theorem}\lb{theorem-ch3.10}
Fix $n\in\bbN$, assume \eqref{ch3.30}, and suppose that
$\{\hat\mu_j\}_{j=1,\dots,n}$ satisfies the stationary Dubrovin equations
\eqref{ch3.31} on an open interval $\Omega_\mu\subseteq\bbR$ such that
$\mu_j$, $j=1,\dots,n$, remain distinct and nonzero on $\Omega_\mu$. Then 
$u\in C^\infty(\Omega_\mu)$ defined by 
\begin{equation}
u(x)=\f12\sum_{j=1}^n\mu_j(x)-\f{1}{4}\sum_{m=1}^{2n+1} E_m \lb{ch3.40}
\end{equation}
satisfies the $n$th stationary CH equation \eqref{ch2.29}, that is,
\begin{equation}
\sCH_n(u)=0 \text{  on $\Omega_\mu$.} \lb{ch3.41}
\end{equation}
\end{theorem}
\begin{proof}
Given the solutions $\hat\mu_j=(\mu_j,y(\hat\mu_j))\in
C^\infty(\Omega_\mu, \calK_n)$, $j=1,\dots,n$ of \eqref{ch3.31} we
introduce 
\begin{align}
F_n(z)&=\prod_{j=1}^n (z-\mu_j), \lb{ch3.42}\\
G_n(z)&=F_n(z)+\f12 F_{n,x}(z) \lb{ch3.43}
\end{align}
on $\bbC\times\Omega_\mu$. The Dubrovin equations imply
\begin{equation}
y(\hat\mu_j)=\f12\mu_j \mu_{j,x}\prod_{\substack{\ell=1\\ \ell\neq
j}}^n(\mu_j-\mu_\ell)=-\f12\mu_j F_{n,x}(\mu_j)=-\mu_j G_{n}(\mu_j).
 \lb{ch3.44}
\end{equation}
Thus
\begin{equation}
R_{2n+2}(\mu_j)-\mu_j^2 G_n(\mu_j)^2=0, \quad j=1,\dots,n. \lb{ch3.45}
\end{equation}
Furthermore $R_{2n+2}(0)=0$, and hence there exists a polynomial $H_n$
such that
\begin{equation}
R_{2n+2}(z)-z^2 G_n(z)^2=z F_n(z) H_n(z). \lb{ch3.45a}
\end{equation}
Computing the coefficient of the term $z^{2n+1}$ in \eqref{ch3.45a} one
finds 
\begin{equation}
 H_n(z)=(4u+2u_x) z^n+\Oh(z^{n-1}) \text{  as $\abs{z}\to\infty$}.
\lb{ch3.45b}
\end{equation}
Next, one defines a polynomial $P_{n-1}$ by
\begin{equation}
P_{n-1}(z)=(4u-u_{xx}) F_n(z)-H_n(z)-z G_{n,x}(z).
\lb{ch3.46}
\end{equation}
Using \eqref{ch3.40}, \eqref{ch3.42}, \eqref{ch3.43}, and \eqref{ch3.45b}
one infers that indeed $P_{n-1}$ has degree at most $n-1$. Multiplying
\eqref{ch3.46} by
$G_n$, and replacing the term $G_n G_{n,x}$ with the result obtained upon
differentiating \eqref{ch3.45a} with respect to $x$, yields
\begin{align}
G_n(z)P_{n-1}(z)&=F_n(z)\big((4u-u_{xx})G_{n}(z)+\f12
H_{n,x}(z)\big)\no\\
&\quad+\big(\f12 F_{n,x}(z)-G_n(z) \big)H_{n}(z), \lb{ch3.47}
\end{align}
and hence
\begin{equation}
G_n(\mu_j)P_{n-1}(\mu_j)=0, \quad j=1,\dots,n \lb{ch3.48}
\end{equation}
on $\Omega_\mu$. Restricting $x\in\Omega_\mu$ temporarily
to $x\in\ti\Omega_\mu$, where 
\begin{align}
\ti\Omega_\mu&=\{x\in\Omega_\mu\mid  F_{n,x}(\mu_j(x),x)=2i
y(\hat\mu_j(x))/\mu_j(x)\neq 0, \, j=1,\dots,n\} \no \\
&=\{x\in\Omega_\mu\mid
\mu_j(x)\notin\{E_0,\dots,E_{2n}\},
\, j=1,\dots,n\} \lb{ch3.49}
\end{align}
one infers that
\begin{equation}
P_{n-1}(\mu_j)=0, \quad j=1,\dots,n  \lb{ch3.50}
\end{equation} 
on $\bbC\times \ti\Omega_\mu$. Since $P_{n-1}(z)$ has degree at most
$n-1$, one concludes 
\begin{equation}
P_{n-1}=0 \text{ on } \bbC\times \ti\Omega_\mu, \lb{ch3.50a}
\end{equation}
and hence \eqref{ch2.18}, that is,
\begin{equation}
z G_{n,x}(z)=(4u-u_{xx}) F_n(z)-H_n(z) \lb{ch3.51}
\end{equation}
on $\bbC\times\ti\Omega_\mu$. Differentiating \eqref{ch3.45a} with
respect to
$x$ and using equations
\eqref{ch3.51} and \eqref{ch3.43} one finds 
\begin{equation}
H_{n,x}(z)=2 F_n(z)-2(4u-u_{xx})G_n(z) \lb{ch3.52}
\end{equation}
on $\bbC\times\ti\Omega_\mu$. In order to extend these results to
$\Omega_\mu$ one must prove that
$\hat\mu_j$ does not pause once it hits $(E_m,0)$. Hence we suppose
\begin{equation}
\mu_{j_0}(x)\to E_{m_0} \text{  as $x\to x_0\in\Omega_\mu$},
\lb{ch3.53}
\end{equation}
for some $j_0\in\{1,\dots,n\}$, $m_0\in\{0,\dots,2n\}$. Introducing 
\begin{equation}
\zeta_{j_0}(x)=\sigma(\mu_{j_0}(x)-E_{m_0})^{1/2}, \quad 
\sigma \in\{1,-1\}, \quad \mu_{j_0}(x)=E_{m_0}+\zeta_{j_0}(x)^2, 
\lb{ch3.54}
\end{equation}
for some $x$ in an open interval centered around $x_0$, the Dubrovin
equation \eqref{ch3.31} for $\mu_{j_0}$ becomes
\begin{equation}
\zeta_{j_0,x}(x)\underset{x\to x_0}{=} -2i\sigma 
\Bigg(\prod_{\substack{m=0\\m\neq m_0}}^{2n} (E_{m_0}-E_m)\Bigg)^{1/2}
\prod_{\substack{k=1\\k\neq j_0}}^{n}
(E_{m_0}-\mu_k(x))\big(1+\Oh\big(\zeta_{j_0}(x)^2\big)\big). \lb{ch3.55}
\end{equation}
Thus, $\hat\mu_{j_0}(x)$ does not pause at $(E_{m_0},0)$ for $x$ in a
small interval centered around $x_0$, and hence relations
\eqref{ch3.50a}--\eqref{ch3.52} extend to
$\Omega_\mu$. We have now established relations
\eqref{ch2.17}--\eqref{ch2.19} on $\bbC\times\Omega_\mu$, and one can
now proceed as in  Section~\ref{chs2} to obtain \eqref{ch3.41}.
\end{proof}

\section{The time-dependent CH formalism} \lb{chs4}

In this section we extend the algebro-geometric  formalism of 
Section \ref{chs3} to the time-dependent CH hierarchy. For most of 
this section we will assume the following hypothesis.
\begin{hypothesis}\label{hypo-ch4.1} 
Suppose that $u\colon \bbR^2\to\bbC$ satisfies
\begin{align}
&u(\dott,t)\in C^\infty(\bbR), \,  
\f{\partial^m u}{\partial x^m}(\dott,t)\in L^\infty(\bbR), \; 
m\in\bbN_{0}, t\in\bbR, \no \\
&u(x,\dott), u_{xx}(x,\dott)\in C^1(\bbR), \; x\in\bbR. \label{ch4.0}
\end{align}
\end{hypothesis}

The basic problem in the analysis of algebro-geometric solutions of 
the CH hierarchy consists in solving the time-dependent $r$th CH
flow with initial data a stationary solution of the $n$th equation
in the hierarchy. More precisely, given $n\in\bbN_0$, consider a solution
$u^{(0)}$ of the $n$th stationary CH equation $\sCH_n(u^{(0)})=0$ 
associated with $\calK_n$ and a given set of integration constants
$\{c_\ell\}_{\ell=1,\dots,n}\subset\bbC$. Next, let $r\in\bbN_0$; we
intend to construct a solution $u$ of the $r$th CH flow $\CH_r(u)=0$
with $u(t_{0,r})=u^{(0)}$ for some $t_{0,r}\in\bbR$. To emphasize that the
integration constants in the definitions of the stationary and the
time-dependent CH equations are independent of each other, we
indicate this by adding a tilde on all the time-dependent quantities.
Hence we shall employ the notation $\ti V_r$, $\ti F_r$,
$\ti G_{r}$, $\ti H_r$, $\tilde f_{s}$, $\tilde g_{s}$, $\tilde h_{s}$,
 $\tilde c_{s}$, etc., in order to distinguish it {}from 
$V_n$, $F_n$, $G_n$, $H_n$, $f_{\ell}$, $g_{\ell}$, $h_{\ell}$, 
$c_{\ell}$, etc., in the following. In addition,  we will follow a more
elaborate notation inspired by Hirota's $\tau$-function approach and
indicate the individual $r$th $\CH$ flow by a separate time variable $t_r
\in \bbR$. 
 
Summing up, we are seeking a solution $u$ of
\begin{align}
&\CH_r(u)= 4u_{t_r}-u_{xxt_r}+(u_{xxx}-4u_{x})\tilde f_{r}
-2(4u-u_{xx})\tilde f_{r,x}=0, \lb{ch4.1} \\
&u(x,t_{0,r})=u^{(0)}(x), \quad x\in\bbR, \no \\
&\sCH_n(u^{(0)})=(u_{xxx}-4u_{x}) f_{n}
-2(4u-u_{xx}) f_{n,x}=0, \lb{ch4.2} 
\end{align}
for some $t_{0,r}\in\bbR$, $n,r\in\bbN_0$, where $u$
satisfies \eqref{ch4.0}. Actually, relying on the isospectral property
of the CH flows, we will go a step further and assume
 \eqref{ch4.2} not only at $t_r=t_{0,r}$ but for all $t_r\in\bbR$.
Hence, we start with
\begin{align}
U_{t_r}(z,x,t_r)-\ti V_{r,x}(z,x,t_r)
+[U(z,x,t_r),\ti V_r(z,x,t_r)]&=0, \quad (z,x,t_r)\in\bbC\times\bbR^2,
\lb{ch4.3} \\ 
-V_{n,x}(z,x,t_r)+[U(z,x,t_r),V_n(z,x,t_r)]&=0,
\quad (z,x,t_r)\in\bbC\times\bbR^2, \lb{ch4.3A}
\end{align}
where (cf. \eqref{ch2.24}--\eqref{ch2.26})
\begin{align}
U(z,x,t_r)&=\begin{pmatrix} -1 &1\\
z^{-1}(4u(x,t_r)-u_{xx}(x,t_r)) &1 \end{pmatrix}, \no\\
\ti V_r(z,x,t_r)&=\begin{pmatrix} -\ti G_{r}(z,x,t_r)& \ti
F_{r}(z,x,t_r) \\ z^{-1} \ti H_{r}(z,x,t_r) & \ti G_{r}(z,x,t_r)
\end{pmatrix}, \lb{ch4.3B} \\
\ti V_n(z,x,t_r)&=\begin{pmatrix} -G_{n}(z,x,t_r)& F_{n}(z,x,t_r)\\
    z^{-1} H_{n}(z,x,t_r) &G_{n}(z,x,t_r) \end{pmatrix}, \no 
\end{align}
and
\begin{align}
F_n(z,x,t_r)&=\sum_{\ell=0}^n f_{n-\ell}(x,t_r)z^\ell=
\prod_{j=1}^n (z-\mu_j(x,t_r)), \lb{ch4.6f} \\
G_{n}(z,x,t_r)&=\sum_{\ell=0}^{n} g_{n-\ell}(x,t_r)z^\ell,  \lb{ch4.6g} \\
H_n(z,x,t_r)&=\sum_{\ell=0}^n h_{n-\ell}(x,t_r)z^\ell=
h_0(x,t_r)\prod_{j=1}^n (z-\nu_j(x,t_r)), \lb{ch4.6h} \\
h_0(x,t_r)&=4u(x,t_r)+2u_x(x,t_r), \lb{ch4.6i} \\
\ti F_r(z,x,t_r)&=\sum_{s=0}^r \tilde f_{r-s}(x,t_r)z^s,
\lb{ch4.6j} \\
\ti G_{r}(z,x,t_r)&=\sum_{s=0}^{r} \tilde g_{r-s}(x,t_r)z^s, 
\lb{ch4.6k} \\ 
\ti H_r(z,x,t_r)&=\sum_{s=0}^r \tilde h_{r-s}(x,t_r)z^s,
\lb{ch4.6l} \\
\tilde h_0(x,t_r)&=4u(x,t_r)+2u_x(x,t_r), \lb{ch4.6m}
\end{align}
for fixed $n,r\in\bbN_0$. Here $f_\ell(x,t_r)$, $\tilde f_s(x,t_r)$,
$g_\ell(x,t_r)$, $\tilde g_s(x,t_r)$, $h_\ell(x,t_r)$, and $\tilde
h_s(x,t_r)$ for $\ell=0,\dots,n$, $s=0,\dots,r$, are defined as in
\eqref{ch2.3}, \eqref{ch2.7}, and \eqref{ch2.8} with $u(x)$ replaced by
$u(x,t_r)$, etc., and with appropriate integration constants. Explicitly,
\eqref{ch4.3},
\eqref{ch4.3A} are equivalent to 
\begin{align}
4u_{t_r}(x,t_r)-&u_{xxt_r}(x,t_r)-\ti H_{r,x}(z,x,t_r)
+2\ti H_r(z,x,t_r)\no \\
-2(4u(x,t_r)&-u_{xx}(x,t_r))\ti G_{r}(z,x,t_r)=0, 
\lb{ch4.6a} \\
\ti F_{r,x}(z,x,t_r) &=2\ti G_{r}(z,x,t_r)-2\ti F_r(z,x,t_r),
\lb{ch4.6b} \\  
z\ti G_{r,x}(z,x,t_r)&=(4u(x,t_r)-u_{xx}(x,t_r))
\ti F_r(z,x,t_r) -\ti H_{r}(z,x,t_r)  \lb{ch4.6c}
\end{align}
and
\begin{align}
F_{n,x}(z,x,t_r)&=2 G_{n}(z,x,t_r)- 2F_n(z,x,t_r),  \lb{ch4.4a}\\
H_{n,x}(z,x,t_r)&=2H_n(z,x,t_r)-2(4u(x,t_r)-u_{xx}(x,t_r))
G_{n}(z,x,t_r),  
\lb{ch4.4b}\\
zG_{n,x}(z,x,t_r)&=(4u(x,t_r)-u_{xx}(x,t_r))  F_n(z,x,t_r)
- H_{n}(z,x,t_r), \lb{ch4.4c} 
\end{align}

First we will assume the existence of a solution of equations 
\eqref{ch4.6a}--\eqref{ch4.4c} and derive an explicit
formula for $u$ in terms of Riemann theta functions. In addition, we will
show in Theorem \ref{theorem-ch4.10} that \eqref{ch4.6a}--\eqref{ch4.4c}
 and hence the algebro-geometric initial value problem \eqref{ch4.1}, 
\eqref{ch4.2} has a solution at least locally, that is, for
$(x,t_r)\in\Omega$ for some open and connected set
$\Omega\subseteq\bbR^2$.

One observes that equations \eqref{ch2.16}--\eqref{ch2.39} apply to 
$F_n$, $G_{n}$, $H_n$, $f_\ell$, $g_\ell$, and $h_\ell$ and  
\eqref{ch2.3}--\eqref{ch2.10}, \eqref{ch2.24}--\eqref{ch2.26}, with 
$n$ replaced by $r$ and $c_{\ell}$ replaced by $\tilde c_{\ell}$, apply
to $\ti F_r$, $\ti G_{r}$, $\ti H_r$, $\tilde f_\ell$, $\tilde g_\ell$,
and $\tilde h_\ell$. In particular, the fundamental identity
\eqref{ch2.23} holds,
\begin{equation}
z^2 G_{n}(z,x,t_r)^2+zF_n(z,x,t_r) H_n(z,x,t_r)=R_{2n+2}(z), \lb{ch4.5}
\end{equation}
and the hyperelliptic curve $\calK_n$ is still given by
\begin{equation}
\calK_n \colon \calF_n(z,y)=y^2-R_{2n+2}(z)=0, \quad 
R_{2n+2}(z)=\prod_{m=0}^{2n+1} (z-E_m), \lb{ch4.7}
\end{equation}
assuming \eqref{ch3.3a} for the remainder of this
section, that is,
\begin{equation}
E_0=0, \; E_1,\dots,E_{2n+1}\in\bbC\setminus\{0\}. \lb{ch4.8}
\end{equation}
In analogy to equations \eqref{ch3.6}, \eqref{ch3.7} we define
\begin{align}
\hat\mu_j(x,t_r)&=(\mu_j(x,t_r),-\mu_j(x,t_r)
G_{n}(\mu_j(x,t_r),x,t_r))\in\calK_n, \lb{ch4.10a}\\
& \hspace*{3.6cm} j=1,\dots,n, \; (x,t_r)\in\bbR^2, \no \\
\hat\nu_j(x,t_r)&=(\nu_j(x,t_r),\nu_j(x,t_r)
G_{n}(\nu_j(x,t_r),x,t_r))\in\calK_n, \lb{ch4.10b} \\
& \hspace*{3.2cm} j=1,\dots,n, \; (x,t_r)\in\bbR^2. \no 
\end{align}
As in Section \ref{chs3}, the regularity assumptions \eqref{ch4.0} on 
$u$ imply
analogous regularity properties of $F_n$, $H_{n}$, $\mu_j$, and $\nu_k$. 

Next, one defines the meromorphic function $\phi(\dott,x,t_r)$ on $\calK_n$ by
\begin{align}
\phi(P,x,t_r)&=\f{y-zG_{n}(z,x,t_r)}{F_n(z,x,t_r)} \lb{ch4.9a} \\
&=\f{zH_{n}(z,x,t_r)}{y+zG_{n}(z,x,t_r)}, 
\quad P=(z,y)\in\calK_n\setminus\{\Pinfpm \}, \;  
(x,t_r)\in\bbR^2. \lb{ch4.9b} 
\end{align}
Assuming \eqref{ch4.8}, the divisor $(\phi(\dott,x,t_r))$ of
$\phi(\dott,x,t_r)$ reads
\begin{equation}
(\phi(\dott,x,t_r))=\calD_{P_0\hat{\ul\nu}(x,t_r)}
-\calD_{\Pinfp\hat{\ul\mu}(x,t_r)},
\lb{ch4.10d}
\end{equation}
with
\begin{equation}
\hat{\ul\mu}=\{\hat\mu_1,\dots,\hat\mu_n\}, \, 
\hat{\ul\nu}=\{\hat\nu_1,\dots,\hat\nu_n\} \in\sigma^n \calK_n.
\lb{ch4.10da}
\end{equation}
The corresponding time-dependent vector $\Psi$, 
\begin{align}
&\Psi(P,x,x_0,t_r,t_{0,r})=
\begin{pmatrix} \psi_1(P,x,x_0,t_r,t_{0,r}) \\
\psi_2(P,x,x_0,t_r,t_{0,r})
\end{pmatrix}, \lb{ch4.11} \\
& \hspace*{1cm} P\in\calK_n\setminus \{\Pinfpm\}, \;
(x,x_0,t_r,t_{0,r})\in\bbR^4 \no
\end{align}
is defined by
\begin{align}
\psi_1(P,x,x_0,t_r,t_{0,r})&=\exp\bigg(-
\int_{t_{0,r}}^{t_r}ds\,\big((1/z)\ti F_r(z,x_0,s)\phi(P,x_0,s)+
\ti G_{r}(z,x_0,s) \big) \no \\
& \qquad \qquad  -(1/z)\int_{x_0}^x
dx'\,\phi(P,x',t_r)-(x-x_0)\bigg),\lb{ch4.12a} \\
\psi_2(P,x,x_0,t_r,t_{0,r})&=-\psi_1(P,x,x_0,t_r,t_{0,r}) \phi(P,x,t_r)/z.
\lb{ch4.12b} 
\end{align}
The properties of $\phi$ can now be summarized as follows.

\begin{lemma} \lb{lemma-ch4.2}
Assume Hypothesis \ref{hypo-ch4.1} and \eqref{ch4.3},
\eqref{ch4.3A}. Moreover, let
$P=(z,y)\in\calK_n\setminus\{\Pinfp,\Pinfm,P_0\}$ and 
$(x,t_r)\in\bbR^2$. Then $\phi$ satisfies
\begin{align}
&\phi_x(P,x,t_r)-z^{-1}\phi(P,x,t_r)^2-2\phi(P,x,t_r)
+4u(x,t_r)-u_{xx}(x,t_r)=0, \lb{ch4.13a} \\
&\phi_{t_r}(P,x,t_r)=(4u(x,t_r)-u_{xx}(x,t_r))\ti
F_r(z,x,t_r) \no\\
&\hspace*{2.2cm} - \ti H_r (z,x,t_r)
 +2(\ti F_r(z,x,t_r)\phi(P,x,t_r))_x\lb{ch4.13b}\\
&\hspace*{1.85cm}=(1/z)\ti F_r(z,x,t_r)\phi(P,x,t_r)^2 +2\ti
G_{r}(z,x,t_r) \phi(P,x,t_r) \no \\
& \hspace*{2.2cm} -\ti H_r (z,x,t_r),  \lb{ch4.13ba} \\
&\phi(P,x,t_r)\phi(P^*,x,t_r)=-\f{zH_{n}(z,x,t_r)}{F_n(z,x,t_r)},
\lb{ch4.13d}\\
&\phi(P,x,t_r)+\phi(P^*,x,t_r)=-2\f{zG_{n}(z,x,t_r)}{F_n(z,x,t_r)},
\lb{ch4.13c}\\
&\phi(P,x,t_r)-\phi(P^*,x,t_r)=\f{2y}{F_n(z,x,t_r)}. \lb{ch4.13e}
\end{align}
\end{lemma}
\begin{proof}
Equations \eqref{ch4.13a} and \eqref{ch4.13d}--\eqref{ch4.13e}
are proved as in Lemma \ref{lemma-ch3.1}.  To prove
\eqref{ch4.13ba} one first observes that
\begin{equation}
\big(\partial_x-((1/z)\phi+1)\big)\big(\phi_{t_r}-(1/z)\ti F_r\phi^2
-2\ti G_{r}
\phi+\ti H_r \big)=0 \lb{ch4.15}
\end{equation}
using \eqref{ch4.13a} and relations \eqref{ch4.6a}--\eqref{ch4.6c}
repeatedly. Thus,
\begin{equation}
\phi_{t_r}-\f{1}{z}\ti F_r\phi^2 -2\ti G_{r}
\phi+\ti H_r
=C\exp\bigg(\int^x dx' \, ((1/z)\phi+1)\bigg), \lb{ch4.16}
\end{equation}
where the left-hand side is meromorphic in a neighborhood of $\Pinfm$,
while the right-hand side is meromorphic near $\Pinfm$ only if $C=0$.
This proves \eqref{ch4.13ba}. 

Using \eqref{ch4.6b} and \eqref{ch4.13a} one obtains
\begin{equation}
(4u-u_{xx})\ti F_r+2(\ti F_r\phi)_x=
2\ti G_r\phi+(1/z)\phi^2\ti F_r. \lb{ch4.17}
\end{equation}
Combining this result with \eqref{ch4.13ba} one concludes that
\eqref{ch4.13b} holds.
\end{proof}
Using relations \eqref{ch4.4a}--\eqref{ch4.4c} and 
\eqref{ch4.6a}--\eqref{ch4.6c}, we next 
determine the time evolution of $F_n$, $G_n$, and $H_n$.
\begin{lemma} \lb{lemma-ch4.3}
Assume Hypothesis \ref{hypo-ch4.1} and \eqref{ch4.3},
\eqref{ch4.3A}. In addition, let $(z,x,t_r)\in\bbC\times\bbR^2$.  Then
\begin{align}
&F_{n,t_r}(z,x,t_r)=2(G_{n}(z,x,t_r)\ti F_r(z,x,t_r)
-F_n(z,x,t_r)\ti G_{r}(z,x,t_r) ), \lb{ch4.18a} \\
&zG_{n,t_r}(z,x,t_r)=F_n(z,x,t_r) \ti H_r(z,x,t_r)
-H_n(z,x,t_r)\ti F_r(z,x,t_r), \lb{ch4.18b} \\
&H_{n,t_r}(z,x,t_r)=2(H_n(z,x,t_r)\ti G_{r}(z,x,t_r)
- G_{n}(z,x,t_r) \ti H_r(z,x,t_r)). \lb{ch4.18c}
\end{align}
Equations \eqref{ch4.18a}--\eqref{ch4.18c} are equivalent to 
\begin{equation}
V_{n,t_r}(z,x,t_r)=[\ti V_r(z,x,t_r),V_n(z,x,t_r)].\lb{ch4.19}
\end{equation}
\end{lemma}
\begin{proof}
We prove \eqref{ch4.18a} by using \eqref{ch4.13e} which shows that
\begin{equation}
(\phi(P)-\phi(P^*))_{t_r}=-2\f{yF_{n,t_r}}{F_n^2}. \lb{ch4.20}
\end{equation}
However, the left-hand side of \eqref{ch4.20} also equals
\begin{equation}
\phi(P)_{t_r}-\phi(P^*)_{t_r}=\f{4y}{F_n^2}(\ti G_{r} F_n-\ti F_r
G_{n}), \lb{ch4.21}
\end{equation}
using \eqref{ch4.13ba}, \eqref{ch4.13c}, and \eqref{ch4.13e}. Combining
\eqref{ch4.20} and \eqref{ch4.21} proves \eqref{ch4.18a}.  Similarly, to
prove \eqref{ch4.18b}, we use \eqref{ch4.13c} to write
\begin{equation}
(\phi(P)+\phi(P^*))_{t_r}=-\f{2z}{F_n^2}\big(G_{n,t_r}F_n
-G_{n}F_{n,t_r}\big). \lb{ch4.22}
\end{equation}
Here  the left-hand side can be expressed as
\begin{equation}
\phi(P)_{t_r}+\phi(P^*)_{t_r}=2\f{zG_{n}}{F_{n}^2}F_{n,t_{r}}+
\f{2}{F_{n}}(\ti F_r H_n-\ti H_r F_n), \lb{ch4.23}
\end{equation}
using \eqref{ch4.13ba}, \eqref{ch4.13d}, and \eqref{ch4.13c}.  Combining
\eqref{ch4.22} and \eqref{ch4.23}, using \eqref{ch4.18a},
proves \eqref{ch4.18b}.  Finally, \eqref{ch4.18c} follows by differentiating
\eqref{ch2.23}, that is, $(z G_{n})^2+z F_n H_n=R_{2n+2}$, with
respect to $t_r$, and using \eqref{ch4.18a} and \eqref{ch4.18b}.
\end{proof}

Lemmas \ref{lemma-ch4.2} and \ref{lemma-ch4.3} permit one to 
characterize $\Psi$.

\begin{lemma} \lb{lemma-ch4.4}
Assume Hypothesis \ref{hypo-ch4.1} and \eqref{ch4.3}, \eqref{ch4.3A}.
Moreover, let $P=(z,y)\in\calK_n\setminus\{\Pinfp,\Pinfm,P_0 \}$ and 
$(x,x_0,t_r,t_{0,r})\in\bbR^4$. Then the Baker--Akhiezer
vector $\Psi$ satisfies
\begin{align}
&\Psi_x(P,x,x_0,t_r,t_{0,r})=U(z,x,t_r)\Psi(P,x,x_0,t_r,t_{0,r}),
\lb{ch4.14a} \\
&-y\Psi(P,x,x_0,t_r,t_{0,r})=zV_n(z,x,t_r)\Psi(P,x,x_0,t_r,t_{0,r}),
\lb{ch4.14ab} \\
&\Psi_{t_r}(P,x,x_0,t_r,t_{0,r})=\ti
V_r(z,x,t_r)\Psi(P,x,x_0,t_r,t_{0,r}),
\lb{ch4.14b} \\
&\psi_1(P,x,x_0,t_r,t_{0,r}) 
=\bigg(\f{F_n(z,x,t_r)}{F_n(z,x_0,t_{0,r})}\bigg)^{1/2}
\times \lb{ch4.24a} \\
& \times \exp\bigg(-(y/z)\int_{t_{0,r}}^{t_r}ds\, \ti
F_r(z,x_0,s)F_n(z,x_0,s)^{-1} -(y/z)\int_{x_0}^x dx'F_n(z,x',t_r)^{-1}
\bigg), \no \\
&\psi_1(P,x,x_0,t_r,t_{0,r})\psi_1(P^*,x,x_0,t_r,t_{0,r})
=\f{F_n(z,x,t_r)}{F_n(z,x_0,t_{0,r})},\lb{ch4.24c} \\
&\psi_2(P,x,x_0,t_r,t_{0,r})\psi_2(P^*,x,x_0,t_r,t_{0,r})
=-\f{H_n(z,x,t_r)}{zF_n(z,x_0,t_{0,r})},\lb{ch4.24d} \\
&\psi_1(P,x,x_0,t_r,t_{0,r})\psi_2(P^*,x,x_0,t_r,t_{0,r}) \no \\
& \quad +\psi_1(P^*,x,x_0,t_r,t_{0,r})\psi_2(P,x,x_0,t_r,t_{0,r})
=2\f{G_{n}(z,x,t_r)}{F_n(z,x_0,t_{0,r})}, \lb{ch4.24e} \\
&\psi_1(P,x,x_0,t_r,t_{0,r})\psi_2(P^*,x,x_0,t_r,t_{0,r}) \no \\
& \quad -\psi_1(P^*,x,x_0,t_r,t_{0,r})\psi_2(P,x,x_0,t_r,t_{0,r})
=\f{2y}{zF_n(z,x_0,t_{0,r})}. \lb{ch4.24f}
\end{align}
In addition, as long as the zeros of $F_n(\dott,x,t_r)$ are all simple for
$(x,t_r), (x_0,t_{0,r})\in\Omega$, $\Omega\subseteq\bbR^2$ open and
connected, $\Psi(\dott,x,x_0,t_r,t_{0,r})$, 
$(x,t_r),(x_0,t_{0,r})\in\Omega$, is meromorphic on $\calK_n
\setminus\{P_0,\Pinfpm\}$.
\end{lemma}
\begin{proof}
By \eqref{ch4.12a}, $\psi_1(\dott,x,x_0,t_r,t_{0,r})$ is meromorphic on 
$\calK_n\setminus\{\Pinfpm\}$ away {}from
the poles $\hat \mu_j(x_0,s)$ of $\phi(\dott,x_0,s)$ and $\hat
\mu_k(x^\prime,t_r)$ of $\phi(\dott,x^\prime,t_r)$. 
That $\psi_1(\dott,x,x_0,t_r,t_{0,r})$ is meromorphic on
$\calK_n\setminus\{\Pinfpm\}$ if 
$F_n(\dott,x,t_r)$ has only simple zeros is a consequence of (cf.\
\eqref{ch3.29})
\begin{equation}
-\f{1}{z}\phi(P,x',t_r)\underset{P\to\hat\mu_j(x',t_r)}{=}
\f{\partial}{\partial x'}\ln\big(F_n(z,x',t_r) \big)+\Oh(1) \text{  as
$z\to\mu_j(x',t_r)$}, \lb{ch4.24fA}
\end{equation}
and 
\begin{equation}
-\f{1}{z}\widetilde F_r(z,x_0,s)\phi(P,x_0,s)
\underset{P\to\hat\mu_j(x_0,s)}{=}
\f{\partial}{\partial s}\ln\big(F_n(z,x_0,s) \big)+\Oh(1) 
\text{  as $z\to\mu_j(x_0,s)$},  \lb{ch4.24fB}
\end{equation}
using \eqref{ch4.10a}, \eqref{ch4.9a}, and \eqref{ch4.18a}. 
This follows {}from \eqref{ch4.12a} by restricting $P$ to
a sufficiently small neighborhood $\calU_j(x_{0})$ of
$\{\hmu_j(x_{0},s)\in\calK_{n}\mid (x_{0},s)\in\Omega, \,s\in
[t_{0,r},t_r] \}$ such that $\hmu_k(x_{0},s)\notin \calU_j(x_{0})$
for all $s\in [t_{0,r},t_r]$ and all
$k\in\{1,\dots,n\}\setminus\{j\}$ and by simultaneously restricting
$P$ to a sufficiently small neighborhood $\calU_j(t_r)$ of
$\{\hmu_j(x^{\prime},t_r)\in\calK_{n}\mid (x^{\prime},t_r)\in\Omega,
\,x^{\prime}\in [x_{0},x] \}$ such that $\hmu_k(x^{\prime},t_r)\notin
\calU_j(t_r)$ for all $x^{\prime}\in [x_{0},x]$ and all
$k\in\{1,\dots,n\}\setminus\{j\}$. 
By \eqref{ch4.12b} and the fact that $\phi$ is meromorphic on $\calK_n$ one
concludes that $\psi_2$ is meromorphic on
$\calK_n\setminus\{\Pinfpm\}$ as well.
Relations \eqref{ch4.14a} and \eqref{ch4.14ab} follow as in Lemma
\ref{lemma-ch3.1}, while the time evolution
\eqref{ch4.14b} is a consequence of the definition of $\Psi$
in \eqref{ch4.12a}, \eqref{ch4.12b} as well as \eqref{ch4.13ba},
rewriting
\begin{equation}
(1/z)\phi_{t_r}=\big((1/z)2\phi\ti F_r+\ti G_{r}\big)_x,
\lb{ch4.17a}
\end{equation}
using \eqref{ch4.6c} and \eqref{ch4.13b}. To prove \eqref{ch4.24a}
we recall the definition \eqref{ch4.12a}, that is, 
\begin{align}
&\psi_1(P,x,x_0,t_r,t_{0,r}) =\exp\bigg(-(x-x_0)-(1/z)\int_{x_0}^x
dx'\,\phi(P,x',t_r) \no \\ 
&\qquad \qquad \quad -
\int_{t_{0,r}}^{t_r}ds\,\big((1/z) \ti F_r(z,x_0,s)\phi(P,x_0,s)+\ti
G_{r}(z,x_0,s)\big) \bigg) \no \\
& \quad =\bigg(\f{F_n(z,x,t_r)}{F_n(z,x_0,t_{r})}\bigg)^{1/2}
\exp\bigg( -(y/z)\int_{x_0}^x dx'F_n(z,x',t_r)^{-1} \lb{ch4.25}\\ 
&\qquad\qquad\quad -\int_{t_{0,r}}^{t_r}ds\,\big((1/z) \ti
F_r(z,x_0,s)\phi(P,x_0,s)+\ti G_{r}(z,x_0,s)\big)\bigg), \no
\end{align}
using the calculation leading to \eqref{ch3.24}.  Equations
\eqref{ch4.9a} and \eqref{ch4.18a} show that
\begin{equation}
\f{1}{z} \ti F_r(z,x_0,s)\phi(P,x_0,s)+\ti
G_{r}(z,x_0,s)=\f{y}{z}\,\f{\ti F_r(z,x_0,s)}{F_n(z,x_0,s)}
-\f12\,\f{F_{n,t_r}(z,x_0,s)}{F_n(z,x_0,s)},\lb{ch4.25a}
\end{equation}
which inserted into \eqref{ch4.25} yields \eqref{ch4.24a}.
 Evaluating \eqref{ch4.24a} at 
the points $P$ and $P^*$ and multiplying
the resulting expressions yields \eqref{ch4.24c}. The remaining statements
are direct consequences of \eqref{ch4.13c}--\eqref{ch4.13e} and \eqref{ch4.24a}.
\end{proof}

Next, we turn to the time evolution of the quantities $\mu_j$ and
$\nu_j$ assuming \eqref{ch3.30}, that is,
\begin{equation}
E_0=0, \; E_m\in\bbC\setminus\{0\}, \; E_m\neq E_{m'} \text{ for } 
m\neq m', \, m,m'=1,\dots,2n.  \lb{ch4.25b}
\end{equation}

\begin{lemma}\lb{lemma-ch4.5}
Assume Hypothesis \ref{hypo-ch4.1}, \eqref{ch4.25b}, and \eqref{ch4.3},
\eqref{ch4.3A} on an open and connected set
$\ti\Omega_\mu\subseteq\bbR^2$. Moreover, suppose that the zeros $\mu_j$,
$j=1,\dots,n$, of $F_n(\dott)$ remain distinct and nonzero on
$\ti\Omega_\mu$. Then $\{\hat\mu_j\}_{j=1,\dots,n}$, defined by
\eqref{ch4.10a}, satisfies the following first-order system of
differential equations
\begin{align}
\mu_{j,x}(x,t_r)&=2\mu_j(x,t_r)^{-1}y(\hat\mu_j(x,t_r))
\prod_{\substack{\ell=1\\ \ell\neq j}}^n(\mu_j(x,t_r)-\mu_\ell(x,t_r))^{-1},
\lb{ch4.26}\\
\mu_{j,t_r}(x,t_r)&=2\ti F_r(\mu_j(x,t_r),x,t_r)\times \no \\
& \quad \times \mu_j(x,t_r)^{-1} y(\hat\mu_j(x,t_r))
\prod_{\substack{\ell=1\\ \ell\neq
j}}^n(\mu_j(x,t_r)-\mu_\ell(x,t_r))^{-1}, \lb{ch4.26a} \\  
& \hspace*{4.5cm} j=1, \dots, n, \, (x,t_r)\in\ti\Omega_\mu. \no
\end{align}
Next, assume $\calK_n$ to be nonsingular and introduce the initial
condition
\begin{equation}
\{\hat\mu_j(x_0,t_{0,r})\}_{j=1, \dots, n}\subset\calK_n \lb{ch4.27}
\end{equation}
for some $(x_0,t_{0,r})\in\bbR^2$, where $\mu_j(x_0,t_{0,r})\neq 0$,
$j=1,\dots,n$, are assumed to be distinct and $\ti
F_r(\mu_j(x_0,t_{0,r}),x_0,t_{0,r})\neq 0$, $j=1,\dots,n$. Then there
exists an open and connected set $\Omega_\mu\subseteq\bbR^2$, with
$(x_0,t_{0,r})\in\Omega_\mu$, such that the initial value
problem \eqref{ch4.26}--\eqref{ch4.27} has a unique
solution $\{\hat\mu_j\}_{j=1,\dots,n}\subset\calK_n$ satisfying
\begin{equation}
\hat\mu_j\in C^\infty(\Omega_\mu,\calK_n),\quad j=1, \dots, n, \lb{ch4.28}
\end{equation}
and $\mu_j$, $j=1,\dots,n$, remain distinct and nonzero on $\Omega_\mu$
with $\ti F_r(\mu_j)\neq 0$ on $\Omega_\mu$, $j=1,\dots,n$. \\ 
For the zeros $\{\nu_j\}_{j=1,\dots,n}$ of $H_n(\dott)$ similar statements
hold with $\mu_j$ and $\Omega_\mu$ replaced by $\nu_j$ and $\Omega_\nu$,
etc. In particular, $\{\hat\nu_j\}_{j=1,\dots,n}$, defined by
\eqref{ch4.10b}, satisfies the system  
\begin{align}
\nu_{j,x}(x,t_r)&=2(4u(x,t_r)-u_{xx}(x,t_r))
(4u(x,t_r)+2u_{x}(x,t_r))^{-1}\times \no \\
&\quad \times
\nu_j(x,t_r)^{-1}y(\hat\nu_j(x,t_r))\prod_{\substack{\ell=1\\ \ell\neq
j}}^n(\nu_j(x,t_r)-\nu_\ell(x,t_r))^{-1},
\lb{ch4.29} \\
\nu_{j,t_r}(x,t_r)&=2\ti H_r(\nu_j(x,t_r),x,t_r)
(4u(x,t_r)+2u_{x}(x,t_r))^{-1} \times\no \\
&\quad \times \nu_j(x,t_r)^{-1}y(\hat\nu_j(x,t_r))
\prod_{\substack{\ell=1\\ \ell\neq j}}^n
(\nu_j(x,t_r)-\nu_\ell(x,t_r))^{-1}, \lb{ch4.29a} \\
& \hspace*{4.4cm} j=1, \dots, n, \, (x,t_r)\in\Omega_\nu. \no 
\end{align}
\end{lemma}
\begin{proof}
It suffices to prove \eqref{ch4.26a} since the argument for 
\eqref{ch4.29a}
is analogous and that for \eqref{ch4.26} and \eqref{ch4.29}
has been given in the proof of Lemma \ref{lemma-ch3.2}. Inserting
$z=\mu_j(x,t_r)$ into \eqref{ch4.18a}, observing \eqref{ch4.10a}, yields
\begin{equation}
F_{n,t_r}(\mu_j)=-\mu_{j,t_r}
\prod_{\substack{\ell=1\\ \ell\neq j}}^n(\mu_j-\mu_{\ell})
=2\ti F_r(\mu_j) G_{n}(\mu_j)
=-2 \f{\ti F_r(\mu_j)}{\mu_j} y(\hat \mu_j). \lb{ch4.30}
\end{equation}
The rest is analogous to the proof of Lemma \ref{lemma-ch3.2}.
\end{proof}

Next we note the following trace formula, the $t_r$-dependent analog
of \eqref{ch3.36}.
\begin{lemma}\lb{lemma-ch4.6}
Assume Hypothesis \ref{hypo-ch4.1}, \eqref{ch4.3}, \eqref{ch4.3A}, and
let $(x,t_r)\in\bbR^2$. Then 
\begin{align}
u(x,t_r)&=\f12\sum_{j=1}^n \mu_{j}(x,t_{r})-\f14\sum_{m=1}^{2n+1}
E_{m}. \lb{ch4.31}
\end{align}
\end{lemma}

We also record the asymptotic properties of $\phi$, the analogs of
\eqref{ch3.38a} and \eqref{ch3.38b}.

\begin{lemma} \lb{lemma-ch4.7}
Assume Hypothesis \ref{hypo-ch4.1}, \eqref{ch4.3}, \eqref{ch4.3A}, and let
$P=(z,y)\in\calK_n\setminus\{\Pinfp,\Pinfm,P_0\}$, $(x,t_r)\in\bbR^2$.
Then
\begin{align}
&\phi(P,x,t_r)\underset{\zeta\to 0}{=}\begin{cases} 
-2\zeta^{-1}-2u(x,t_r)+u_x(x,t_r)+\Oh(\zeta), & P\to\Pinfp, \\
2u(x,t_r)+u_x(x,t_r)+\Oh(\zeta), & P\to\Pinfm, \end{cases} 
\quad \zeta=z^{-1},  \lb{ch4.31a} \\
&\phi(P,x,t_r)\underset{\zeta\to 0}{=}\f{\Big(\prod_{m=1}^{2n+1}
E_m\Big)^{1/2}}{f_n(x,t_r)}\zeta +\Oh(\zeta^2),
\quad  P\to P_0, \quad \zeta=z^{1/2}. \lb{ch4.31b} 
\end{align}
\end{lemma}

Since the proofs of Lemmas \ref{lemma-ch4.6} and \ref{lemma-ch4.7} are
identical to the corresponding stationary results in Lemmas
\ref{lemma-ch3.3} and \ref{lemma-ch3.4} we omit the corresponding
details. 

Next, recalling the definition of $\ti F_r(\mu_j)$ introduced in 
\eqref{tF} and also the definition of $\ul{\hatt B}_{Q_0}$ and
$\ul{\hatt\beta}_{Q_0}$ in \eqref{ch3.38m} and \eqref{ch3.38n},
respectively,  we now state the analog of Lemma \ref{lemma-ch3.6}, thereby
underscoring the marked differences between the CH hierarchy and other
completely integrable systems such as the KdV and  AKNS hierarchies. 

\begin{lemma} \lb{lemma-ch4.8}
Assume \eqref{ch4.25b} and suppose that
$\{\hat\mu_j\}_{j=1,\dots,n}$ satisfies the Dubrovin equations
\eqref{ch4.26}, \eqref{ch4.26a} on an open set $\Omega_\mu\subseteq\bbR^2$
such that $\mu_j$, $j=1,\dots,n$, remain distinct and nonzero on
$\Omega_\mu$ and that $\ti F_r(\mu_j)\neq 0$ on $\Omega_\mu$,
$j=1,\dots,n$. Introducing the associated divisor
$\calD_{\humu}\in\sigma^n\hatt\calK_n$,
$\humu=\{\hat\mu_1,\dots,\hat\mu_n\}\in\sigma^n\hatt\calK_n$, one computes
\begin{align}
\f{\partial}{\partial x}\ul \al_{Q_0}
(\calD_{\humu(x,t_r)})&=-\f{2}{\Psi_n(\ul\mu(x,t_r))}\ul c(1), \quad
(x,t_r)\in\Omega_\mu, \lb{ch4.80} \\
\f{\partial}{\partial t_r}\ul \al_{Q_0}
(\calD_{\humu(x,t_r)})&= -\f{2}{\Psi_n(\ul\mu(x,t_r))}
\bigg(\sum_{k=0}^{r\mini n} d_{r,k}(\ul E) \Psi_k(\ul\mu (x,t_r))\bigg)
\ul c(1) \no \\
& \quad +2 \bigg(\sum_{\ell=1\maxi (n+1-r)}^n d_{r,n+1-\ell}(\ul E)
 \ul c(\ell)\bigg), \quad (x,t_r)\in\Omega_\mu. \lb{ch4.81}
\end{align}
In particular, the Abel map does not linearize the divisor
$\calD_{\humu(\dott,\dott)}$ on $\Omega_\mu$. In addition,
\begin{align}
&\f{\partial}{\partial x}\sum_{j=1}^n \int_{Q_0}^{\hat\mu_j(x,t_r)}
\eta_1=-\f{2}{\Psi_n(\ul\mu (x,t_r))}, \quad (x,t_r)\in\Omega_\mu,
\lb{ch4.82}
\\ 
&\f{\partial}{\partial x}\ul{\hatt\beta}(\calD_{\ul{\hat\mu}(x,t_r)})
=2(0,\dots,0,1), \quad (x,t_r)\in\Omega_\mu, \lb{ch4.83} \\
&\f{\partial}{\partial t_r}\sum_{j=1}^n \int_{Q_0}^{\hat\mu_j(x,t_r)}
\eta_1=-\f{2}{\Psi_n(\ul\mu (x,t_r))}\sum_{k=0}^{r\mini n} 
d_{r,k}(\ul E) \Psi_k(\ul\mu (x,t_r))+2
d_{r,n}\delta_{n,r\mini n}, \no \\
& \hspace*{9.2cm} (x,t_r)\in\Omega_\mu, \lb{ch4.84} \\
&\f{\partial}{\partial t_r}\ul{\hatt\beta}(\calD_{\ul{\hat\mu}(x,t_r)})
=2\bigg(\sum_{s=0}^r \tilde c_{r-s}c_{s+1-n}(\ul E),\dots,\sum_{s=0}^r
\tilde c_{r-s}c_{s+1}(\ul E),\sum_{s=0}^r \tilde c_{r-s}c_{s}(\ul
E)\bigg), \no \\ 
& \hspace*{6.2cm} c_{-\ell}(\ul E)=0, \, \ell\in\bbN, \;
(x,t_r)\in\Omega_\mu. \lb{ch4.85} 
\end{align}
\end{lemma}
\begin{proof}
Let $(x,t_r)\in\Omega_\mu$. Since \eqref{ch4.80}, \eqref{ch4.82}, and
\eqref{ch4.83} are proved as in in the stationary context of Lemma
\ref{lemma-ch3.6}, we focus on the proofs of \eqref{ch4.81},
\eqref{ch4.84}, and \eqref{ch4.85}. Then, using \eqref{ch4.26a},
\eqref{ch3.38k}, \eqref{A8}, and \eqref{A1c}, \eqref{B18}, and \eqref{B19}
one obtains 
\begin{align}
&\f{\partial}{\partial t_r}\bigg(\sum_{j=1}^n \int_{Q_0}^{\hat\mu_j}
\ul\omega\bigg) =\sum_{j=1}^n \mu_{j,t_r}\sum_{k=1}^n \ul
c(k)\f{\mu_j^{k-1}}{y(\hat\mu_j)} =2\sum_{j=1}^n\sum_{k=1}^n \ul
c(k)\f{\mu_j^{k-1}}{\prod_{\substack{\ell=1\\ \ell\neq
j}}^n(\mu_j-\mu_\ell)}\f{\ti F_r(\mu_j)}{\mu_j} \no \\
&=2 \sum_{j=1}^n\sum_{k=1}^n \ul c(k)
\f{\mu_j^{k-1}}{\prod_{\substack{\ell=1\\ \ell\neq j}}^n
(\mu_j-\mu_\ell)}\bigg(-\sum_{m=0}^{r\mini n} d_{r,m}(\ul E)
\Psi_m(\ul\mu)\f{\Phi^{(j)}_{n-1}(\ul\mu)}{\Psi_n(\ul\mu)} \no \\
&\hspace*{5.1cm} +\sum_{m=1}^{r\mini n} d_{r,m}(\ul E)
\Phi^{(j)}_{m-1}(\ul\mu)\bigg) \no \\  
&=-2 \sum_{m=0}^{r\mini n} d_{r,m}(\ul E)
\f{\Psi_m(\ul\mu)}{\Psi_n(\ul\mu)}\sum_{k=1}^n\sum_{j=1}^n
\ul c(k) (U_n(\ul\mu))_{k,j} (U_n(\ul\mu))^{-1}_{j,1} \no \\
&\quad +2 \sum_{m=1}^{r\mini n} d_{r,m}(\ul E)
\sum_{k=1}^n\sum_{j=1}^n
\ul c(k) (U_n(\ul\mu))_{k,j} (U_n(\ul\mu))^{-1}_{j,n-m+1} \no \\
&=-\f{2}{\Psi_n(\ul\mu)}\sum_{m=0}^{r\mini n}d_{r,m}(\ul E)
\Psi_m(\ul\mu) \ul c(1)
+2\sum_{m=1}^{r\mini n} d_{r.m}(\ul E)\ul c(n-m+1) \no \\
&=-\f{2}{\Psi_n(\ul\mu)}\sum_{m=0}^{r\mini n}d_{r,m}(\ul E)
\Psi_m(\ul\mu) \ul c(1) 
+2 \sum_{m=1\maxi (n+1-r)}^n d_{r,n+1-m}(\ul E)\ul c(m). \lb{ch4.86}
\end{align}
Equation \eqref{ch4.84} is just a special case of \eqref{ch4.81} and
\eqref{ch4.85} follows as in \eqref{ch4.86} using again \eqref{A1c}.
\end{proof}
The analogous results hold for the corresponding divisor
$\calD_{\hunu (x,t_r)}$ associated with $\phi(\dott,x,t_r)$.

The fact that the Abel map does not effect a linearization of the divisor
$\calD_{\humu(x,t_r)}$ in the CH context is well-known and discussed
(using different approaches) by Constantin and McKean
\cite{ConstantinMcKean:1999}, Alber, Camassa, Fedorov, Holm, and Marsden
\cite{AlberCamassaFedorovHolmMarsden:2001}, Alber and Fedorov
\cite{AlberFedorov:2000}, \cite{AlberFedorov:2001}. A change of the 
variable $t_1$ in analogy to that in \eqref{ch3.38L} in the stationary 
context, which avoids the use of a meromorphic differential (cf.\ \eqref{ch3.38m},
\eqref{ch3.38n}) and linearizes the Abel map when considering the $\CH_1$ flow, is
discussed in \cite{Alber:2000}. That change of variables corresponds to the case
$r=1$ in \eqref{ch4.89}.

Next we turn to one of the principal results of this section, the 
representations of $\phi$ and $u$ in terms of the Riemann theta function
associated with $\calK_n$, assuming $\calK_n$ to be nonsingular. Recalling
\eqref{ch3.38q}--\eqref{ch3.38ya}, the analog of Theorem
\ref{theorem-ch3.7} in the stationary case then reads as follows. 

\begin{theorem}\lb{theorem-ch4.9} 
Suppose Hypothesis \ref{hypo-ch4.1} and \eqref{ch4.1}, \eqref{ch4.2} on
$\Omega$ subject to the constraint \eqref{ch4.25b}. In addition, let
$P\in\calK_n\setminus\{\Pinfp,P_0\}$ and 
$(x,t_r), (x_0,t_{0,r})\in\Omega$, where $\Omega\subseteq\bbR^2$ is open
and connected. Moreover, suppose that $\calD_{\hat{\ul\mu}(x,t_r)}$, or
equivalently, $\calD_{\hat{\ul\nu}(x,t_r)}$, is nonspecial for
$(x,t_r)\in\Omega$. Then $\phi$ and $u$ admit the representations 
\begin{align}
\phi (P,x,t_r) &= -2\frac{\theta (\ul z(\Pinfp,\humu(x,t_r)))\theta (\ul
z(P,\hunu(x,t_r)))}{\theta(\ul z(\Pinfp,\hunu(x,t_r)))\theta
(\ul z(P,\humu(x,t_r)))} \exp \bigg(-\int_{Q_0}^P
\omega_{\Pinfp,P_0}^{(3)}+e_0\bigg), \lb{ch4.87} \\ 
u(x,t_r)&=\f{1}{2}\sum_{j=1}^n \lambda_j -\f{1}{4}\sum_{m=1}^{2n+1} E_m  
 \no\\
&\quad +\f{1}{2}\sum_{j=1}^n U_j \f{\partial}{\partial w_j}\ln\bigg(
\f{\theta\big(\uz(\Pinfp,\hat{\ul\mu}(x,t_r))+\ul w\big)} 
{\theta\big(\uz(\Pinfm,\hat{\ul\mu}(x,t_r))+\ul w\big)}
\bigg)\bigg|_{\ul w=0}. \lb{ch4.88}
\end{align}
Moreover, let $\ti\Omega\subseteq\Omega$ be such that $\mu_j$,
$j=1,\dots,n$, are nonvanishing on $\ti\Omega$, Then, the constraint 
\begin{align}
&2(x-x_0)+2(t_r-t_{0,r})\sum_{s=0}^r \tilde c_{r-s}c_s(\ul E) \no \\
&=\bigg(-2 \int_{x_0}^x \f{dx'}{\prod_{k=1}^n \mu_k(x',t_r)} \no \\
&\qquad -2\sum_{k=0}^{r\mini n} d_{r,k}(\ul E) \int_{t_{0,r}}^{t_r} 
\f{\Psi_k(\ul\mu(x_0,t')}{\Psi_n(\ul\mu(x_0,t')}dt'\bigg)
\sum_{j=1}^n\bigg(\int_{a_j}\ti\omega^{(3)}_{\Pinfp,\Pinfm}\bigg) c_j(1)
\no \\
& \quad + 2(t_r-t_{0,r})\sum_{\ell=1\maxi (n+1-\ell)}^n 
d_{r,n+1-\ell}(\ul E)
\sum_{j=1}^n\bigg(\int_{a_j}\ti\omega^{(3)}_{\Pinfp,\Pinfm} \bigg)
c_j(\ell) \no \\ 
& \quad +\ln\bigg(\f{\theta\big(\ul z(\Pinfp,\humu(x,t_r))\big)
\theta\big(\ul z(\Pinfm,\humu(x_0,t_{0,r}))\big)} 
{\theta\big(\ul z(\Pinfm,\humu(x,t_r))\big)
\theta\big(\ul z(\Pinfp,\humu(x_0,t_{0,r}))\big)}\bigg), \lb{ch4.89} \\
& \hspace*{5.05cm} (x,t_r), (x_0,t_{0,r})\in\ti\Omega \no
\end{align}
holds, with
\begin{align}
\ul{\hat z} (\Pinfpm,\humu (x,t_r)))&=\ul{\hatt\Xi}_{Q_0}-\ul{\hatt
A}_{Q_0}(\Pinfpm) +\ul{\hatt\alpha}_{Q_0}(\calD_{\humu (x,t_r)}) \no \\
&=\ul{\hatt\Xi}_{Q_0}-\ul{\hatt
A}_{Q_0}(\Pinfpm) +\ul{\hatt\alpha}_{Q_0}(\calD_{\humu (x_0,t_{r})}) 
\no \\
& \quad -2\bigg(\int_{x_0}^x \f{dx'}{\Psi_n(\ul\mu(x',t_r)}\bigg)\,
\ul c(1) \lb{ch4.90} \\ 
&=\ul{\hatt\Xi}_{Q_0}-\ul{\hatt A}_{Q_0}(\Pinfpm)
+\ul{\hatt\alpha}_{Q_0}(\calD_{\humu (x,t_{0,r})}) \no \\
&\quad -2 \bigg(\sum_{k=0}^{r\mini n} d_{r,k}(\ul E) 
\int_{t_{0,r}}^{t_r} 
\f{\Psi_k(\ul\mu(x,t'))}{\Psi_n(\ul\mu(x,t'))}dt'\bigg)
\ul c(1) \no \\
& \quad +2 (t_r-t_{0,r})\bigg(\sum_{\ell=1\maxi (n+1-r)}^n
d_{r,n+1-\ell}(\ul E) \ul c(\ell)\bigg),  \lb{ch4.91} \\
&\hspace*{4.2cm} (x,t_r), (x_0,t_{0,r})\in\ti\Omega. \no
\end{align}
\end{theorem}
\begin{proof}
First, let $\ti{\ti\Omega}\subseteq\Omega$ be defined by requiring that
$\mu_j$, $j=1,\dots,n$, are distinct and nonvanishing on $\ti{\ti\Omega}$
and $\ti F_r(\mu_j)\neq 0$ on $\ti{\ti\Omega}$, $j=1,\dots,n$. The
representation \eqref{ch4.87} for $\phi$ on $\ti{\ti\Omega}$ then follows
by combining \eqref{ch4.10d}, \eqref{ch4.31a}, \eqref{ch4.31b}, and Theorem
\ref{taa17a} since $\calD_{\hat{\ul\mu}}$ and $\calD_{\hat{\ul\nu}}$ are
simultaneously nonspecial as discussed in the proof of Theorem
\ref{theorem-ch3.7}. The representation \eqref{ch4.88} for $u$ on
$\ti{\ti\Omega}$ follows {}from the trace formula \eqref{ch4.31} and
\eqref{g.62} (taking $k=1$). By continuity, \eqref{ch4.87} and
\eqref{ch4.88} extend from $\ti{\ti\Omega}$ to $\Omega$. The constraint
\eqref{ch4.89} then holds on $\ti{\ti\Omega}$ by combining
\eqref{ch4.82}--\eqref{ch4.85}, and \eqref{g.61c}. Equations
\eqref{ch4.90} and \eqref{ch4.91} are clear {}from \eqref{ch4.80} and
\eqref{ch4.81}. Again by continuity, \eqref{ch4.89}--\eqref{ch4.91}
extend {}from $\ti{\ti\Omega}$ to $\ti\Omega$.
\end{proof}

As discussed by Alber, Camassa, Fedorov, Holm, and Marsden 
\cite{AlberCamassaFedorovHolmMarsden:2001}, Alber and Fedorov
\cite{AlberFedorov:2000},
\cite{AlberFedorov:2001}, the algebro-geometric CH solution $u$ in
\eqref{ch4.88} is not meromorphic with respect to $x,t_r$, in general. In
more geometrical terms, the $\CH_r$ flows evolve on a nonlinear subvariety
(corresponding to the constraint \eqref{ch4.89}) of a generalized
Jacobian, topologically given by $J(\calK_n)\times \bbC^*$
($\bbC^*=\bbC\setminus\{0\}$). For discussions of generalized Jacobians in
this context we refer, for instance, to
\cite{Fedorov:1999},
\cite{GagnonHarnadWinternitzHurtubise:1985}, \cite{Gavrilov:1999}. Smooth
(i.e., $C^1$ with respect to $t_1$ and $C^3$ and hence $C^\infty$ with
respect to $x$) spatially periodic $\CH_1$ solutions $u$ are 
quasi-periodic in $t_1$ as shown by Constantin \cite{Constantin:1998}.

Without going into details we mention that our approach extends in a
straightforward manner to the Dym-type equation,
\begin{equation}
v_{xxt}+2vv_{xxx}+4v_xv_{xx}-4\kappa v=0, \quad
\kappa\in\bbR, \; (x,t)\in\bbR^2. \lb{ch4.92}
\end{equation}
The corresponding zero-curvature formalism leads to a trace formula
analogous to \eqref{ch4.31} (cf.\ \cite{AlberCamassaFedorovHolmMarsden:2001},
\cite{AlberFedorov:2000}, \cite{AlberFedorov:2001}). One needs to replace the
polynomial $R_{2n+2}(z)$ by $R_{2n+1}(z)=\prod_{m=0}^{2n} (z-E_m)$, which results
in a branch point $P_\infty$ at infinity, and replaces the (non-normalized)
differential $\tilde \omega^{(3)}_{\Pinfp,\Pinfm}$ of the third kind by the
(non-normalized) differential $\tilde \omega^{(2)}_{P_\infty}=z^ndz/y$ of
the second kind, etc. This approach (applied to the Dym equation 
$4\rho_t=\rho^3 \rho_{xxx}$, related to \eqref{ch4.92} by proper variable
transformations) was first realized by Novikov \cite{Novikov:1999} and
inspired our treatment of the CH hierarchy.

Expressing $\ti F_r$ in terms of $\Psi_k(\ul\mu)$ and hence in terms of
the theta function associated with $\calK_n$, one can use \eqref{ch4.24a}
to derive a theta function representation of $\psi_j$, $j=1,2$, in analogy
to the stationary case discussed in Remark \ref{remark-ch3.8}. We omit
further details. 

Up to this point we assumed Hypothesis \ref{hypo-ch4.1} together with the
basic equations \eqref{ch4.3} and \eqref{ch4.3A}. Next, we will show that
solvability of the Dubrovin equations \eqref{ch4.26} and \eqref{ch4.26a}
on $\Omega_\mu\subseteq\bbR^2$ in fact implies
equations \eqref{ch4.3} and \eqref{ch4.3A} on $\Omega_\mu$ and hence 
solves the algebro-geometric initial value problem \eqref{ch4.1},
\eqref{ch4.2} on $\Omega_\mu$. In this context we recall the definition of
$\ti F_r(\mu_j)$ in terms of $\mu_1,\dots,\mu_\g$, introduced in 
\eqref{tF}, \eqref{tid},
\begin{align}
\ti F_r(\mu_j)&=\sum_{k=0}^{r\mini n}
\tilde d_{r,k}(\ul E) \Phi^{(j)}_k(\ul \mu), \quad r\in\bbN_0, \;
\tilde c_0=1, \lb{chtF} \\
\tilde d_{r,k}(\ul E)&=\sum_{s=0}^{r-k} \tilde c_{r-k-s} c_s(\ul E),
\quad  k=0,\dots,r\mini n, \lb{chtd} 
\end{align}
in terms of a given set of integration constants $\{\tilde
c_1,\dots,\tilde c_r\}\subset\bbC$.

\begin{theorem}\lb{theorem-ch4.10}
Fix $n\in\bbN$ and assume \eqref{ch4.25b}. Suppose that 
$\{\hat\mu_j\}_{j=1,\dots,n}$ satisfies the Dubrovin equations
\eqref{ch4.26}, \eqref{ch4.26a} on an open and connected set
$\Omega_\mu\subseteq\bbR^2$, with $\ti F_r(\mu_j)$ in \eqref{ch4.26a}
expressed in terms of $\mu_k$, $k=1,\dots,n$, by \eqref{chtF},
\eqref{chtd}. Moreover, assume that $\mu_j$, $j=1,\dots,n$, remain 
distinct and nonzero  on $\Omega_\mu$ and that $\ti F_r(\mu_j)\neq 0$ on
$\Omega_\mu$, $j=1,\dots,n$. Then $u\in C^\infty(\Omega_\mu)$ defined by
\begin{equation}
u(x,t_r)=\f12\sum_{j=1}^n \mu_{j}(x,t_{r})-\f14\sum_{m=1}^{2n+1}
E_{m}, \lb{ch4.32}
\end{equation}
satisfies the $r$th CH equation \eqref{ch4.1}, that is,
\begin{equation}
\CH_r(u)=0 \text{  on $\Omega_\mu$},\lb{ch4.33}
\end{equation} 
with initial values satisfying the  $n$th stationary CH equation
\eqref{ch4.2}.
\end{theorem}
\begin{proof}
Given solutions $\hat\mu_j=(\mu_j,y(\hat\mu_j))\in
C^\infty(\Omega_\mu,\calK_n)$, $j=1,\dots,n$ of \eqref{ch4.26} and
\eqref{ch4.26a}, we define polynomials $F_n$, $G_{n}$, and $H_n$ on
$\Omega_\mu$ as in the stationary case, cf.\ Theorem \ref{theorem-ch3.10},
with properties
\begin{align}
F_n(z)&=\prod_{j=1}^n (z-\mu_j), \lb{ch4.34}\\
G_n(z)&=F_n(z)+\f12 F_{n,x}(z), \lb{ch4.35} \\
zG_{n,x}(z)&=(4u-u_{xx}) F_n(z)-H_n(z), \lb{ch4.36} \\
H_{n,x}(z)&=2H_n(z)-2(4u-u_{xx})G_n(z), \lb{ch4.37} \\
R_{2n+2}(z)&=z^2 G_n(z)^2+z F_n(z) H_n(z), \lb{ch4.38} 
\end{align}
treating $t_r$ as a parameter.  Define polynomials $\ti G_r$ and
$\ti H_r$ by 
\begin{align}
\ti G_r(z)&=\ti F_r(z)+\f12 \ti F_r(z), \lb{ch4.39} \\
\ti H_r(z)&=(4u-u_{xx})\ti F_r(z)-z\ti G_{r,x}(z),
\lb{ch4.40}
\end{align}
respectively.  We claim that
\begin{equation}
F_{n,t_r}(z)=2\big(G_n(z) \ti F_r(z)-F_n(z) \ti G_r(z) \big).
\lb{ch4.41}
\end{equation}
To prove \eqref{ch4.41}  we compute {}from  \eqref{ch4.26}
and \eqref{ch4.26a} that
\begin{align}
F_{n,t_r}(z)&=-F_n(z)\sum_{j=1}^n \ti F_{r}(\mu_j)\mu_{j,x}
(z-\mu_j)^{-1},
\lb{ch4.42} \\
F_{n,x}(z)&=-F_n(z)\sum_{j=1}^n \mu_{j,x} (z-\mu_j)^{-1}.
\lb{ch4.43}
\end{align}
Using \eqref{ch4.35} and \eqref{ch4.39} we see that \eqref{ch4.41} is
equivalent to 
\begin{equation}
\ti F_{r,x}(z)=\sum_{j=1}^n\big(\ti F_r(z)-\ti F_r(\mu_j)
\big)\mu_{j,x}(z-\mu_j)^{-1}. \lb{ch4.44}
\end{equation}
Equation \eqref{ch4.44} is proved in Lemma \ref{lemmaH.6}. This in turn
proves \eqref{ch4.41}. Next, taking the derivative of \eqref{ch4.41} with
respect to $x$ and inserting \eqref{ch4.36} and \eqref{ch4.35}, yields
\begin{align}
F_{n,t_rx}(z)=2\bigg(&\f1{z}(4u-u_{xx})F_n(z)+G_n(z)\ti
F_{r,x}(z)\no \\
&-2(G_n(z)-F_n(z))\ti G_r(z)-F_n(z)\ti G_{r,x}(z)\bigg).
\lb{ch4.45}
\end{align}
On the other hand, by differentiating \eqref{ch4.35} with respect to
$t_r$, using \eqref{ch4.41} we obtain
\begin{align}
F_{n,t_rx}(z)=2\big(G_{n,t_r}(z)
-2(G_n(z)\ti F_r(z)-F_n(z)\ti G_r(z)\big).
\lb{ch4.46}
\end{align}
Combining \eqref{ch4.45} and \eqref{ch4.46} we conclude
\begin{equation}
zG_{n,t_r}(z)=F_n(z)\ti H_r(z)-\ti F_r(z) H_n(z).\lb{ch4.47}
\end{equation}
Next, we take the derivative of \eqref{ch4.38} with respect to $t_r$ and
use the expressions \eqref{ch4.41} and \eqref{ch4.47} for
$F_{n,t_r}$ and $G_{n,t_r}$, respectively, to obtain
\begin{equation}
H_{n,t_r}(z)=2\big(\ti G_{r}(z) H_n(z)-G_n(z)\ti H_r(z) \big).
\lb{ch4.48}
\end{equation}
Finally, we compute $G_{n,xt_r}$ in two different ways. Differentiating 
\eqref{ch4.47} with respect to $x$, using \eqref{ch4.35},
\eqref{ch4.39}, and \eqref{ch4.37}, one finds
\begin{align}
zG_{n,xt_r}(z)&=\ti H_{r,x}(z) F_n(z)+2(G_n(z) \ti H_r(z)-\ti G_r(z)
H_n(z))\no \\
&\qquad\qquad+2(4u-u_{xx})G_n(z) \ti F_r(z)-2F_n(z) \ti H_r(z).
\lb{ch4.49}
\end{align}
Differentiating \eqref{ch4.36} with respect to $t_r$, using
\eqref{ch4.41} and \eqref{ch4.48}, results in 
\begin{align}
zG_{n,xt_r}(z)&=(u_{t_r}-u_{xxt_r})F_n(z)
-2(\ti G_r(z) H_n(z)-G_n(z)\ti H_r(z))\no \\
&\qquad\qquad+2(4u-u_{xx})(G_n(z)\ti F_r(z)-F_n(z)\ti G_r(z)).
\lb{ch4.50}
\end{align}
Combining \eqref{ch4.49} and \eqref{ch4.50} one concludes
\begin{equation}
u_{t_r}-u_{xxt_r}=\ti H_r(z)+2(4u-u_{xx})\ti G_r(z)-\ti H_r(z)
\lb{ch4.51}
\end{equation}
which is equivalent to \eqref{ch4.33}. 
\end{proof}

\appendix
\section{Hyperelliptic curves and their theta functions}\lb{A}
\renewcommand{\theequation}{A.\arabic{equation}}
\renewcommand{\thetheorem}{A.\arabic{theorem}}
\setcounter{theorem}{0}
\setcounter{equation}{0}

We provide a brief summary of some of the fundamental properties
and notations needed {}from the
theory of hyperelliptic curves.  More details can be found in
some of the standard textbooks
\cite{FarkasKra:1992} and \cite{Mumford:1984}, as well as monographs
dedicated to integrable systems such as
\cite[Ch.\ 2]{BelokolosBobenkoEnolskiiItsMatveev:1994},
\cite[App.\ A--C]{GesztesyHolden:2000}.

Fix $n\in\bbN$. The hyperelliptic curve $\calK_n$
of genus $n$ used in Sections~\ref{chs3} and \ref{chs4} is
defined by
\begin{align}
&\calK_n\colon \calF_n(z,y)=y^2-R_{2n+2}(z)=0, \quad
R_{2n+2}(z)=\prod_{m=0}^{2n+1}(z-E_m), \lb{b0} \\
& \{E_m\}_{m=0,\dots,2n+1}\subset\bbC, \quad
E_m \neq E_{n} \text{ for } m \neq n, \, m,n=0,\dots,2n+1. \label{b1}
\end{align}
The curve \eqref{b1} is compactified by adding the
points $P_{\infty_+}$ and $P_{\infty_-}$,
$P_{\infty_+} \neq P_{\infty_-}$, at infinity.
One then introduces an appropriate set of
$n+1$ nonintersecting cuts $\calC_j$ joining
$E_{m(j)}$ and $E_{m^\prime(j)}$. We denote
\begin{equation}
\calC=\bigcup_{j=1}^{n+1}\calC_j,
\quad
\calC_j\cap\calC_k=\emptyset,
\quad j\neq k.\label{b2}
\end{equation}
Define the cut plane
\begin{equation}
\Pi=\bbC\setminus\calC, \label{b3}
\end{equation}
and introduce the holomorphic function
\begin{equation}
R_{2n+2}(\dott)^{1/2}\colon \Pi\to\bbC, \quad
z\mapsto \left(\prod_{m=0}^{2n+1}(z-E_m) \right)^{1/2}\label{b4}
\end{equation}
on $\Pi$ with an appropriate choice of the square root
branch in \eqref{b4}. Define
\begin{equation}
\calM_{n}=\{(z,\sigma R_{2n+2}(z)^{1/2}) \mid
z\in\bbC,\; \sigma\in\{1,-1\}
\}\cup\{P_{\infty_+},P_{\infty_-}\} \label{b5}
\end{equation}
by extending $R_{2n+2}(\dott)^{1/2}$ to $\calC$. The
hyperelliptic curve $\calK_n$ is then the set
$\calM_{n}$ with its natural complex structure obtained
upon gluing the two sheets of $\calM_{n}$
crosswise along the cuts. The set of branch points
$\calB(\calK_n)$ of $\calK_n$ is given by
\begin{equation}
\calB(\calK_n)=\{(E_m,0)\}_{m=0,\dots,2n+1} \lb{5a}
\end{equation}
and finite points $P$ on $\calK_n$ are denoted by
$P=(z,y)$, where $y(P)$ denotes the meromorphic function
on $\calK_n$ satisfying $\calF_n(z,y)=y^2-R_{2n+2}(z)=0$.
Local coordinates near $P_0=(z_0,y_0)\in\calK_n\setminus
\{\calB(\calK_n)\cup\{P_{\infty_+},P_{\infty_-}\}\}$ are
given by $\zeta_{P_0}=z-z_0$, near $P_{\infty_\pm}$ by
$\zeta_{P_{\infty_\pm}}=1/z$, and near branch points
$(E_{m_0},0)\in\calB(\calK_n)$ by
$\zeta_{(E_{m_0},0)}=(z-E_{m_0})^{1/2}$. The Riemann surface
$\calK_n$ defined in this manner has topological genus $n$. 
Moreover, we introduce the holomorphic sheet exchange map 
(involution)
\begin{equation}
* \colon \calK_n \to \calK_n,\quad
P=(z,y)\mapsto P^*=(z,-y),\;
P_{\infty_\pm} \mapsto P_{\infty_\pm}^*=P_{\infty_\mp}
\lb{a12}
\end{equation}

One verifies that $dz/y$ is a holomorphic differential
on $\calK_n$ with zeros of order $n-1$ at $P_{\infty_\pm}$
and hence
\begin{equation}
\eta_j=\frac{z^{j-1}dz}{y}, \quad j=1,\dots,n
\lb{b24}
\end{equation}
form a basis for the space of holomorphic differentials
on $\calK_n$.  Introducing the
invertible matrix $C$ in $\bbC^n$,
\begin{align}
C & =(C_{j,k})_{j,k=1,\dots,n}, \quad C_{j,k}
= \int_{a_k} \eta_j, \lb{A.7}\\
\underline{c} (k) & = (c_1(k), \dots,
c_n(k)), \quad c_j (k) = C_{j,k}^{-1}, \quad j,k=1,\dots,n, \lb{A.7a}
\end{align}
the corresponding basis of normalized holomorphic
differentials $\omega_j$, $j=1,\dots,n$ on $\calK_n$ is given by
\begin{equation}
\omega_j = \sum_{\ell=1}^n c_j (\ell) \eta_\ell,
\quad \int_{a_k} \omega_j =
\delta_{j,k}, \quad j,k=1,\dots,n. \lb{b26}
\end{equation}
Here $\{a_j,b_j\}_{j=1,\dots,n}$ is a homology basis for
$\calK_n$ with intersection matrix of the cycles satisfying
\begin{equation}
a_j \circ b_k=\delta_{j,k}, \quad j,k=1,\dots,n. \lb{c26}
\end{equation}

Near $P_{\infty_\pm}$ one infers
\begin{align}
\ul\omega & = (\omega_1,\dots,\omega_n)=
\pm \bigg( \sum_{j=1}^n \f{\ul c (j)
\zeta^{n-j}}{\big(\prod_{m=0}^{2n+1}
(1-E_m \zeta) \big)^{1/2}} \bigg) d\zeta \no \\
& \underset{\zeta \to 0}{=} \pm \bigg( \ul c (n) +
\bigg( \frac12 \ul c (n)
\sum_{m=0}^{2n+1} E_m +\ul c
(n-1) \bigg) \zeta + \Oh(\zeta^2) \bigg)d\zeta
\text{ as } P\to\Pinfpm, \no \\
& \hspace*{9cm} \zeta=1/z, \lb{b27}
\end{align}
and
\begin{equation}
y(P) \underset{\zeta \to 0}{=} \mp \bigg(1-\frac12
\bigg( \sum_{m=0}^{2n+1} E_m \bigg)\zeta +
\Oh(\zeta^2)\bigg)\zeta^{-n-1} \text{ as }
P\to P_{\infty_\pm}. \lb{b27a}
\end{equation}
Similarly, near $P_0$ one computes
\begin{align}
&\ul\omega \underset{\zeta\to 0}{=} -2i \big(\ti Q^{-1/2}\ul c(1)
+ \Oh(\zeta^2) \big)d\zeta \text{ as } P\to P_0, \lb{b27ab} \\
& \ti Q^{1/2}=\bigg(\prod_{m=1}^{2n+1}E_m\bigg)^{1/2},
\quad \zeta=\sigma z^{1/2}, \; \sigma\in\{1,-1\}, \no
\end{align}
using
\begin{equation}
y(P)\underset{\zeta\to 0}{=} i{\ti Q}^{1/2}\zeta + \Oh(\zeta^3)
\text{ as } P\to P_0, \quad \zeta=\sigma z^{1/2}, \; \sigma\in\{1,-1\},
\lb{b27ac}
\end{equation}
with the sign of ${\ti Q}^{1/2}$ determined by the compatibility
of charts.

Associated with the homology basis
$\{a_j, b_j\}_{j=1,\dots,n}$ we
also recall the canonical dissection of $\calK_n$
along its cycles yielding
the simply connected interior $\hatt \calK_n$ of the
fundamental polygon $\partial {\hatt \calK}_n$ given by
\begin{equation}
\partial  {\hatt \calK}_n =a_1 b_1 a_1^{-1} b_1^{-1}
a_2 b_2 a_2^{-1} b_2^{-1} \cdots a_n^{-1} b_n^{-1}. \lb{aa19}
\end{equation}
Let $\calM (\calK_n)$ and $\calM^1 (\calK_n)$ denote the
set of meromorphic
functions (0-forms) and meromorphic
differentials (1-forms)
on $\calK_n$. The residue of a meromorphic differential
$\nu\in \calM^1 (\calK_n)$ at a
point $Q \in \calK_n$ is defined by
\begin{equation}
\text{res}_{Q}(\nu)
=\frac{1}{2\pi i} \int_{\gamma_{Q}} \nu,
\lb{a33}
\end{equation}
where $\gamma_{Q}$ is a counterclockwise oriented
smooth simple closed
contour encircling $Q$ but no other pole of
$\nu$.  Holomorphic
differentials are also called Abelian differentials
of the first kind (dfk). Abelian differentials of the
second kind
(dsk) $\omega^{(2)} \in \calM^1 (\calK_n)$ are characterized
by the property that all their residues vanish.  They will 
usually be normalized by demanding that all their $a$-periods
vanish, that is,
\begin{equation}
\int_{a_j} \omega^{(2)} =0, \quad  j=1,\dots,n.
\lb{a34}
\end{equation}
If $\omega_{P_1, n}^{(2)}$ is a dsk on $\calK_n$ whose
only pole is $P_1 \in \hatt \calK_n$ with principal part
$\zeta^{-n-2}\,d\zeta$, $n\in\bbN_0$ near
$P_1$ and $\omega_j =
 (\sum_{m=0}^\infty d_{j,m} (P_1) \zeta^m)\, d\zeta$
near $P_1$, then
\begin{equation}
\frac{1}{2\pi i} \int_{b_j} \omega_{P_1, m}^{(2)} =
 \frac{d_{j,m} (P_1)}{m+1}, \quad m=0,1,\dots
\lb{a35}
\end{equation}

Any meromorphic differential $\omega^{(3)}$ on
$\calK_n$ not of the first or
second kind is said to be of the third
kind (dtk).
A dtk $\omega^{(3)} \in \calM^1 (\calK_n)$
is usually normalized by the vanishing of its
$a$-periods, that is,
\begin{equation}
\int_{a_j} \omega^{(3)} =0, \quad  j=1,\dots, n.
\lb{a36}
\end{equation}
A normal dtk $\omega_{P_1, P_2}^{(3)}$ associated
with two points $P_1$,
$P_2 \in \hatt \calK_n$, $P_1 \neq P_2$ by definition
has simple poles at
$P_j$ with residues $(-1)^{j+1}$, $j=1,2$ and
vanishing $a$-periods.  If $\omega_{P,Q}^{(3)}$ is a
normal dtk associated
with $P$, $Q\in\hatt \calK_n$, holomorphic on
$\calK_n \setminus \{ P,Q\}$, then
\begin{equation}
\frac{1}{2\pi i} \int_{b_j} \omega_{P,Q}^{(3)} = \int_{Q}^P \omega_j,
\quad  j=1,\dots,n,
\lb{a37}
\end{equation}
where the path {}from $Q$ to $P$ lies in
$\hatt \calK_n$ (i.e.,
does not touch any of the cycles $a_j$, $b_j$). Explicitly,
one obtains
\begin{align}
\omega^{(3)}_{P_{\infty_+},P_{\infty_-}}&=
\f{z^n d z}{y}+\sum_{j=1}^n \gamma_j\omega_j
=\f{\prod_{j=1}^n (z -\lambda_j)\, d z}{y}, \lb{a37a} \\
\omega^{(3)}_{P_1,P_{\infty_+}}&=\f{1}{2}\f{y+y_1}
{z-z_1}\f{d z}{y}-
\f{\prod_{j=1}^n (z -\tilde\lambda_j)\,d z}
{2y}, \lb{a37b} \\
\omega^{(3)}_{P_1,P_{\infty_-}}&=\f{1}{2}\f{y+y_1}
{z-z_1}\f{d z}{y}+
\f{\prod_{j=1}^n (z -\lambda_j^\prime)\,d z}
{2y}, \lb{a37c} \\
\omega^{(3)}_{P_1,P_2}&=\bigg(\f{y+y_1}
{z-z_1}-\f{y+y_2}{z-z_2}\bigg)\f{d
z}{2y}+\lambda_n''\f{\prod_{j=1}^{n-1}(z-\lambda_j'')dz}{y}, 
\lb{a37d} \\
& \hspace*{3.2cm} P_1,P_2\in\calK_n\setminus\{P_{\infty_+},P_{\infty_-}\},
\no
\end{align}
where $\gamma_j, \lambda_j, \tilde\lambda_j, \lambda_j^\prime, \lambda_j''
\in\bbC$, $j=1,\dots,n$, are uniquely determined by the requirement of
vanishing $a$-periods and we abbreviated $P_j=(z_j,y_j)$,
$j=1,2$. (If $n=0$, we use the standard convention that the
product over an empty index set is replaced by $1$.)

We shall always assume (without loss of generality)
that all poles of
dsk's and dtk's on $\calK_n$ lie on $\hatt \calK_n$ (i.e.,
not on $\partial \hatt \calK_n$).

Define the matrix $\tau=(\tau_{j,\ell})_{j,\ell=1,\dots,n}$ by
\begin{equation}
\tau_{j,\ell}=\int_{b_j}\omega_\ell, \quad j,\ell=1,
\dots,n. \label{b8}
\end{equation}
Then
\begin{equation}
\Im(\tau)>0, \quad \text{and} \quad \tau_{j,\ell}=\tau_{\ell,j},
\quad j,\ell =1,\dots,n.  \lb{a18a}
\end{equation}
Associated
with $\tau$ one introduces the period lattice
\begin{equation}
L_n = \{ \ul z \in\bbC^n \mid \ul z = \ul m +\tau \ul n,
\; \ul m, \ul n \in\bbZ^n\}
\lb{a28}
\end{equation}
and the Riemann theta function associated with $\calK_n$ and
the given homology basis $\{a_j,b_j\}_{j=1,\dots,n}$,
\begin{equation}
\theta(\ul z)=\sum_{\ul n\in\bbZ^n}\exp\big(2\pi
i(\ul n,\ul z)+\pi
i(\ul n,\tau \ul n)\big),
\quad \ul z\in\bbC^n, \label{b9}
\end{equation}
where $(\ul u, \ul v)=\sum_{j=1}^n \overline{u}_j v_j$
denotes the
scalar product
in $\bbC^n$. It has the fundamental properties
\begin{align}
& \theta(z_1, \ldots, z_{j-1}, -z_j, z_{j+1},
\ldots, z_n) =\theta
(\ul z), \lb{a27}\\
& \theta (\ul z +\ul m +\tau \ul n)
=\exp \big(-2 \pi i (\ul n,\ul z) -\pi i (\ul n, \tau
\ul n) \big) \theta (\ul z), \quad \ul m, \ul n \in\bbZ^n.
\lb{aa51}
\end{align}

Next, fix a base point $Q_0\in\calK_n\setminus
\{\Pzeropm,\Pinfpm\}$, denote by
$J(\calK_n) = \bbC^n/L_n$ the Jacobi variety of $\calK_n$,
and define the
Abel map $\underline{A}_{Q_0}$ by
\begin{equation}
\underline{A}_{Q_0} \colon \calK_n \to J(\calK_n), \quad
\underline{A}_{Q_0}(P)=
\bigg(\int_{Q_0}^P \omega_1,\dots,\int_{Q_0}^P \omega_n \bigg)
\pmod{L_n}, \quad P\in\calK_n. \label{b10}
\end{equation}
Similarly, we introduce
\begin{equation}
\ul \alpha_{Q_0}  \colon
\Div(\calK_n) \to J(\calK_n),\quad
\calD \mapsto \ul \alpha_{Q_0} (\calD)
=\sum_{P \in \calK_n} \calD (P) \ul A_{Q_0} (P),
\label{aa47}
\end{equation}
where $\Div(\calK_n)$ denotes the set of
divisors on $\calK_n$. Here $\calD \colon \calK_n \to \bbZ$
is called a divisor on $\calK_n$ if $\calD(P)\neq0$ for only
finitely many $P\in\calK_n$. (In the main body of this paper
we will choose $Q_0$ to be one of the branch points, i.e.,
$Q_0\in\calB(\calK_n)$, and for simplicity we will always choose
the same path of integration {}from $Q_0$ to $P$ in all Abelian
integrals.) For subsequent use in Remark \ref{raa26a} we also introduce
\begin{align}
\hua_{Q_0} & \colon\hatt{\calK}_n\to\bbC^n, \lb{aa52} \\
&P\mapsto\hua_{Q_0}(P)
=\big(\hatt A_{Q_0,1}(P),\dots,\hatt A_{Q_0,n}(P)\big)
=\bigg(\int_{Q_0}^P\omega_1,\dots,\int_{Q_0}^P\omega_n\bigg) \no 
\end{align}
and
\begin{equation}
\hatt {\ul \al}_{Q_0}  \colon
\Div(\hatt\calK_n) \to \bbC^n, \quad
\calD \mapsto \hatt {\ul \al}_{Q_0} (\calD)
=\sum_{P \in \hatt\calK_n} \calD (P) \hua_{Q_0} (P). \lb{aa52a}
\end{equation}

In connection with divisors on $\calK_n$ we shall employ the
following (additive) notation,
\begin{align} 
&\calD_{Q_0\ul Q}=\calD_{Q_0}+\calD_{\ul Q}, \quad \calD_{\ul
Q}=\calD_{Q_1}+\cdots +\calD_{Q_m}, \lb{A.17} \\
& {\ul Q}=\{Q_1, \dots ,Q_m\} \in \sigma^m \calK_n,
\quad Q_0\in\calK_n, \; m\in\bbN, \no
\end{align}
where for any $Q\in\calK_n$,
\begin{equation} \lb{A.18}
\calD_Q \colon  \calK_n \to\bbN_0, \quad
P \mapsto  \calD_Q (P)=
\begin{cases} 1 & \text{for $P=Q$},\\
0 & \text{for $P\in \calK_n\setminus \{Q\}$}, \end{cases}
\end{equation}
and $\sigma^m \calK_n$ denotes the $m$th symmetric product of
$\calK_n$. In particular, $\sigma^m \calK_n$ can be
identified with
the set of nonnegative
divisors $0 \leq \calD \in \Div(\calK_n)$ of degree $m\in\bbN$.

For $f\in \calM (\calK_n) \setminus \{0\}$,
$\omega \in \calM^1 (\calK_n) \setminus \{0\}$ the
divisors of $f$ and $\omega$ are denoted
by $(f)$ and
$(\omega)$, respectively.  Two
divisors $\calD$, $\calE\in \Div(\calK_n)$ are
called equivalent, denoted by
$\calD \sim \calE$, if and only if $\calD -\calE
=(f)$ for some
$f\in\calM (\calK_n) \setminus \{0\}$.  The divisor class
$[\calD]$ of $\calD$ is
then given by $[\calD]
=\{\calE \in \Div(\calK_n)\mid\calE \sim \calD\}$.  We
recall that
\begin{equation}
\deg ((f))=0,\, \deg ((\omega)) =2(n-1),\,
f\in\calM (\calK_n) \setminus
\{0\},\,  \omega\in \calM^1 (\calK_n) \setminus \{0\},
\lb{a38}
\end{equation}
where the degree $\deg (\calD)$ of $\calD$ is given
by $\deg (\calD)
=\sum_{P\in \calK_n} \calD (P)$.  It is customary to call
$(f)$ (respectively,
$(\omega)$) a principal (respectively, canonical)
divisor.

Introducing the complex linear spaces
\begin{align}
\calL (\calD) & =\{f\in \calM (\calK_n)\mid f=0
 \text{ or } (f) \geq \calD\}, \;
r(\calD) =\dim_\bbC \calL (\calD),
\lb{a39}\\
\calL^1 (\calD) & =
 \{ \omega\in \calM^1 (\calK_n)\mid \omega=0
 \text{ or } (\omega) \geq
\calD\},\; i(\calD) =\dim_\bbC \calL^1 (\calD),
\lb{a40}
\end{align}
($i(\calD)$ the index of speciality of $\calD$) one
infers that $\deg
(\calD)$, $r(\calD)$, and $i(\calD)$ only depend on
the divisor class
$[\calD]$ of $\calD$.  Moreover, we recall the
following fundamental
facts.

\begin{theorem} \lb{thm1}
Let $\calD \in \Div(\calK_n)$,
$\omega \in \calM^1 (\calK_n) \setminus \{0\}$. Then
\begin{equation}
 i(\calD) =r(\calD-(\omega)), \quad n\in\bbN_0.
\lb{a41}
\end{equation}
The Riemann-Roch theorem reads
\begin{equation}
r(-\calD) =\deg (\calD) + i (\calD) -n+1,
\quad n\in\bbN_0.
\lb{a42}
\end{equation}
By Abel's theorem, $\calD\in \Div(\calK_n)$,
$n\in\bbN$, is principal
if and only if
\begin{equation}
\deg (\calD) =0 \text{ and } \ul \alpha_{Q_0} (\calD)
=\ul{0}.
\lb{a43}
\end{equation}
Finally, assume
$n\in\bbN$. Then $\ul \alpha_{Q_0}
: \Div(\calK_n) \to J(\calK_n)$ is surjective
$($Jacobi's inversion theorem$)$.
\end{theorem}

Next we introduce
\begin{equation}
\ul W_0=\{0\}\subset J(\calK_n), \quad
\ul W_m=\ul \alpha_{Q_0} (\sigma^m \calK_n), \,\, m\in\bbN
\lb{a43a}
\end{equation}
and note that while $\sigma^m\calK_n\not\subset\sigma^n\calK_n$
for $m<n$, one has $\ul W_m\subseteq\ul W_n$ for $m<n$. Thus
$\ul W_m=J(\calK_n)$ for $m\geq n$ by Jacobi's inversion theorem.

Denote by $\ul \Xi_{Q_0}=(\Xi_{Q_{0,1}}, \dots,
\Xi_{Q_{0,n}})$ the vector of Riemann constants,
\begin{equation}
\Xi_{Q_{0,j}}=\frac12(1+\tau_{j,j})-
\sum_{\substack{\ell=1 \\ \ell\neq j}}^n\int_{a_\ell}
\omega_\ell(P)\int_{Q_0}^P\omega_j,
\quad j=1,\dots,n. \lb{aa55}
\end{equation}

\begin{theorem} \lb{thm2}
The set $\ul W_{n-1}+\ul {\Xi}_{Q_0}\subset J(\calK_n)$ is the complete
set of zeros of $\theta$ on $J(\calK_n)$, that is,
\begin{equation}
\theta (X)=0 \text{ if and only if } X\in\ul W_{n-1}+\ul {\Xi}_{Q_0}
\lb{a43b}
\end{equation}
$($i.e., if and only if $X=(\ul \alpha_{Q_0} (\calD)+\ul {\Xi}_{Q_0})
\pmod{L_n}$ for some $\calD\in\sigma^{n-1}\calK_n$$)$. The set
$\ul W_{n-1}+\ul{\Xi}_{Q_0}$ has complex dimension $n-1$.
\end{theorem}

\begin{theorem} \lb{thm3}
Let $\calD_{\ul Q} \in \sigma^n \calK_n$,
$\ul Q=\{Q_1, \ldots, Q_n\}$.  Then
\begin{equation}
1 \leq i (\calD_{\ul Q} ) =s
\lb{a46}
\end{equation}
if and only if there are $s$ pairs of the type $\{P, P^*\}\subseteq 
\{Q_1,\ldots, Q_n\}$ $($this includes, of course, branch points for which
$P=P^*$$)$. Obviously, one has $s\leq n/2$.
\end{theorem}

\begin{remark} \lb{raa19}
While $\theta(\ul z)$ is well-defined (in fact, entire)
for $\ul z\in\bbC^n$, it is not well-defined on
$J(\calK_n)=\bbC^n/L_n$ because of  \eqref{aa51}.
Nevertheless, $\theta$ is a ``multiplicative
function'' on $J(\calK_n)$ since the multipliers in
\eqref{aa51} cannot vanish.  In particular, if
$\ul z_1=\ul z_2\pmod{L_n}$, then $\theta(\ul z_1)=0$
if and only
if  $\theta(\ul z_2)=0$.  Hence it is
meaningful to state that $\theta$ vanishes at points of
$J(\calK_n)$. Since the Abel map
$\ul A_{Q_0}$ maps $\calK_n$ into  $J(\calK_n)$, the function
$\theta(\ul A_{Q_0}(P)-\ul{\xi})$ for
$\ul\xi\in\bbC^n$, becomes a multiplicative function on
$\calK_n$. Again it makes sense to say that
$\theta(\ul A_{Q_0}(\dott)-\ul{\xi})$ vanishes at points of
$\calK_n$.
\end{remark}

\begin{theorem} \lb{taa17a}
Let $\ul Q =\{Q_1,\dots,Q_n\}\in \sigma^n \calK_n$ and
assume $\calD_{\ul Q}$ to be nonspecial, that is,
$i(\calD_{\ul Q})=0$. Then
\begin{equation}
\theta(\ul {\Xi}_{Q_0} -\ul {A}_{Q_0}(P) + \alpha_{Q_0}
(\calD_{\ul Q}))=0 \text{ if and only if }
P\in\{Q_1,\dots,Q_n\}. \lb{aa55a}
\end{equation}
\end{theorem}

\begin{theorem} \lb{taa20}
Suppose $\calD_{\humu}\in\sigma^n\calK_{n}$ is nonspecial,
$\humu=\{\hmu_{1},\dots,\hmu_{n}\}$, and $\hmu_{n+1}\in\calK_{n}$
with $\hmu^{*}_{n+1}\not\in\{\hmu_{1},\dots,\hmu_{n}\}$.  Let
$\{\hlam_{1},\dots,\hlam_{n+1}\}\subset\calK_{n}$ with
$\calD_{\hulam\hlam_{n+1}}\sim\calD_{\humu\hmu_{n+1}}$
$($i.e., $\calD_{\hulam\hlam_{n+1}}\in[\calD_{\humu\hmu_{n+1}}]$$)$.
Then any $n$ points
$\hnu_{j}\in\{\hlam_{1},\dots,\hlam_{n+1}\}$, $j=1,\dots,n$
define a nonspecial divisor $\calD_{\hunu}\in\sigma^n\calK_{n}$,
$\hunu=\{\hnu_{1},\dots,\hnu_{n}\}$.
\end{theorem}
\begin{proof}
Since $i(\calD_{P})=0$ for all $P\in\calK_{1}$, there is nothing to
prove in the special case $n=1$.  Hence we assume $n\ge 2$. Let
$Q_{0}\in\calB(\calK_{n})$ be a fixed branch point of
$\calK_{n}$ and suppose that $\calD_{\hunu}$ is special. Then
by Theorem
\ref{thm3} there is a pair
$\{\hnu,\hnu^{*}\}\subset\{\hnu_{1},\dots,\hnu_{n}\}$ such that
\begin{equation}
    \ual_{Q_{0}}(\calD_{\hunu})=\ual_{Q_{0}}(\calD_{\huunu}),
    \lb{aa100}
\end{equation}
where $\huunu=\{\hnu_{1},\dots,\hnu_{n}\}\setminus\{\hnu,\hnu^{*}\}
\in \sigma^{n-2}\calK_{n}$.  Let
$\hnu_{n+1}\in\{\hlam_{1},\dots,\hlam_{n+1}\}
\setminus\{\hnu_{1},\dots,\hnu_{n}\}$ so that
$\{\hnu_{1},\dots,\hnu_{n+1}\}
=\{\hlam_{1},\dots,\hlam_{n+1}\}\subset\sigma^{n+1}\calK_{n}$. Then
\begin{align}
    \ual_{Q_{0}}(\calD_{\huunu\hnu_{n+1}})
    &=\ual_{Q_{0}}(\calD_{\hunu\hnu_{n+1}})
    =\ual_{Q_{0}}(\calD_{\hulam\hlam_{n+1}}) \no \\
    &=\ual_{Q_{0}}(\calD_{\humu\hmu_{n+1}})
    =-\ua_{Q_{0}}(\hmu_{n+1}^{*})+\ual_{Q_{0}}(\calD_{\humu}),
    \lb{aa101}
\end{align}
and hence by Theorem \ref{thm2} and \eqref{aa101},
\begin{equation}
0=\theta(\uxi_{Q_{0}}+\ual_{Q_{0}}(\calD_{\huunu\hnu_{n+1}}))
   =\theta(\uxi_{Q_{0}}-\ua_{Q_{0}}(\hmu^{*}_{n+1})
   +\ual_{Q_{0}}(\calD_{\humu})). \lb{aa102}
\end{equation}
Since by hypothesis $\calD_{\humu}$ is nonspecial and
$\hmu_{n+1}^{*}\not\in\{\hmu_{1},\dots,\hmu_{n}\}$, \eqref{aa102}
contradicts Theorem \ref{taa17a}. Thus,  $\calD_{\hunu}$ is nonspecial.
\end{proof}

\begin{remark} \lb{raa26a}
In Sections \ref{chs3} and \ref{chs4} we frequently deal with theta
function expressions of the type 
\begin{equation}
\phi(P)=\f{\theta(\ul\Xi_{Q_0}-\ul A_{Q_0}(P)+\ul\alpha_{Q_0}(\calD_1))}
{\theta(\ul\Xi_{Q_0}-\ul A_{Q_0}(P)+\ul\alpha_{Q_0}(\calD_2))}
\exp\bigg(\int_{Q_0}^P \omega^{(3)}_{Q_1,Q_2}\bigg), \quad P\in\calK_n
\lb{aa76a}
\end{equation}
and
\begin{equation}
\psi(P)=\f{\theta(\ul\Xi_{Q_0}-\ul A_{Q_0}(P)+\ul\alpha_{Q_0}(\calD_1))}
{\theta(\ul\Xi_{Q_0}-\ul A_{Q_0}(P)+\ul\alpha_{Q_0}(\calD_2))}
\exp\bigg(-c \int_{Q_0}^P \Omega^{(2)}\bigg), \quad P\in\calK_n,
\lb{aa76b}
\end{equation}
where $\calD_j\in\sigma^n\calK_n$, $j=1,2$, are nonspecial positive
divisors of degree $n$, $c\in\bbC$ is a constant,
$Q_j\in\calK_n\setminus\{P_{\infty_1},\dots,P_{\infty_N}\}$, where
$\{P_{\infty_1},\dots,P_{\infty_N}\}$, $N\in\bbN$, denotes the set of
points of $\calK_n$ at infinity, $\omega^{(3)}_{Q_1,Q_2}$ is a normal
differential of the third kind, and $\Omega^{(2)}$ a normalized
differential of the second kind with singularities contained in
$\{P_{\infty_1},\dots,P_{\infty_N}\}$. In particular, one has
\begin{equation}
\int_{a_j} \omega^{(3)}_{Q_1,Q_2}=\int_{a_j} \Omega^{(2)}=0, \quad
j=1,\dots,n. \lb{aa76c}
\end{equation}
Even though we agree to always choose identical paths of integration
{}from $P_0$ to $P$ in all Abelian integrals \eqref{aa76a} and
\eqref{aa76b}, this is not sufficient to render $\phi$ and $\psi$
single-valued on $\calK_n$. To achieve single-valuedness, one needs to
replace $\calK_n$ by its simply connected canonical dissection
$\hatt\calK_n$ and then replace $\ul A_{Q_0}$, $\ul \alpha_{Q_0}$ 
in \eqref{aa76a} and \eqref{aa76b}, with ${\hua}_{Q_0}$, $\hatt{\ul
\alpha}_{Q_0}$ as introduced in \eqref{aa52} and \eqref{aa52a}. In 
particular, one regards $a_j,b_j$, $j=1,\dots,n$, as curves (being a 
part of $\partial\hatt\calK_n$, cf. \eqref{aa19}) and not as homology
classes. Moreover, to render $\phi$ single-valued on $\hatt\calK_n$ one
needs to assume in addition that 
\begin{equation}
\hatt{\ul\alpha}_{Q_0}(\calD_1)-\hatt{\ul\alpha}_{Q_0}(\calD_2)=0
\lb{aa76d}
\end{equation}
(as opposed to merely
$\ul\alpha_{Q_0}(\calD_1)-\ul\alpha_{Q_0}(\calD_2)=0 \pmod {L_n}$).
Similarly, in connection with $\psi$, one introduces the vector of
$b$-periods $\ul U^{(2)}$ of $\Omega^{(2)}$ by
\begin{equation}
\ul U^{(2)}=(U_1^{(2)},\dots,U_g^{(2)}), \quad U_j^{(2)}=\f{1}{2\pi
i}\int_{b_j} \Omega^{(2)}, \quad j=1,\dots,n, \lb{aa76e}
\end{equation}
and then renders $\psi$ single-valued on $\hatt\calK_n$ by requiring 
\begin{equation}
\hatt{\ul\alpha}_{Q_0}(\calD_1)-\hatt{\ul\alpha}_{Q_0}(\calD_2)
=c \,\ul U^{(2)} \lb{aa76f}
\end{equation}
(as opposed to merely
$\ul\alpha_{Q_0}(\calD_1)-\ul\alpha_{Q_0}(\calD_2)=c \,\ul U^{(2)}
\pmod {L_n}$). These statements easily follow {}from \eqref{a37} and
\eqref{aa51} in the case of $\phi$ and simply {}from \eqref{aa51} in the
case of $\psi$. In fact, by \eqref{aa51},
\begin{equation}
\hatt{\ul\alpha}_{Q_0}(\calD_1+\calD_{Q_1})-
\hatt{\ul\alpha}_{Q_0}(\calD_2+\calD_{Q_2})\in\bbZ^n,
\lb{aa76g}
\end{equation}
respectively,
\begin{equation}
\hatt{\ul\alpha}_{Q_0}(\calD_1)-\hatt{\ul\alpha}_{Q_0}(\calD_2)
- c\, \ul U^{(2)}\in\bbZ^n, \lb{aa76h}
\end{equation}
suffice to guarantee single-valuedness of $\phi$, respectively, $\psi$ on
$\hatt\calK_n$. Without the replacement of $\ul A_{Q_0}$ and  
$\ul \alpha_{Q_0}$ by ${\hua}_{Q_0}$ and $\hatt{\ul \alpha}_{Q_0}$  in
\eqref{aa76a} and \eqref{aa76b} and without the assumptions \eqref{aa76d}
and \eqref{aa76f} (or \eqref{aa76g} and \eqref{aa76h}), $\phi$
and $\psi$ are multiplicative (multi-valued) functions on $\calK_n$, and
then most effectively discussed by introducing the notion of characters
on $\calK_n$ (cf.\ \cite[Sect.\ III.9]{FarkasKra:1992}). For
simplicity, we decided to avoid the latter possibility and throughout
this paper will tacitly always assume \eqref{aa76d} and \eqref{aa76f}
without particularly emphasizing this convention each time it is used.
\end{remark}

\section{High-Energy Expansions} \lb{B}
\renewcommand{\theequation}{B.\arabic{equation}}
\renewcommand{\thetheorem}{B.\arabic{theorem}}
\setcounter{theorem}{0}
\setcounter{equation}{0}

In this appendix we study the relationship between the homogeneous
coefficients $\hat f_\ell$ and nonhomogeneous coefficients $f_\ell$ of the
polynomial $F_n$, discuss the high-energy expansion of $F_n/y$, and use it
to derive a nonlinear recursion relation for $\hat f_\ell$,
$\ell\in\bbN_0$. 

Let
\begin{align}
&\{E_m\}_{m=0,\dots,2n+1} \; \text{ for some } \; n\in\bbN_0 \lb{ch100} \\
& \text{and } \; \eta\in\bbC \; \text{ such that } \; 
|\eta|<\min\{|E_0|^{-1},\dots, |E_{2n+1}|^{-1}\}. \lb{ch101}
\end{align}
Then 
\begin{equation}
\bigg(\prod_{m=0}^{2n+1} \big(1-E_m\eta \big)
\bigg)^{-1/2}=\sum_{k=0}^{\infty}\hat c_k(\ul E)\eta^{k}, \lb{ch102}
\end{equation}
where
\begin{align}
\hat c_0(\ul E)&=1,\no \\
\hat c_k(\ul E)&=\sum_{\substack{j_0,\dots,j_{2n+1}=0\\
 j_0+\cdots+j_{2n+1}=k}}^{k}
\f{(2j_0-1)!!\cdots(2j_{2n+1}-1)!!}
{2^k j_0!\cdots j_{2n+1}!}E_0^{j_0}\cdots E_{2n+1}^{j_{N}},
\,\, k\in\bbN. \label{ch103}
\end{align}
The first few coefficients explicitly read
\begin{equation}
\hat c_0(\ul E)=1, \; 
\hat c_1(\ul E)=\f12\sum_{m=0}^{2n+1} E_m,
\; \hat c_2(\ul E)=\f14\sum_{\substack{m_1,m_2=0\\ m_1< m_2}}^{2n+1}
E_{m_1} E_{m_2}+\f38 \sum_{m=0}^{2n+1} E_m^2, 
\quad \text{etc.} \lb{ch104}
\end{equation}
Similarly,
\begin{equation}
\bigg(\prod_{m=0}^{2n+1} \big(1-E_m \eta \big)
\bigg)^{1/2}=\sum_{k=0}^{\infty}c_k(\ul E)\eta^{k}, \lb{ch105}
\end{equation}
where
\begin{align}
c_0(\ul E)&=1,\no \\
c_k(\ul E)&=\sum_{\substack{j_0,\dots,j_{2n+1}=0\\
 j_0+\cdots+j_{2n+1}=k}}^{k}
\f{(2j_0-3)!!\cdots(2j_{2n+1}-3)!!}
{2^k j_0!\cdots j_{2n+1}!}E_0^{j_0}\cdots E_{2n+1}^{j_{2n+1}},
\,\, k\in\bbN. \label{ch106}
\end{align}
The first few coefficients explicitly are given by 
\begin{align}
c_0(\ul E)=1, \; 
c_1(\ul E)=-\f12\sum_{m=0}^{2n+1} E_m, \; 
c_2(\ul E)=\f14\sum_{\substack{m_1,m_2=0\\ m_1< m_2}}^{2n+1}
E_{m_1} E_{m_2}-\f18 \sum_{m=0}^{2n+1} E_m^2, 
\quad \text{etc.} \lb{ch107}
\end{align}
Here we used the abreviations
\begin{equation}
(2q-1)!!=1\cdot3\cdots (2q-1), \quad q\in\bbN, \quad (-3)!!=-1, 
\quad (-1)!!=1. \lb{ch108}
\end{equation}


\begin{theorem} \lb{tE.6}
Assume $u\in C^\infty(\bbR)$, $d^mu/dx^m \in L^\infty(\bbR)$,
$m\in\bbN_0$, $\sCH_n(u)=0$ and suppose 
$P=(z,y)\in\calK_n\setminus\{\Pinfp,\Pinfm\}$. Then $F_n/y$
has the following convergent expansion as $P\to \Pinfpm$,
\begin{equation}
\f{F_n(z)}{y} \underset{\zeta \to 0}{=} \mp \sum_{\ell=0}^\infty
\hat f_\ell \, \zeta^{\ell+1}, \lb{ch109}
\end{equation}
with $\zeta =1/z$ the local coordinate near $\Pinfpm$
described in 
Appendix \ref{A} and $\hat f_\ell$ the homogeneous 
coefficients $f_\ell$ in 
\eqref{ch2.10}. In particular, $\hat f_\ell$ can be computed from the
nonlinear recursion  relation 
\begin{align}
\hat f_0 & = 1, \quad \hat f_1 = -2u, \no \\
\hat f_{\ell+1} & = \Green\bigg(\sum_{k=1}^\ell \big(\hat f_{\ell+1-k,xx} 
\hat f_{k} -\tfrac12 \hat f_{\ell+1-k,x}\hat f_{k,x} 
-2\hat f_{\ell+1-k} \hat f_{k}\big) \no \\
& \qquad \;\, +2(u_{xx}-4u) \sum_{k=0}^\ell \hat f_{\ell-k} \hat
f_{k}\bigg), \quad \ell\in\bbN, \lb{ch110}
\end{align}
assuming
\begin{equation}
\hat f_\ell \in L^\infty(\bbR), \quad \ell\in\bbN. \lb{ch110a}
\end{equation}
Moreover, one infers for the $E_m$-dependence of the integration constants
$c_\ell$, $\ell=0,\dots,n$, in $F_n$, 
\begin{equation}
c_\ell=c_\ell(\ul E), \quad \ell=0,\dots,n \lb{ch111}
\end{equation}
and
\begin{align}
f_\ell&=\sum_{k=0}^\ell c_{\ell-k}(\ul E) \hat f_k, \quad
\ell=0,\dots,n, \lb{ch112}
\\
\hat f_\ell&=\sum_{k=0}^\ell \hat c_{\ell-k}(\ul E) f_k, \quad
\ell=0,\dots,n. \lb{ch113} 
\end{align}
\end{theorem}
\begin{proof}
Dividing $F_n$  by $R_{2n+1}^{1/2}$ (temporarily fixing the  branch of
$R_{2n+1}(z)^{1/2}$ as $z^{n+1}$ near infinity), one obtains 
\begin{equation}
\f{F_n(z)}{R_{2n+1}(z)^{1/2}}\underset{|z|\to\infty}{=} 
\bigg(\sum_{k=0}^\infty \hat c_k(\ul
E)z^{-k}\bigg) \bigg(\sum_{\ell=0}^{n} f_\ell z^{-\ell-1}\bigg)
=\sum_{\ell=0}^\infty {\check f}_\ell z^{-\ell-1} \lb{ch116} 
\end{equation}
for some coefficients $\check f_\ell$ to be determined next. Dividing
\eqref{ch2.39a} by $R_{2n+1}$, and inserting the 
expansion \eqref{ch116} into the resulting equation then yields the
recursion relation \eqref{ch110} (with $\hat f_\ell$ replaced by 
${\check f}_\ell$). More precisely, for
$\check f_1$ one originally obtains the relation
\begin{equation}
-\check f_{1,xx}+4 \check f_1 =2(u_{xx}-4u), \text{ that is, }
\bigg(-\f{d^2}{dx^2}+4\bigg)\bigg(\check f_1 +2u\bigg)=0.
\lb{ch116a} 
\end{equation}
Thus,
\begin{equation}
\check f_1(x)=-2u(x)+a_1 e^{2x} + b_1 e^{-2x} \lb{ch116b}
\end{equation}
for some $a_1, b_1 \in\bbC$, and hence the requirement $\check f_1 \in
L^\infty(\bbR)$ then yields $a_1=b_1=0$. The open sign of $\check f_0$  has
been chosen such that $\check f_0=\hat f_0=1$. For $\ell\geq 2$ one 
obtains similarly
\begin{align}
-\check f_{\ell+1,xx} +4\check f_{\ell+1}&= \bigg(\sum_{k=1}^\ell
\big(\check f_{\ell+1-k,xx} 
\check f_{k} -\tfrac12 \check f_{\ell+1-k,x}\check f_{k,x} 
-2\check f_{\ell+1-k} \check f_{k}\big) \no \\
& \qquad +2(u_{xx}-4u) \sum_{k=0}^\ell \check f_{\ell-k} \check
f_{k}\bigg), \quad \ell\geq 1,
\end{align}
and hence,
\begin{align}
\check f_{\ell+1}&= \Green \bigg(\sum_{k=1}^\ell
\big(\check f_{\ell+1-k,xx} 
\check f_{k} -\tfrac12 \check f_{\ell+1-k,x}\check f_{k,x} 
-2\check f_{\ell+1-k} \check f_{k}\big) \no \\
& \qquad +2(u_{xx}-4u) \sum_{k=0}^\ell \check f_{\ell-k} \check
f_{k}\bigg)+a_{\ell+1} e^{2x} + b_{\ell+1} e^{-2x}, \quad \ell\geq 1
\end{align}
for some $a_{\ell+1}, b_{\ell+1} \in\bbC$. Again the requirement 
$\check f_{\ell+1} \in L^\infty(\bbR)$ then yields $a_{\ell+1}=
b_{\ell+1}=0$, $\ell\geq 1$. Introducing
$\hat f_\ell$ by \eqref{ch2.10} with $c_k=0$, $k\in\bbN$, and $\check
f_\ell$ by \eqref{ch110}, a straightforward computation shows that
\begin{align}
\check
f_{\ell,x}&=\Green\bigg(\sum_{k=1}^{\ell-1}\big(f_{\ell-k,xxx}
-4f_{\ell-k,x}\big)f_k-\sum_{k=0}^{\ell-1}2\big(-2(u_{xx}-4u)
f_{\ell-k-1,x} \no \\
& \qquad \;\, +(4u_{x}-u_{xxx})f_{\ell-k-1}\big)\bigg)f_k \no \\ 
&=\Green\bigg(-\sum_{k=1}^{\ell-1}\Green^{-1}
f_{\ell-k,x}f_k+\sum_{k=0}^{\ell-1}\big(\Green^{-1}f_{\ell-k,x}\big)
f_k\bigg) \no \\ 
&=\hat f_{\ell,x}, \quad \ell\in\bbN. 
\end{align}
Hence,
\begin{equation}
\check f_\ell=\hat f_\ell +d_\ell, \quad \ell\in\bbN \lb{ch117}
\end{equation}
for some constants $d_\ell\in\bbC$, $\ell\in\bbN$. Since $d_0=d_1=0$ by
inspection, we next proceed by induction on $\ell$ and suppose that
\begin{equation}
d_k=0 \text{ and hence $\check f_k=\hat f_k$ for $k=0,\dots,\ell$.}
\end{equation}
Thus, \eqref{ch110} and \eqref{ch117} imply
\begin{equation}
\check f_{\ell+1}=\Green \{\dots\}=\hat f_{\ell+1} + d_{\ell+1},
\end{equation} 
where $\{\dots\}$ denotes the expression on the right-hand side of
\eqref{ch110} in terms of $\check f_k =\hat f_k$, $k=0,\dots,\ell$. Hence, 
\begin{equation}
\{\dots\}-\hat f_{\ell+1}+\alpha_{\ell+1} e^{2x} + \beta_{\ell+1}e^{-2x}
=\Green^{-1}d_{\ell+1}=4d_{\ell+1}
\end{equation}
for some constants $\alpha_{\ell+1}, \beta_{\ell+1}\in\bbC$. Since 
$\{\dots\}-\hat f_{\ell+1}\in L^\infty (\bbR)$, one concludes once more
that $\alpha_{\ell+1}=\beta_{\ell+1}=0$. Moreover, since 
$\{\dots\}-\hat f_{\ell+1}$ contains no constants by construction, one
concludes $d_{\ell+1}=0$ and hence 
\begin{equation}
\check f_\ell=\hat f_\ell \text{ for all } \ell\in\bbN_0. \lb{ch118}
\end{equation}
Thus, we proved
\begin{equation}
\f{F_n(z)}{R_{2n+1}(z)^{1/2}}\underset{|z|\to\infty}{=} 
\bigg(\sum_{k=0}^\infty \hat c_k(\ul
E)z^{-k}\bigg)\bigg(\sum_{\ell=0}^{n} f_\ell z^{-\ell-1}\bigg)
= \sum_{\ell=0}^\infty {\hat f}_\ell z^{-\ell-1} \lb{ch119} 
\end{equation}
and hence \eqref{ch109}. A comparison of coefficients in \eqref{ch119} 
then proves \eqref{ch113}. Next, multiplying \eqref{ch102}
and \eqref{ch105}, a comparison of coefficients of $z^{-k}$ yields
\begin{equation}
\sum_{\ell=0}^k \hat c_{k-\ell}(\ul E)c_\ell(\ul E) =\delta_{k,0}, 
\quad k\in\bbN_0. \lb{ch120}
\end{equation}
Thus, one computes
\begin{align}
&\sum_{m=0}^\ell c_{\ell-m}(\ul E)\hat f_m=\sum_{m=0}^\ell \sum_{k=0}^m
c_{\ell-m}(\ul E)\hat c_{m-k}(\ul E) f_k
=\sum_{k=0}^\ell \sum_{p=k}^{\ell} c_{\ell-p}(\ul E)\hat c_{p-k}(\ul
E)f_k \no \\
&=\sum_{k=0}^\ell\bigg(\sum_{m=0}^{\ell-k} c_{\ell-k-m}(\ul E)
\hat c_{m}(\ul E)\bigg)f_k=f_\ell, \quad \ell=0,\dots,n, \lb{ch121}
\end{align}
applying \eqref{ch120}. Hence one obtains \eqref{ch112} and thus
\eqref{ch111}. 
\end{proof}

\section{Symmetric Functions and their \\
Theta Function Representations} \lb{C}
\renewcommand{\theequation}{C.\arabic{equation}}
\renewcommand{\thetheorem}{C.\arabic{theorem}}
\setcounter{theorem}{0}
\setcounter{equation}{0}

In this appendix we consider Dubrovin-type equations for auxiliary
divisors $\calD_{\humu}$ of degree $n$ on $\calK_n$ and study in detail
elementary symmetric functions associated with the projections $\mu_j$ of
$\hat\mu_j$, $j=1,\dots,n$. In addition to various applications of
Lagrange interpolation formulas we derive explicit theta function
representations of elementary symmetric functions of $\mu_j$,
$j=1,\dots,n$. While some of the material of this appendix is classical,
some parts are taken {}from \cite{GesztesyHolden:1999} (cf.\ also
\cite[App.\ F and G]{GesztesyHolden:2000}), and \cite{Novikov:1999}.
Proofs are only presented for results that do not appear to belong to
the standard arsenal of the literature on hierarchies of soliton equations.
Our principal results on theta function representations derived in
Sections \ref{chs3} and \ref{chs4} are based on Theorem \ref{tg.6}. The 
results of this appendix apply to a  variety of soliton equations and hence 
are of independent interest.

Assuming $\g\in\bbN$ to be fixed and introducing
\begin{align}
\mathcal S_k&=\{\ul\ell=(\ell_1,\dots,\ell_k)\in\bbN^k \mid
\ell_1<\cdots<\ell_k\leq \g \}, \quad 1\le k\le \g,\lb{inds} \\
\mathcal I^{(j)}_k&=\{\ul\ell=
(\ell_1,\dots,\ell_k)\in\mathcal S_k\mid
\ell_m\neq j\}, \quad 1\le k\le \g-1,  \lb{indt}
\end{align}
one defines
\begin{align}
\Psi_{0}(\ul\mu)&=1, \quad \Psi_{k}(\ul\mu)=
(-1)^k\sum_{\ul\ell\in\mathcal
S_k}\mu_{\ell_1}\cdots\mu_{\ell_k},
\quad 1\le k\le \g,\lb{functpsi} \\
\Phi_0^{(j)}(\ul\mu)&=1, \quad
\Phi_k^{(j)}(\ul\mu)=(-1)^k\sum_{\ul\ell\in\mathcal
I^{(j)}_k}\mu_{\ell_1}\cdots\mu_{\ell_k},
\quad 1\le k\le \g-1, \quad
\Phi_\g^{(j)}(\ul\mu)=0,
\lb{functphi}
\end{align}
where $\ul\mu=(\mu_{1},\dots,\mu_{\g})\in\bbC^\g$.
Explicitly, one verifies
\begin{align}
\Psi_{1}(\ul\mu)&=-\sum_{\ell=1}^\g \mu_\ell,
\quad \Psi_{2}(\ul\mu)=
\sum_{\substack{\ell_1,\ell_2=1\\ \ell_1<\ell_2}}^\g
\mu_{\ell_1}\mu_{\ell_2}, \text{ etc.,} \lb{functspsi} \\
\Phi_1^{(j)}(\ul\mu)&=-
\sum_{\substack{\ell=1\\ \ell\neq j}}^\g \mu_\ell, \quad
\Phi_2^{(j)}(\ul\mu)=
\sum^\g_{\substack{\ell_1, \ell_2=1\\ \ell_1, \ell_2\neq j\\
\ell_1<\ell_2}}\mu_{\ell_1}
\mu_{\ell_2}, \text{ etc.} \lb{functsphi}
\end{align}

Introducing
\begin{equation}
F_\g(z)=\prod_{j=1}^\g (z-\mu_j)=\sum_{\ell=0}^n f_{n-\ell}z^\ell
=\sum_{\ell=0}^\g\Psi_{\g-\ell}(\ul \mu)
z^\ell,
\label{A1a}
\end{equation}
one infers ($F_\g^\prime(z)=\partial F_\g(z)/\partial z$)
\begin{equation}
F_\g^\prime(\mu_k)=\prod_{\substack{j=1\\ j\neq k}}^\g (\mu_k-\mu_j).
\label{A1b}
\end{equation}
The general form of Lagrange's interpolation theorem (cf., e.g.,
\cite[App.\ F]{GesztesyHolden:2000}, \cite[App.~E]{Toda:1989}) then
reads  as follows. 
\begin{theorem}\label{lagrange}
Assume that $\mu_1,\dots,\mu_\g$ are $\g$ distinct complex numbers. Then
\begin{align}
&\sum_{j=1}^\g
\f{\mu_j^{m-1}}{F_\g^\prime(\mu_j)}\Phi_k^{(j)}(\ul\mu)
=\delta_{m,\g-k}-\Psi_{k+1}(\ul\mu)\delta_{m,\g+1}, \label{A1c} \\ 
& \hspace*{1.7cm} m=1,\dots, \g+1,\quad k=0,\dots, \g-1. \no
\end{align}
\end{theorem}

The simplest Lagrange interpolation formula reads in the case $k=0$,
\begin{equation}
\sum_{j=1}^\g \f{\mu_j^{m-1}}{F_\g^\prime(\mu_j)}=\delta_{m,\g}, 
\quad
m=1,\dots,\g. \label{A5}
\end{equation}

For  use in the main text we also recall the following results. 
\begin{lemma}[\cite{GesztesyHolden:1999}, {\cite[App.~F]{GesztesyHolden:2000}} ]
\lb{lemmaG.2} Assume that $\mu_1,\dots,\mu_\g$ are $\g$ distinct complex
numbers.  Then
\begin{align}
\text{$(i)$} \quad 
&\Psi_{k+1}(\ul\mu)+\mu_j\Phi_{k}^{(j)}(\ul\mu)=\Phi_{k+1}^{(j)}(\ul\mu),
\quad k=0,\dots,\g-1, \, j=1,\dots,\g. \label{A8} \\
\text{$(ii)$} \quad 
&\sum_{\ell=0}^k \Psi_{k-\ell}(\ul\mu) \mu_j^\ell =\Phi_{k}^{(j)}(\ul\mu), 
\quad k=0,\dots,\g, \, j=1,\dots,\g. \label{A9} \\
\text{$(iii)$} \quad  
&\sum_{\ell=0}^{k-1} \Phi_{k-1-\ell}^{(j)}(\umu) z^\ell =\frac{1}{z-\mu_j}
\bigg(\sum_{\ell=0}^k \Psi_{k-\ell}(\umu) z^\ell-\Phi_k^{(j)}(\umu)\bigg), 
\lb{A20} \\
& \hspace*{4.5cm} k=0,\dots,\g,\, j=1,\dots,\g. \no
\end{align}
\end{lemma}

Next, assuming $\mu_j \neq \mu_{j'}$ for $j \neq j'$,
introduce the $\g\times \g$ matrix $U_\g(\ul\mu)$ by
\begin{equation}
U_1(\ul\mu)=1, \quad U_\g(\ul\mu)=\left(\frac{\mu_k^{j-1}}
{\prod_{\substack{m=1\\ m\neq k}}^\g(\mu_k-\mu_m)} \right)_{j,k=1}^\g, 
\lb{B18}
\end{equation}
where $\ul \mu =(\mu_1,\dots,\mu_\g)\in\bbC^\g$.

\begin{lemma} [\cite{GesztesyHolden:1999}, {\cite[App.\ G]{GesztesyHolden:2000}}]
\label{lemma2} Suppose $\mu_j\in\bbC$, $j=1,\dots,\g$, are $\g$ distinct
complex  numbers. Then
\begin{equation}
U_\g(\ul\mu)^{-1}=
\left(\Phi_{\g-k}^{(j)}(\ul\mu)\right)_{j,k=1}^\g. \lb{B19}
\end{equation}
\end{lemma}

Next, we  express $f_\ell$, $F_n(\mu_j)$, and $\ti F_r(\mu_j)$ in terms of
elementary symmetric functions of $\mu_1,\dots,\mu_\g$. We start with 
the homogeneous expressions denoted by $\hat f_\ell$ and $\hatt F_r
(\mu_j)$, where  $c_k=0$, $k=0,\dots,\ell$ and $\tilde c_s=0$,
$s=1,\dots,r$. Let $\hat c_k(\ul E)$, $k\in\bbN_0$, be defined as in
\eqref{ch103} and suppose $r\in\bbN_0$. Then, combining \eqref{ch113} and
\eqref{A1a} one infers
\begin{equation}
\hat f_\ell=\sum_{k=0}^{\ell}
\hat c_{\ell-k}(\ul E) \Psi_k(\ul \mu), \quad \ell=0,\dots,\g. \lb{f}
\end{equation}
Next, we turn to $\hatt F_r(\mu_j)$. 
\begin{lemma} \label{lemma1}
Let $r\in\bbN_0$. Then\footnote{$m\maxi n=\max\{m,n\}$.},
\begin{equation}
\hatt F_r(\mu_j)=\sum_{s=(r-\g)\maxi 0}^r \hat c_s(\ul E)
\Phi_{r-s}^{(j)}(\ul\mu). \lb{KdV-F}
\end{equation}
\end{lemma}
\begin{proof}
By definition
\begin{equation}
\hatt F_r(z)=\sum_{\ell=0}^r \hat f_{r-\ell} z^\ell=
\sum_{\ell=0}^r z^\ell
\sum_{m=0}^{(r-\ell)\mini \g}\Psi_m(\ul\mu) \hat c_{r-\ell-m}(\ul E).
\end{equation}
Consider first the case  $r\le \g$. Then
\begin{equation}
\hatt F_r(z)=\sum_{s=0}^r \hat c_s(\ul E) \sum_{\ell=0}^{r-s} 
\Psi_{r-\ell-s}(\ul\mu) z^\ell \lb{H.22}
\end{equation}
and hence
\begin{equation}
\hatt F_r(\mu_j)=\sum_{s=0}^r \hat c_s(\ul E)
\Phi_{r-s}^{(j)}(\ul\mu),\lb{H.23}
\end{equation}
using \eqref{A9}.  In the case where $r \geq \g+1$ we find (applying 
\eqref{A1a})
\begin{align}
\hatt F_r(z)&=\sum_{m=0}^\g \Psi_m(\ul \mu)
\sum_{s=0}^{r-m} z^{r-m-s} \hat c_s(\ul E) \no \\
&=\sum_{s=0}^{r-\g} \hat c_s(\ul
E)\bigg(\sum_{\ell=0}^\g\Psi_\ell(\ul\mu)z^{\g-\ell}\bigg)z^{r-\g-s}
+\sum_{s=r-\g+1}^r \hat c_s(\ul E)\sum_{\ell=0}^{r-s}
\Psi_\ell(\ul\mu) z^{r-s-\ell}\no \\
&=F_\g(z)\sum_{s=0}^{r-\g} \hat c_s(\ul E)z^{r-\g-s}+
\sum_{s=r-\g+1}^r \hat c_s(\ul E)
\sum_{\ell=0}^{r-s} \Psi_\ell(\ul\mu) z^{r-s-\ell} \lb{H.24}  \\
&=F_\g(z)\sum_{s=0}^{r-\g} \hat c_s(\ul E)z^{r-\g-s}+
\sum_{s=r-\g+1}^r \hat c_s(\ul E)
\sum_{\ell=0}^{r-s}\Psi_{r-s-\ell}(\ul\mu) z^\ell. \no
\end{align}
Hence
\begin{equation}
\hatt F_r(\mu_j)=\sum_{s=r-\g+1}^r \hat c_s(\ul E)
\Phi_{r-s}^{(j)}(\ul\mu),\lb{H.25} 
\end{equation}
using \eqref{A9} again.
\end{proof}

Introducing
\begin{align}
d_{\ell,k}(\ul E)&=\sum_{m=0}^{\ell-k} c_{\ell-k-m}(\ul E) \hat c_m(\ul E),
\quad  k=0,\dots,\ell, \; \ell=0,\dots,n, \lb{d} \\
\tilde d_{r,k}(\ul E)&=\sum_{s=0}^{r-k} \tilde c_{r-k-s}  
\hat c_s(\ul E), \quad  k=0,\dots,r\mini n, \lb{tid} 
\end{align}
for a given set of constants $\{\tilde c_s\}_{s=1,\dots,r}\subset\bbC$, 
the corresponding nonhomogeneous quantities $f_\ell$, $F_n(\mu_j)$, and
$\ti F_r(\mu_j)$ are then given by\footnote{$m\mini n=\min\{m,n\}$.}
\begin{align}
f_\ell&=\sum_{k=0}^\ell c_{\ell-k}(\ul E) \hat f_k=\sum_{k=0}^\ell 
d_{\ell,k}(\ul E) \Psi_k(\ul \mu), \quad \ell=0,\dots,n, \lb{tf} \\
F_n(\mu_j)&=\sum_{\ell=0}^n c_{n-\ell}(\ul E) \hatt F_\ell(\mu_j)=
\sum_{\ell-0}^n d_{n,\ell}(\ul E) \Phi^{(j)}_\ell(\ul\mu), \quad c_0=1,
\lb{F} \\
\ti F_r(\mu_j)&=\sum_{s=0}^r \tilde c_{r-s} \hatt F_s(\mu_j)
=\sum_{k=0}^{r\mini n} \tilde d_{r,k}(\ul E) \Phi^{(j)}_k(\ul \mu), \quad
r\in\bbN_0, \; \tilde c_0=1, \lb{tF}
\end{align}
using \eqref{ch111} and \eqref{ch112}. Here $c_k(\ul E)$, $k\in\bbN_0$, is
defined by \eqref{ch106}.

Next, we prove a result needed in the proof of Theorem
\ref{theorem-ch4.10}. 

\begin{lemma} \lb{lemmaH.6}
Suppose $r\in\bbN_0$, $(x,t_r)\in\Omega_\mu$, where 
$\Omega_\mu\subseteq\bbR^2$ is open and connected, and assume
$\mu_j\neq \mu_{j^\prime}$ on $\Omega_\mu$ for $j\neq j^\prime$,
$j,j^\prime=1,\dots,\g$. Then,
\begin{equation}
\ti F_{r,x}(z,x,t_r)= \sum_{j=1}^\g\big(\ti F_r(\mu_j(x,t_r),x,t_r)
- \ti F_r(z,x,t_r)\big)\f{\mu_{j,x}(x,t_r)}{(z-\mu_j(x,t_r))}. \lb{H.31a} 
\end{equation}
\end{lemma}
\begin{proof}
It suffices to prove \eqref{H.31a} for the homogeneous case where $\ti F_r$ is replaced by 
$\hatt F_r$.  Using
\begin{equation}
\Psi_{k,x}(\umu)=-\sum_{j=1}^\g \mu_{j,x} \Phi_{k-1}^{(j)}(\umu), \quad
k=0, \dots, \g, \lb{H.31c}
\end{equation}
with the convention
\begin{equation}
\Phi_{-1}^{(j)}(\umu)=0, \quad 
j=1, \dots, \g, \lb{H.31d}
\end{equation}
one computes for $r\le \g$,
\begin{align}
\hatt F_{r,x}(z)&= \sum_{s=0}^r \hat c_s(\ul E)\sum_{\ell=0}^{r-s} 
\Psi_{r-s-\ell,x}(\umu) z^\ell \no \\
&= -\sum_{j=1}^\g \mu_{j,x}\sum_{s=0}^r \hat c_s(\ul E)\sum_{\ell=0}^{r-s}
\Phi_{r-s-\ell-1}^{(j)}(\umu) z^\ell \no \\
&=\sum_{j=1}^\g \mu_{j,x}(z-\mu_j)^{-1} \sum_{s=0}^r \hat c_s(\ul E) 
\bigg(\Phi_{r-s}^{(j)}(\umu)-\sum_{\ell=0}^{r-s}
\Psi_{r-s-\ell}(\umu)z^\ell\bigg) \no\\ 
&=\sum_{j=1}^\g \big(\hatt F_r(\mu_j)-\hatt F_r(z)
\big)\mu_{j,x}(z-\mu_j)^{-1}, \lb{H.31e}
\end{align}
applying \eqref{A20}, \eqref{H.22}, and \eqref{H.23}.  For $r\ge \g+1$ 
one obtains from \eqref{A20}, \eqref{H.24}, and \eqref{H.25},
\begin{align}
\hatt F_{r,x}(z)&= F_{\g,x}(z) \sum_{s=0}^{r-\g} 
\hat c_s(\ul E) z^{r-\g-s}+
\sum_{s=r-\g+1}^ r \hat c_s(\ul E)\sum_{\ell=0}^{r-s}
\Psi_{r-s-\ell,x}(\umu) z^\ell \no \\
&= -F_\g(z) \sum_{j=1}^\g \mu_{j,x} (z-\mu_j)^{-1} 
\sum_{s=0}^{r-\g} \hat c_s(\ul E) z^{r-\g-s} \no \\
&\quad -\sum_{j=1}^\g\mu_{j,x}  \sum_{s=r-\g+1}^r 
\hat c_s(\ul E) \sum_{\ell=0}^{r-s}
\Phi_{r-s-\ell-1}^{(j)}(\umu) z^\ell \no \\
&= -F_\g(z) \sum_{j=1}^\g \mu_{j,x} (z-\mu_j)^{-1} 
\sum_{s=0}^{r-\g} \hat c_s(\ul E) z^{r-\g-s} \no \\
&\quad +\sum_{j=1}^\g\mu_{j,x}(z-\mu_j)^{-1} 
\sum_{s=r-\g+1}^r \hat c_s(\ul E)
\bigg(\Phi_{r-s}^{(j)}(\umu)-\sum_{\ell=0}^{r-s}
\Psi_{r-s-\ell}(\umu) z^\ell \bigg) \no \\
&= \sum_{j=1}^\g\big(\hatt F_r(\mu_j)
-\hatt F_r(\umu) \big)\mu_{j,x}(z-\mu_j)^{-1}.
 \lb{H.31f}
\end{align}
\end{proof}

Next we turn to a detailed discussion of elementary symmetric functions 
of $\{\mu_1,\dots,\mu_\g\}$. Given the nonsingular hyperelliptic curve
$\calK_\g$ in \eqref{b0}, \eqref{b1}, we introduce the first-order Dubrovin-type
system 
\begin{align}
&\f{\partial\mu_j (\ul v)}{\partial v_k}= 
\Phi^{(j)}_{\g-k}(\ul\mu(\ul v))
\f{y(\hat\mu_j (\ul v))}{\prod_{\substack{m=1\\ m\neq j}}^\g
(\mu_j (\ul v)-\mu_m (\ul v))}, \quad j,k=1,\dots,\g, \lb{g.30} \\
&\hspace*{6.4cm} \ul v=(v_1,\dots,v_\g)\in \calV, \no 
\end{align}
with initial conditions
\begin{equation}
\{\hat{\mu}_j(\ul v_0)\}_{j=1,\dots,\g}\subset\calK_\g \lb{g.30a}
\end{equation} 
for some $\ul v_0\in\calV$, where $\calV\subseteq\bbC^\g$ is an open
connected set such that $\mu_j$ remain distinct on $\calV$,
$\mu_j\neq\mu_{j'}$ for $j\neq j'$, $j,j'=1,\dots,\g$, and
$\Phi^{(j)}_{\g-k}(\ul\mu) \neq 0$, $j,k=1,\dots,\g$. One then obtains, using
\eqref{g.30} and \eqref{A1c},
\begin{align}
\f{\partial}{\partial v_k}\sum_{j=1}^\g \int_{Q_0}^{\hat\mu_j(\ul v)}
\f{z^{k-1}}{y}&=\sum_{j=1}^\g \f{\mu_j(\ul v)^{k-1}}{y(\hat\mu_j(\ul v))}
\f{\partial \mu_j(\ul v)}{\partial v_k}\no \\
&=\sum_{j=1}^\g \f{\mu_j(\ul v)^{k-1}}{y(\hat\mu_j(\ul v))}
\Phi^{(j)}_{\g-k}(\ul\mu(\ul v))
\f{y(\hat\mu_j (\ul v))}{\prod_{\substack{m=1\\ m\neq j}}^\g
(\mu_j (\ul v)-\mu_m (\ul v))} \no \\
&=\sum_{j=1}^\g 
\Phi^{(j)}_{\g-k}(\ul\mu(\ul v))
\f{\mu_j(\ul v)^{k-1}}{\prod_{\substack{m=1\\ m\neq j}}^\g
(\mu_j (\ul v)-\mu_m (\ul v))} =1, \lb{g.30b}
\end{align}
implying
\begin{align}
&\sum_{j=1}^\g \int_{Q_0}^{\hat\mu_j(\ul v)} \f{z^{k-1}dz}{y}
-\sum_{j=1}^\g \int_{Q_0}^{\hat\mu_j(\ul v_0)} \f{z^{k-1}dz}{y}
=(\ul v)_k-(\ul v_0)_k, \lb{g.31} \\
& \hspace*{4.9cm} k=1,\dots,\g, \quad \ul v, \ul v_0\in\calV. \no 
\end{align}
Moreover, introducing
\begin{equation}
v_{\g+1}(\ul v)=\sum_{j=1}^\g \int_{Q_0}^{\hat\mu_j(\ul v)}
\f{z^{\g}dz}{y}, \lb{g.32}
\end{equation}
one then computes as in \eqref{g.30b}
\begin{equation}
\f{\partial v_{\g+1}(\ul v)}{\partial v_k}=-\Psi_{\g+1-k}(\ul \mu(\ul v)),
\quad  k=1,\dots,\g, \lb{g.34}
\end{equation}
using 
\begin{equation}
\sum_{\ell=1}^\g \Phi^{(\ell)}_{\g-p}(\ul\mu)
\f{\mu_\ell^{\g}}{\prod_{\substack{q=1 \\q\neq \ell}}^\g
(\mu_\ell-\mu_q)}=-\Psi_{\g+1-p}(\ul\mu), \quad p=1,\dots,\g \lb{g.35}
\end{equation}
(cf. \eqref{A1c}). Thus, one concludes
\begin{equation}
\prod_{j=1}^\g(z-\mu_j(\ul v))=\sum_{\ell=0}^\g
\Psi_{\g-\ell}(\ul\mu(\ul v))z^\ell= z^\g-\sum_{k=1}^\g \f{\partial
v_{\g+1}(\ul v)}{\partial v_k}z^{k-1}, \quad \ul v\in\calV, \lb{g.36}
\end{equation}
whenever $\ul \mu$ satisfies \eqref{g.30}. 

In order to derive theta function representations of the elementary
symmetric functions $\Psi_k(\ul\mu)$ of
$\mu_1(\ul v),\dots,\mu_\g(\ul v)$, $k=1,\dots,\g$ we recall that $\calK_\g$
corresponds to  the curve $y^2=\prod_{m=0}^{2\g+2}(z-E_m)$ with pairwise distinct
$E_m\in\bbC$, $m=0,\dots,2\g+2$ (cf. \eqref{b0} and \eqref{b1}). Using the
notation established in Appendix \ref{A}, $v_{\g+1}(\ul v)$ can be
written as 
\begin{equation}
v_{\g+1}(\ul v)=\sum_{j=1}^\g \int_{Q_0}^{\hat\mu_j(\ul v)}
\f{z^{\g}dz}{y}=\sum_{j=1}^\g \int_{Q_0}^{\hat\mu_j(\ul v)}
\ti\omega^{(3)}_{\Pinfp,\Pinfm}, \lb{g.60}
\end{equation}
where 
\begin{equation}
\ti\omega^{(3)}_{\Pinfp,\Pinfm}=z^{\g}dz/y \lb{g.61} 
\end{equation}
represents a differential of the third kind with simple poles at $\Pinfp$
and $\Pinfm$ and corresponding residues $+1$ and $-1$, respectively. This
differential is not normalized, that is, the $a$-periods of
$\ti\omega^{(3)}_{\Pinfp,\Pinfm}$ are not all vanishing. We also
introduce the  notation
\begin{equation}
\ul z(P, \ul \mu)=\ul \Xi_{Q_0} -\ul A_{Q_0}(P)+
\ul\alpha_{Q_0}(\calD_{\ul \mu}), \quad P\in\calK_\g, \; \ul
\mu=\{\mu_1,\dots,\mu_\g\}\in\sigma^\g\calK_\g \lb{g.61a}
\end{equation}
in connection with $\calK_\g$, and similarly, 
\begin{equation}
\ul {\hat z}(P, \ul \mu)=\ul {\hatt \Xi}_{Q_0} -\ul {\hatt A}_{Q_0}(P)+
\ul{\hatt \alpha}_{Q_0}(\calD_{\ul \mu}), \quad P\in\calK_\g, \; \ul
\mu=\{\mu_1,\dots,\mu_\g\}\in\sigma^\g\hatt\calK_\g \lb{g.61b}
\end{equation}
in connection with $\hatt \calK_\g$. Moreover, we conveniently choose
$Q_0\in\partial\hatt\calK_\g$ (e.g., the initial point of the curve
$a_1\subset\partial\hatt\calK_\g$).
\begin{theorem} \label{tg.6}
Suppose $\calD_{\humu}\in\sigma^\g\hatt\calK_\g$ is nonspecial,
$\ul{\hat\mu}=\{\hat\mu_1,\dots,\hat\mu_n\}\in\sigma^n\hatt\calK_n$. Then,
\begin{align}
\sum_{j=1}^\g \int_{Q_0}^{\hat\mu_j} \ti\omega^{(3)}_{\Pinfp,\Pinfm}&=
\sum_{j=1}^\g \bigg(\int_{a_j} \ti\omega^{(3)}_{\Pinfp,\Pinfm}\bigg)
\bigg(\sum_{k=1}^\g \int_{Q_0}^{\hat\mu_k} \omega_j -\sum_{k=1}^\g
\int_{a_k} 
\big(\ul{\hatt A}_{Q_0}\big)_j \omega_k\bigg) \no \\
& \quad + \ln\bigg(\f{\theta(\ul {\hat z}(\Pinfp,\ul{\hat\mu})}
{\theta(\ul {\hat z}(\Pinfm,\ul{\hat\mu})}\bigg) \lb{g.61c}
\end{align} 
and 
\begin{align}
\Psi_{\g+1-k}(\ul\mu)&= \Psi_{\g+1-k}(\ul\lambda) -\sum_{j=1}^\g c_j(k)
\f{\partial}{\partial w_j} 
\ln\bigg(\f{\theta(\ul z(\Pinfp,\humu)+\ul w)}
{\theta(\ul z(\Pinfm,\humu)+\ul w)}\bigg)\bigg|_{\ul w=0}, \lb{g.62} \\
& \hspace*{7.2cm} k=1,\dots,\g, \no
\end{align}
with $\ul\lambda=(\lambda_1,\dots,\lambda_n)\in\bbC^n$ introduced in 
\eqref{a37a}.
\end{theorem}
\begin{proof}
Let $\calD_{\ul {\hat \mu}}\in\sigma^n\hatt\calK_n$ be a nonspecial divisor on
$\hatt\calK_\g$, $\ul{\hat \mu}=\{{\hat
\mu}_1,\dots,{\hat \mu}_\g\}\in\sigma^\g\hatt\calK_\g$. Introducing 
\begin{equation}
\ti\Omega^{(3)}(P)=\int_{Q_0}^P \ti\omega^{(3)}_{\Pinfp,\Pinfm}, 
\quad P\in\calK_\g\setminus\{\Pinfp,\Pinfm\}, \lb{g.63}
\end{equation}
we can render $\ti\Omega^{(3)}(\dott)$ single-valued on 
\begin{equation}
\hatt{\hatt\calK}_\g=\hatt\calK_\g\setminus\Sigma, \lb{g.64}
\end{equation}
where $\Sigma$ denotes
the union of cuts
\begin{equation}
\Sigma=\Sigma (\Pinfp)\cup\Sigma(\Pinfm), \quad 
\Sigma (\Pinfp)\cap \Sigma (\Pinfm)=\{Q_0\}, \lb{g.65}
\end{equation}
with $\Sigma (\Pinfp)$ (resp., $\Sigma (\Pinfm)$) a cut connecting $Q_0$
and $\Pinfp$ (resp., $\Pinfm$) through the open interior $\hatt\calK_\g$
(i.e., avoiding all curves $a_j,b_j,a_j^{-1},b_j^{-1}$, $j=1,\dots,\g$, 
with the exception of the point $Q_0\in\partial\hatt\calK_\g$), avoiding
the points ${\hat \mu}_j$, $j=1,\dots,\g$. The left and right side of the cut
$\Sigma(\Pinfpm)$ is denoted by $\Sigma(\Pinfpm)_\ell$ and
$\Sigma(\Pinfpm)_r$. The oriented boundary $\partial\hatt{\hatt\calK}_\g$
of $\hatt{\hatt\calK}_\g$, in obvious notation, is then given by 
\begin{equation}
\partial\hatt{\hatt\calK}_\g=\Sigma(\Pinfp)_\ell\cup\Sigma(\Pinfp)_r\cup 
\Sigma(\Pinfm)_\ell\cup\Sigma(\Pinfm)_r\cup\partial\hatt\calK_\g,
\lb{g.65a}
\end{equation}
that is, it consists of $\partial\hatt\calK_\g$ together
with the piece {}from $Q_0$ to $\Pinfp$ along the left side of the cut
$\Sigma(\Pinfp)$ and then back to
$Q_0$ along the right side of $\Sigma(\Pinfp)$, plus the corresponding
pieces {}from $Q_0$ to $\Pinfm$ and back to $Q_0$ along the cut
$\Sigma(\Pinfm)$, preserving orientation. Introducing the meromorphic
differential, 
\begin{equation}
\nu=d\ln(\theta(\ul z(\dott,\ul \mu)), \lb{g.66}
\end{equation}
the residue theorem applied to $\ti\Omega^{(3)}\nu$ yields
\begin{align}
\int_{\partial\hatt{\hatt\calK}_\g} \ti\Omega^{(3)}\nu&=
\sum_{j=1}^\g\bigg(\bigg(\int_{a_j}
\ti\omega^{(3)}_{\Pinfp,\Pinfm}\bigg)\bigg(\int_{b_j} \nu\bigg)-
\bigg(\int_{b_j}\ti\omega^{(3)}_{\Pinfp,\Pinfm}\bigg)\bigg(\int_{a_j}
\nu\bigg)\bigg) \no \\
& \quad + \int_\Sigma \ti\Omega^{(3)} \nu =2\pi i
\sum_{P\in\hatt{\hatt\calK}_\g} \res_{P}\big(\ti\Omega^{(3)}\nu\big).
\lb{g.67}
\end{align}
Investigating separately the items occurring in \eqref{g.67} then yields
the following facts: 
\begin{align}
&\sum_{P\in\hatt{\hatt\calK}_\g} \res_{P}\,(\ti\Omega^{(3)}\nu) 
=\sum_{j=1}^\g \ti\Omega^{(3)} ({\hat \mu}_j)=\sum_{j=1}^\g \int_{Q_0}^{\hat\mu_j}
\ti\omega^{(3)}_{\Pinfp,\Pinfm}, \lb{g.68} \\
&\int_{a_j} \nu =0, \quad j=1,\dots,\g, \lb{g.69} \\
&\int_{b_j} \nu =2\pi i\big(\big(\ul {\hatt\Xi}_{Q_0}\big)_j-
\big(\ul {\hatt A}_{Q_0}(R(a_j))\big)_j +\big(\ul {\hatt \alpha}_{Q_0}
(\calD_{\ul \mu})\big)_j\big) -i\pi\tau_{j,j}, \quad j=1,\dots,\g, \lb{g.70}
\end{align}
applying \eqref{aa51} in \eqref{g.69} and \eqref{g.70}. Here $R(a_j)$
denotes the end point of $a_j\subset\partial\hatt\calK_\g$,
$j=1,\dots,\g$. In addition, the cut $\Sigma$
produces the contribution  
\begin{equation}
\int_\Sigma \ti\Omega^{(3)} \nu=2\pi i \bigg(\int_{Q_0}^{\Pinfp} \nu -
\int_{Q_0}^{\Pinfm} \nu\bigg)= 2\pi i \int_{\Pinfm}^{\Pinfp} \nu 
=2\pi i \ln\bigg(\f{\theta(\ul{\hat z}(\Pinfp,\ul \mu))}
{\theta(\ul{\hat z}(\Pinfm,\ul \mu))}\bigg), \lb{g.71}
\end{equation}
since (by an application of the residue theorem) 
\begin{equation}
\ti\Omega^{(3)}({\hat \mu}_\ell)-\ti\Omega^{(3)}({\hat \mu}_r)=\pm 2\pi i, \quad
{\hat \mu}_\ell\in\Sigma(\Pinfpm)_\ell, \, {\hat \mu}_r\in\Sigma(\Pinfpm)_r,
\lb{g.71a}
\end{equation}
where ${\hat \mu}_\ell\in\Sigma(\Pinfpm)_\ell$ and ${\hat \mu}\in\Sigma(\Pinfpm)_r$
are on opposite sides of the cut $\Sigma(\Pinfpm)$. Recalling the well-known
results,
\begin{align}
\big(\ul{\hatt A}_{Q_0}(R(a_j))\big)_j&=\f{1}{2}+\int_{a_j} \big(\ul{\hatt
A}_{Q_0}\big)_j\omega_j, \quad j=1,\dots,\g, \lb{g.72} \\
\big(\ul {\hatt\Xi}_{Q_0}\big)_j &= \f{1}{2}(1+\tau_{j,j})- 
\sum_{\substack{k=1\\k\neq j}}^\g 
\int_{a_k}\big(\ul{\hatt A}_{Q_0}\big)_j\omega_k, \quad j=1,\dots,\g, 
\lb{g.73}
\end{align}
equations \eqref{g.67}--\eqref{g.73} imply
\begin{align}
\sum_{j=1}^\g \int_{Q_0}^{\hat\mu_j} \ti\omega^{(3)}_{\Pinfp,\Pinfm}&=
\sum_{j=1}^\g \bigg(\int_{a_j} \ti\omega^{(3)}_{\Pinfp,\Pinfm}\bigg)
\bigg(\sum_{k=1}^\g \int_{Q_0}^{\hat\mu_k} \omega_j -\sum_{k=1}^\g \int_{a_k} 
\big(\ul{\hatt A}_{Q_0}\big)_j \omega_k\bigg) \no \\
& \quad + \ln\bigg(\f{\theta(\ul {\hat z}(\Pinfp,\ul \mu))}
{\theta(\ul {\hat z}(\Pinfm,\ul \mu))}\bigg). \lb{g.74}
\end{align}
This proves \eqref{g.61c}. 

In the following we will apply \eqref{g.74} to $\hat\mu_j$, $j=1,\dots,\g$
satisfying the first-order system \eqref{g.30}, \eqref{g.30a} on some open
connected set $\calV$ such that $\mu_j$, $j=1,\dots,\g$, remain
distinct on $\calV$ and $\Phi^{(j)}_{\g-k}(\ul\mu)\neq 0$ on
$\calV$, $j,k=1,\dots,\g$.
Using \eqref{b26}, \eqref{g.30b}, and \eqref{A1c} one computes
\begin{align}
\f{\partial}{\partial v_k}
\big(\ul{\hatt\alpha}_{Q_0}(\calD_{\humu(\ul v)})\big)_j
&=\f{\partial}{\partial v_k}\sum_{\ell=1}^\g \int_{Q_0}^{\hat\mu_\ell(\ul
v)}\omega_j =\sum_{\ell,m=1}^\g \f{\partial}{\partial v_k}
\int_{Q_0}^{\hat\mu_\ell(\ul v)}c_j(m)\eta_m \no\\
&=\sum_{\ell,m=1}^\g c_j(m)\f{\mu_\ell(\ul v)^{m-1}}{y(\hat\mu_\ell(\ul
v))}\f{\partial}{\partial v_k}\mu_\ell(\ul v)\no \\
&=\sum_{\ell,m=1}^\g c_j(m) \Phi^{(\ell)}_{n-k}
\f{\mu_\ell(\ul v)^{m-1}}{\prod_{\substack{\ell'=1\\ \ell'\neq\ell}}^\g
(\mu_\ell(\ul v)-\mu_{\ell'}(\ul v))} \no\\
&=c_j(k), \quad \ul v \in\calV. \lb{g.76}
\end{align}
Thus, \eqref{g.36} and  \eqref{g.76} imply 
\begin{align}
\Psi_{\g+1-k}(\ul\mu(\ul v))&=-\f{\partial v_{\g+1}(\ul v)}{\partial v_k}=
-\sum_{j=1}^\g c_j(k)\bigg(\int_{a_j}
\ti\omega^{(3)}_{\Pinfp,\Pinfm}\bigg) \no \\
& \quad -\sum_{j=1}^\g c_j(k) 
\f{\partial}{\partial w_j} 
\ln\bigg(\f{\theta(\ul z(\Pinfp,\humu(\ul v))+\ul w)}
{\theta(\ul z(\Pinfm,\humu(\ul v))+\ul w)}\bigg)\bigg|_{\ul w=0},
\lb{g.77} \\
& \hspace*{4.65cm} \ul v\in\calV, \; k=1,\dots,\g. \no 
\end{align}
We replaced $\hat {\ul z}$ by $\ul z$ to arrive at 
\eqref{g.77} using properties \eqref{aa51} of $\theta$. If
$\hat\mu_j$, $j=1,\dots,\g$, are distinct and
$\Phi^{(j)}_{\g-k}(\ul\mu)\neq 0$, $j,k=1,\dots,\g$, we can choose 
$\hat\mu_j(\ul v_0)=\hat\mu_j$, $j=1,\dots,\g$, and obtain \eqref{g.62}.
The general case where $\calD_{\humu}$ is nonspecial, then follows {}from
\eqref{g.77} by continuity, choosing $\calV$ such that there exists a
sequence $\ul v_p\in\calV$ with $\ul{\hat\mu}(\ul v_p)\to\ul{\hat\mu}$ as
$p\to\infty$. Finally, invoking the normal differential of the third kind
in \eqref{a37a},
$\omega^{(3)}_{\Pinfp,\Pinfm} = \prod_{j=1}^n(z-\lambda_j)dz/y$,
corresponding to $\ti\omega^{(3)}_{\Pinfp,\Pinfm}=z^n dz/y$, a simple
computation, combining \eqref{b24}, \eqref{A.7}, \eqref{A.7a},
\eqref{a37a}, and the normalization $\int_{a_j}
\omega^{(3)}_{\Pinfp,\Pinfm}=0$, $j=1,\dots,n$, yields
\begin{equation}
\sum_{j=1}^n c_j(k)\bigg(\int_{a_j} \ti\omega^{(3)}_{\Pinfp,\Pinfm}\bigg)
=\Psi_{\g+1-k}(\ul\lambda), \quad k=1,\dots,\g. \lb{g.78}
\end{equation}
Equations \eqref{g.77} and \eqref{g.78} complete the proof of
\eqref{g.62}. 
\end{proof}

Formulas \eqref{g.36}, \eqref{g.61c}, and \eqref{g.62} (without explicit
proofs and without the explicit form of the constant terms on the
right-hand sides of \eqref{g.61c} and \eqref{g.62}) have been used in
\cite{Novikov:1999} in the course of deriving algebro-geometric solutions
of the Dym equation. Our approach based on the Dubrovin-type system
\eqref{g.30} appears to be new. It can easily be adapted to the case of 
KdV-type hyperelliptic curves branched at infinity (cf.\ 
\cite[App.\ B]{GesztesyHolden:2000}). Since solutions of $1+1$-dimensional 
soliton equations typically can be expressed in terms of trace formulas 
involving elementary symmetric functions of (projections of) auxiliary 
divisors, results of the type of \eqref{g.62} are of general interest in 
this context. 

{\bf Acknowledgments.}
We are indebted to Mark Alber, Darryl Holm, and Jerry Marsden for helpful 
comments and many hints regarding the literature. 


\begin{thebibliography}{10}

\bibitem{Alber:2000}
M.~S. Alber.
\newblock {$N$}-component integrable systems and geometric asymptotics. In:
  {\em {I}ntegrability: {T}he {S}eiberg-{W}itten and {W}hitham equations}
  (H.~W. Braden and I.~M. Krichever, editors), Gordon and Breach Science
  Publishers, Singapore, 2000, pp. 213--228.

\bibitem{AlberCamassaFedorovHolmMarsden:1999}
M.~S. Alber, R. Camassa, Yu.~N. Fedorov, D.~D. Holm, and J.~E. Marsden.
\newblock On billiard solutions of nonlinear {PDE}s.
\newblock Phys. Lett. A {\bf 264} (1999),  171--178.

\bibitem{AlberCamassaFedorovHolmMarsden:2001}
M.~S. Alber, R. Camassa, Yu.~N. Fedorov, D.~D. Holm, and J.~E. Marsden.
\newblock The complex geometry of weak piecewise smooth solutions of 
integrable nonlinear pde's of shallow water and Dym type.
\newblock arXiv:nlin.CD/0105025, Comm. Math. Phys., to appear.

\bibitem{AlberCamassaGekhtman:2000}
M.~S. Alber, R.~Camassa, and M.~Gekhtman.
\newblock Billiard weak solutions of nonlinear {PDE}'s and {T}oda flows. In:
  {\em {SIDE III}--{S}ymmetries and {I}ntegrability of {D}ifference
  {E}quations} (D.~Levi and O.~Ragnisco, editors), volume~25 of {\em {CRM}
  {P}roceedings and {L}ecture {N}otes}, {A}mer. {M}ath. {S}oc., {P}rovidence,
  {RI}, 2000, pp. 1--11.

\bibitem{AlberCamassaHolmMarsden:1994}
M.~S. Alber, R.~Camassa, D.~D. Holm, and J.~E. Marsden.
\newblock The geometry of peaked solitons and billiard solutions of a class of
  integrable {PDE}'s.
\newblock Lett. Math. Phys. {\bf 32} (1994),  137--151.

\bibitem{AlberCamassaHolmMarsden:1995}
M.~S. Alber, R.~Camassa, D.~D. Holm, and J.~E. Marsden.
\newblock On the link between umbilic geodesics and soliton solutions of
  nonlinear {PDE}'s.
\newblock Proc. Roy. Soc. London Ser. A {\bf 450} (1995),  677--692.

\bibitem{AlberFedorov:2000}
M.~S. Alber and Yu.~N. Fedorov.
\newblock Wave solutions of evolution equations and {H}amiltonian flows on
  nonlinear subvarieties of generalized {J}acobians.
\newblock J. Phys. A {\bf 33} (2000),  8409--8425.

\bibitem{AlberFedorov:2001}
M.~S. Alber and Yu.~N. Fedorov.
\newblock Algebraic geometrical solutions for certain evolution equations and
  {H}amiltonian flows on nonlinear subvarieties of generalized {J}acobians.
\newblock Inverse Problems {\bf 17} (2001), to appear.

\bibitem{AlberLutherMiller:2000}
M.~S. Alber, G.~G. Luther, and C.~A. Miller.
\newblock On soliton-type solutions of equations associated with
  {$N$}-component systems.
\newblock J. Math. Phys. {\bf 41} (2000),  284--316.

\bibitem{AlberMiller:2001}
M.~S. Alber and C.~Miller.
\newblock Peakon solitons of the shallow water equation.
\newblock Appl. Math. Lett {\bf 14} (2001),  93--98.

\bibitem{BealsSattingerSzmigielski:1998}
R.~Beals, D.~H. Sattinger, and J.~Szmigielski.
\newblock Acoustic scattering and the extended {K}orteweg-de {V}ries hierarchy.
\newblock Adv. in Math. {\bf 140} (1998),  190--206.

\bibitem{BealsSattingerSzmigielski:1999}
R.~Beals, D.~H. Sattinger, and J.~Szmigielski.
\newblock Multi-peakons and a theorem of {S}tieltjes.
\newblock Inverse Problems {\bf 15} (1999),  {L}1--{L}4.

\bibitem{BealsSattingerSzmigielski:2000}
R.~Beals, D.~H. Sattinger, and J.~Szmigielski.
\newblock Multipeakons and the classical moment problem.
\newblock Adv. in Math. {\bf 154} (2000),  229--257.

\bibitem{BealsSattingerSzmigielski:2001}
R.~Beals, D.~H. Sattinger, and J.~Szmigielski.
\newblock Peakons, strings, and the finite {T}oda lattice.
\newblock Comm. Pure Appl. Math. {\bf 54} (2001),  91--106.

\bibitem{BelokolosBobenkoEnolskiiItsMatveev:1994}
E.~D. Belokolos, A.~I. Bobenko, V.~Z. Enol'skii, A.~R. Its, and V.~B. Matveev.
\newblock {\em Algebro-{G}eometric {A}pproach to {N}onlinear {I}ntegrable
  {E}quations}.
\newblock Springer, Berlin, 1994.

\bibitem{BullaGesztesyHoldenTeschl:1997}
W.~Bulla, F.~Gesztesy, H.~Holden, and G.~Teschl.
\newblock Algebro-geometric quasi-periodic finite-gap solutions of the {T}oda
  and {K}ac-van {M}oerbeke hierarchy.
\newblock Mem. Amer. Math. Soc. {\bf 135} (1998), no.~641,  1--79.

\bibitem{CamassaHolm:1993}
R.~Camassa and D.~D. Holm.
\newblock An integrable shallow water equation with peaked solitons.
\newblock Phys. Rev. Lett. {\bf 71} (1993),  1661--1664.

\bibitem{CamassaHolmHyman:1994}
R.~Camassa, D.~D. Holm, and J.~M. Hyman.
\newblock A new integrable shallow water equation.
\newblock Adv. Appl. Mech. {\bf 31} (1994),  1--33.

\bibitem{ClebschGordan:1866}
A.~Clebsch and P.~Gordan.
\newblock {\em Theorie der {A}belschen {F}unktionen}.
\newblock Teubner, Leipzig, 1866.

\bibitem{Constantin:1997}
A.~Constantin.
\newblock On the {C}auchy problem for the periodic {C}amassa-{H}olm 
equation.
\newblock J. Differential Equations {\bf 141} (1997),  218--235.

\bibitem{Constantin:1998a}
A.~Constantin.
\newblock On the inverse spectral problem for the {C}amassa-{H}olm equation.
\newblock J. Funct. Anal. {\bf 155} (1998),  352--363.

\bibitem{Constantin:1998}
A.~Constantin.
\newblock Quasi-periodicity with respect to time of spatially periodic
  finite-gap solutions of the {C}amassa-{H}olm equation.
\newblock Bull. Sci. Math. {\bf 122} (1998),  487--494.

\bibitem{Constantin:2000}
A.~Constantin.
\newblock Existence of permanent and breaking waves for a shallow water
  equation: a geometric approach.
\newblock Ann. Inst. Fourier (Grenoble) {\bf 50} (2000),  321--362.

\bibitem{ConstantinEscher2:1998}
A.~Constantin and J.~Escher.
\newblock Global existence and blow-up for a shallow water equation.
\newblock Ann. Scuola Norm. Sup. Pisa Cl. Sci.(4) {\bf 26} (1998),  303--328.

\bibitem{ConstantinEscher3:1998}
A.~Constantin and J.~Escher.
\newblock Global weak solutions for a shallow water equation.
\newblock Indiana Univ. Math. J. {\bf 47} (1998),  1527--1545.

\bibitem{ConstantinEscher1:1998}
A.~Constantin and J.~Escher.
\newblock Wave breaking for nonlinear nonlocal shallow water equations.
\newblock Acta Math. {\bf 181} (1998),  229--243.

\bibitem{ConstantinEscher:1998}
A.~Constantin and J.~Escher.
\newblock Well-posedness, global existence, and blow-up phenomena for a
  periodic quasi-linear hyperbolic equation.
\newblock Comm. Pure Appl. Math. {\bf 51} (1998),  475--504.

\bibitem{ConstantinEscher:2000}
A.~Constantin and J.~Escher.
\newblock On the blow-up rate and the blow-up set of breaking waves for a
  shallow water equation.
\newblock Math. Z. {\bf 233} (2000),  75--91.

\bibitem{ConstantinMcKean:1999}
A.~Constantin and H.~P. McKean.
\newblock A shallow water equation on the circle.
\newblock Comm. Pure Appl. Math. {\bf 52} (1999),  949--982.

\bibitem{ConstantinMolinet:2000}
A.~Constantin and L.~Molinet.
\newblock Global weak solutions for a shallow water equation.
\newblock Comm. Math. Phys. {\bf 211} (2000),  45--61.

\bibitem{DicksonGesztesyUnterkofler1:2000}
R.~Dickson, F.~Gesztesy, and K.~Unterkofler.
\newblock Algebro-geometric solutions of the {B}oussinesq hierarchy.
\newblock Rev. Math. Phys. {\bf 11} (1999),  823--879.

\bibitem{DicksonGesztesyUnterkofler:2000}
R.~Dickson, F.~Gesztesy, and K.~Unterkofler.
\newblock A new approach to the {B}oussinesq hierarchy.
\newblock Math. Nachr. {\bf 198} (1999),  51--108.

\bibitem{Dmitrieva:1993}
L.~A. Dmitrieva.
\newblock Finite-gap solutions of the {H}arry {D}ym equation.
\newblock Phys. Lett. A {\bf 182} (1993),  65--70.

\bibitem{DullinGottwaldHolm:2001}
H.~R.~Dullin, G.~Gottwald, and  D.~D.~Holm.
\newblock An integrable shallow water equation with linear and nonlinear 
dispersion.
\newblock Preprint, arXiv:nlin.CD/0104004.

\bibitem{EnolskiiGesztesyHolden:1999}
V.~Z. Enolskii, F.~Gesztesy, and H.~Holden.
\newblock The classical massive {T}hirring system revisited. In: {\em
  {S}tochastic {P}rocesses, {P}hysics and {G}eometry: {N}ew {I}nterplays. {I}:
  {A} {V}olume in {H}onor of {S}ergio {A}lbeverio} ({F}. {G}esztesy, {H}.
  {H}olden, {J}. {J}ost, {S}. {P}aycha, {M}.{R}{\"o}ckner, and {S}.
  {S}carlatti, editors), volume~28 of {\em CMS Conference Proceedings}, Amer.
  Math. Soc., Providence, RI, 2000, pp. 163--200.

\bibitem{FarkasKra:1992}
H.~M. Farkas and I.~Kra.
\newblock {\em Riemann {S}urfaces}, second edition.
\newblock Springer, New {Y}ork, 1992.

\bibitem{Fedorov:1999}
Yu. Fedorov.
\newblock Classical integrable systems and billiards related to generalized
  {J}acobians.
\newblock Acta Appl. Math. {\bf 55} (1999),  251--301.

\bibitem{FisherSchiff:1999}
M.~Fisher and J.~Schiff.
\newblock The {C}amassa {H}olm equation: conserved quantities and the initial
  value problem, preprint, 1999.

\bibitem{FoiasHolmTiti:2001}
C.~Foias, D.~D.~Holm, and E.~S.~Titi.
\newblock The three dimensional viscous {C}amassa-{H}olm equations, and their
relation to the {N}avier-{S}tokes equations and turbulence theory.
\newblock Preprint, arXiv:nlin.CD/0103039.

\bibitem{FornbergWitham:1978}
B.~Fornberg and G.~B.~Witham.
\newblock A numerical and theoretical study of certain nonlinear wave phenomena.
\newblock Phil. Trans. Roy. Soc. London A {\bf 289} (1978), 373--404.

\bibitem{Fuchssteiner:1996}
B.~Fuchssteiner.
\newblock Some tricks from the symmetry-toolbox for nonlinear equations: 
generalizations of the {C}amassa-{H}olm equation.
\newblock Phys. D {\bf 95} (1996),  229--243.

\bibitem{FuchssteinerFokas:1981}
B.~Fuchssteiner and A.~S. Fokas.
\newblock Symplectic structures, their {B}{\"a}cklund transformations and
  hereditary symmetries.
\newblock Phys. D {\bf 4} (1981),  47--66.

\bibitem{GagnonHarnadWinternitzHurtubise:1985}
L.~Gagnon, J.~Harnad, P.~Winternitz, and J. Hurtubise.
\newblock Abelian integrals and the reduction method for an integrable
  {H}amiltonian system.
\newblock J. Math. Phys. {\bf 26} (1985),  1605--1612.

\bibitem{Gavrilov:1999}
L.~Gavrilov.
\newblock Generalized {J}acobians of spectral curves and completely integrable
  systems.
\newblock Math. Z. {\bf 230} (1999),  487--508.

\bibitem{GesztesyHolden:1999}
F.~Gesztesy and H.~Holden.
\newblock Dubrovin equations and integrable systems on hyperelliptic curves.
\newblock ArXiv:solv-int/9807006, Math. Scand., to appear.

\bibitem{GesztesyHolden:2000}
F.~Gesztesy and H.~Holden.
\newblock {\em Hierarchies of {S}oliton {E}quations and {T}heir
  {A}lgebro-{G}eometric {S}olutions. Vol. I: $(1+1)$-Dimensional 
Continuous Models}.
\newblock Cambridge Studies in Advanced Mathematics, Cambridge Univ. Press, 
in preparation.

\bibitem{GesztesyHolden2:1997}
F.~Gesztesy and H.~Holden.
\newblock A combined sine-{G}ordon and modified {K}orteweg--de {V}ries
  hierarchy and its algebro-geometric solutions. In: {\em {D}ifferential
  {E}quations and {M}athematical {P}hysics} ({R}. Weikard and {G}. Weinstein,
  editors), volume~16 of {\em Studies in {A}dvanced {M}athematics}, Amer. Math.
  Soc. and International Press, Providence and Boston, 2000, pp. 133--173.

\bibitem{GesztesyHolden:1999b}
F.~Gesztesy and H.~Holden.
\newblock Darboux-type transformations and hyperelliptic curves.
\newblock J. Reine Angew. Math. {\bf 527} (2000),  151--183.

\bibitem{GesztesyRatnaseelan:1996}
F.~Gesztesy and R.~Ratnaseelan.
\newblock An alternative approach to algebro-geometric solutions of the {AKNS}
  hierarchy.
\newblock Rev. Math. Phys. {\bf 10} (1998),  345--391.

\bibitem{GesztesyRatnaseelanTeschl:1996}
F.~Gesztesy, R.~Ratnaseelan, and G.~Teschl.
\newblock The {K}d{V} hierarchy and associated trace formulas. In: {\em Recent
  {D}evelopments in {O}perator {T}heory and {I}ts {A}pplications}
  (I.~{G}ohberg, {P}. {L}ancaster, and {P}.~{N}. {S}hivakumar, editors),
  volume~87 of {\em Operator {T}heory: {A}dvances and {A}pplications},
  Birkh\"auser, Basel, 1996, pp. 125--163.

\bibitem{MarsdenRatiuShkoller:2000}
J.~E.~Marsden, T.~S.~Ratiu, and S.~Shkoller.
\newblock The geometry and analysis of the averaged {E}uler equations and a new
diffeomorphism group.
\newblock Geom. Funct. Anal {\bf 10} (2000), 582--599.

\bibitem{MarsdenShkoller:2000}
J.~E.~Marsden and S.~Shkoller.
\newblock Global well-posedness for the Lagrangian averaged {N}avier-{S}tokes 
({LANS}-$\alpha$) equations on bounded domains.
\newblock Preprint, 2000.

\bibitem{Misiolek:1998}
G.~Misio{\l}ek.
\newblock A shallow water equation as a geodesic flow on the {B}ott-{V}irasoro 
group.
\newblock J. Geom. Phys. {\bf 24} (1998),  203--208.

\bibitem{Mumford:1984}
D.~Mumford.
\newblock {\em Tata {L}ectures on {T}heta {II}}, volume~43 of {\em {P}rogress
  in {M}athematics}.
\newblock Birkh{\"a}user, Boston, 1984.

\bibitem{Novikov:1999}
D.~P. Novikov.
\newblock Algebraic-geometric solutions of the {H}arry {D}ym equation.
\newblock Siberian Math. J. {\bf 40} (1999),  136--140.

\bibitem{Schiff:1996}
J.~Schiff.
\newblock Zero curvature formulations of dual hierarchies.
\newblock J. Math. Phys. {\bf 37} (1996),  1928--1938.

\bibitem{Shkoller:1998}
S.~Shkoller.
\newblock Geometry and curvature of diffeomorphism groups with $H^1$ metric 
and mean hydrodynamics.
\newblock J. Funct. Anal. {\bf 160} (1998),  337--365.

\bibitem{Shkoller:2000}
S.~Shkoller.
\newblock On incompressible averaged {L}agrangian hydrodynamics. 
\newblock Preprint, arXiv:math.AP/9908109.

\bibitem{Toda:1989}
M.~Toda.
\newblock {\em Theory of {N}onlinear {L}attices}, second edition, volume~20 of
  {\em Solid-{S}tate {S}ciences}.
\newblock Springer, Berlin, 1989.

\bibitem{XinZhang:2000}
Z.~Xin and P.~Zhang.
\newblock On the weak solutions to a shallow water equation.
\newblock Comm. Pure Appl. Math. {\bf 53} (2000),  1411--1433.

\end{thebibliography}

\end{document}